\documentclass[aps,twocolumn,nofootinbib,preprintnumbers,superscriptaddress]{revtex4-1}

\usepackage{psfrag}
\usepackage{mathrsfs}

\usepackage{amsfonts}
\usepackage{amsmath}
\usepackage{empheq}		
\usepackage{cancel}     
\usepackage[toc,page]{appendix}

\usepackage{amssymb}
\usepackage{graphicx}
\usepackage{epsfig}
\usepackage{float}
\usepackage{color,array,subfigure}
\usepackage{epstopdf}
\usepackage{esint}

\usepackage{geometry}
\RequirePackage{CJKutf8}

\usepackage[colorlinks=true,  
            citecolor=red,
            urlcolor= blue,                     %
            filecolor=black,
            linktocpage=true,  
            linkcolor=blue,
             ]{hyperref}

\newcommand{\beq}{\begin{equation}}
\newcommand{\eeq}{\end{equation}}
\newcommand{\bq}{\begin{equation}}
\newcommand{\eq}{\end{equation}}
\newcommand{\ba}{\begin{array}}
\newcommand{\ea}{\end{array}}
\newcommand{\beqa}{\begin{eqnarray}}
\newcommand{\eeqa}{\end{eqnarray}}
\newcommand{\bsel}{{\begin{subequations}\begin{empheq}}}
\newcommand{\bse}{{\begin{subequations}\begin{empheq}[left={\ii}\empheqlbrace]{align}}}
\newcommand{\ese}{{\end{empheq}\end{subequations}}}
\def\bc{\begin{center}}
\def\ec{\end{center}}
\def\bnum{\begin{enumerate} }
\def\enum{\end{enumerate}}
\def\nn{\nonumber}
\def\ii{\!\!\!\!\!\!}  
\def\3i{\!\!\!}
\def\2i{\!\!}

\def\ea{{e_a}}

\def\ec{{e_c}}

\def\log{\ln}

\def\nn{\nonumber}

\def\[{\left[}
\def\]{\right]}
\def\({\left(}
\def\){\right)}

\def\abs#1{\left|#1\right|}
\def\vev#1{\left\langle#1\right\rangle}
\def\>{\rightarrow}

\def\Diracslash#1{\not{\hbox{\kern-4pt $#1$}}}

\def\Dslash{\not{\hbox{\kern-4pt $D$}}}
\def\pslash{\not{\hbox{\kern-4pt $p$}}}
\def\qslash{\not{\hbox{\kern-4pt $q$}}}

\def\lv{\not{\hbox{\kern-4pt $L$}}}
\def\lsim{\mathrel{\raise.3ex\hbox{$<$\kern-.75em\lower1ex\hbox{$\sim$}}}}
\def\gsim{\mathrel{\raise.3ex\hbox{$>$\kern-.75em\lower1ex\hbox{$\sim$}}}}
\def\ifmath#1{\relax\ifmmode #1\else $#1$\fi}

\evensidemargin=\oddsidemargin

\numberwithin{equation}{section}

\def\V{\mathcal{V}}
\def\W{\mathcal{W}}
\def\X{\mathcal{X}}
\def\Y{\mathcal{Y}}
\def\Z{\mathcal{Z}}

\begin{document}
\begin{CJK}{UTF8}{gbsn}

\begin{titlepage}
\begin{flushright}
\end{flushright}

\begin{center}
 \vspace*{10mm}

{\LARGE\bf
Schwarzian correction to quantum correlation in SYK model
}\\
\medskip
\bigskip\vspace{0.6cm}
{\large {Yong-Hui Qi$^{\dag,\star}$,}  \,{Yunseok Seo$^{\dag,\ast}$,} \, {Sang-Jin Sin$^{\dag}$,} \, {Geunho Song$^{\dag}$}
 }
\\[7mm]
{\it
$^{\dag}$Department of Physics, Hanyang University, Seoul, 04763, Korea\\
$^{\star}$Center for High Energy Physics, Peking University, Beijing, 100871, Peoples Republic of China\\
$^{\ast}$GIST College, Gwangju Institute of Science and Technology, Gwangju 500-712, Korea
}

\vspace*{0.3cm}
 {\tt yhqi@pku.edu.cn, 
     ~yseo@gist.ac.kr, 
     ~sjsin@hanyang.ac.kr,
     ~sgh8774@gmail.com
 }
\bigskip\bigskip\bigskip

{
\centerline{\large\bf Abstract}
\begin{quote}
We study a class of SYK-type models in large N limit from the gravity dual side  in terms of Schwarzian action analytically. The quantum correction to two point correlation function due to the Schwarzian action produces transfer of degree of freedom from the quasiparticle peak to Hubbard band in density of states (DOS), a signature strong correlation. In Schwinger-Keldysh (SK) formalism, we calculate higher point thermal out-of-time order correlation (OTOC) functions, which indicate quantum chaos by having Lyapunov exponent. Higher order local spin-spin correlations are also calculated, which can be related to the dynamical local susceptibility of quantum liquids such as spin glasses, disordered metals.
\bigskip \\
\end{quote}}
\end{center}
\end{titlepage}


\tableofcontents


\section{Introduction}

Recently,   Sachdev-Ye-Kitaev (SYK) model~\cite{Kitaev:2015,Sachdev:1992fk} attracted a lot of   interests  ~\cite{Polchinski:2016xgd,Jevicki:2016bwu,Maldacena:2016hyu,Jensen:2016pah,Maldacena:2016upp,Engelsoy:2016xyb,Bagrets:2016cdf,Cvetic:2016eiv,Gu:2016oyy,Gross:2016kjj,Witten:2016iux,Cotler:2016fpe,Berkooz:2016cvq,Davison:2016ngz,Turiaci:2017zwd,Garcia-Garcia:2017pzl,Bi:2017yvx,Mandal:2017thl,Gross:2017hcz,Stanford:2017thb,Cai:2017nwk,Mertens:2017mtv,Chen:2017dbb,Gross:2017vhb,Peng:2017kro,Das:2017pif,Gu:2017njx,Gross:2017aos,Jian:2017unn,Kitaev:2017awl,Haehl:2017pak,Wu:2018vua,Gaikwad:2018dfc}. It is a $(0+1)$-dimensional quantum mechanical system composed of $N$ Majorana fermions, with a random, all-to-all quartic interaction.

There are three novel features of SYK model. The first one is the solvability at large N in the strong coupling limit. The second one is the emergence of conformal symmetry at IR limit, as well as its spontaneous  breaking which results in soft modes as the pseudo Nambu-Goldstone bosons (pNGBs)  ~\cite{Maldacena:2016upp,Stanford:2017thb,Kitaev:2017awl}: we know it is dynamically broken since the Virasoro symmetry of the boundary is broken to SL(2,R) in its bulk dual. The third one is the quantum chaos behavior  in $4$-point correlation functions.

The Schwarzian action is determined by the pattern of spontaneous breaking of reparametrization symmetry~\cite{Maldacena:2016upp}. It has been conjectured that the gravity dual of the SYK model can be described by a $2$-D dilaton gravity~\cite{Jensen:2016pah}: one example is the Jakiew-Teitelboim (JT) model \cite{Jackiw:1984je,Teitelboim:1983ux}~\cite{Jensen:2016pah,Maldacena:2016upp}. Another example is the Almheiri-Polchinsky (AP) model\cite{Almheiri:2014cka}~\cite{Engelsoy:2016xyb}. The low energy quantum description of SYK model is proposed to be holographically dual to a $(1+1)$-dimensional model of black hole~\cite{Maldacena:2016hyu,Stanford:2017thb}, although it is not completely conventional AdS/CFT.

\emph{Higher dimensional generalization of SYK model}--
There are several way to generalize the SYK model, one is its generalization in higher dimensional spacetime~\cite{Gu:2016oyy,Berkooz:2016cvq,Davison:2016ngz,Das:2017pif,Jian:2017unn}. It has been argued that the SYK model has a $3$-D interpretation of a bulk massive scalar, which is subject to a delta function potential at the center of the interval along the extra/third dimension being a dilaton~\cite{Das:2017pif}. After standard $S^1/Z_2$ compactification, an infinite series of Kaluza-Klein (KK) tower~\cite{Perelstein:2010yd} mass spectrum of a bulk massive scalar exactly matches the strong coupling ($J\to \infty$) spectrum of the SYK model~\cite{Das:2017pif}, while higher point correlation functions might still not be matching~\cite{Gross:2017aos}. To be specific, it has been shown that the asymptotic three-point function of SYK model has a string-like bulk interaction~\cite{Gross:2017hcz}. It turns out that the interaction grows much faster than that of massive KK scalar with cubic coupling, which is overlaps of the wavefunctions along the $S^1$ in a AdS$_3\simeq$ AdS$_2\times S^1$ bulk~\cite{Gross:2017aos}.


As a result, the KK description from bulk might be broken down in dealing with strongly correlated system.

It has also been claimed that the bulk dual to the soft mode sector of SYK model might be realized through the KK reduction from $3$-D Maxwell-Einstein gravity~\cite{Cvetic:2016eiv,Gaikwad:2018dfc}, which leads to the $2$-D JT model, where the dilaton is proportional to the KK radius. However, it is still insufficient to confirm whether there is a local bulk dual for the full SYK model.


\emph{Flavor symmetry generalization of SYK model}--
An alternative generalization is one in flavor symmetry spacetime, which generalize the four-majorana fermion interactions of SYK-like model to fermion with $p$-body interactions~\cite{Gross:2016kjj}. Since in $(0+1)$ dimensional SYK model, the Majorana field $\psi$ is dimensionless ($[\psi]=0$), the coupling $J$ is always dimensionful ($[J]=1$). Thus, the $p \ge 4$-fermion interaction is always \emph{relevant} at UV. While in general this is not true in $D\ge 2$, since the UV relevance of the $p$-fermions interactions depends on the dimension of the spacetime. For example, in a generalization of $(1+1)$-dimensional SYK model, in analogy to the $2$-D Gross-Neveu model, it can be obtained by integrating out the tensor field that coupled with a new vector field~\cite{Peng:2017kro}. In this case, the fermion field is dimensionful ($[\psi]=1/2$), meanwhile the coupling is dimensionless ($[J]=0$). Thus, the four fermion interaction term is \emph{marginal}. Generally speaking, the generalization to more higher dimension ($D \ge 3$), will inevitably lead to \emph{irrelevant} $p$-fermion interactions.

Therefore, it might not be necessary to stick to the original SYK model~\cite{Wu:2018vua}, but instead focus on a general class of SYK-like models in $(0+1)$-dimensional spacetime with three most notable universal features: solvability in the IR and large N limit, maximal chaotic behavior and emergent conformal theory spontaneously broken at IR. Few studies, however, have examined the effects of $1$-D Schwarzian correction to the quantum correlations functions
and none, to our knowledge, have compared the DOS of quantum liquid such as spin glasses, non-Fermi liquid.

In this paper, motivated by the novel features of a class of quantum spin glass or disordered metals depicted by SYK-like models~\cite{Grempel:1998,Grempel:1999,Grempel:2001}, we study a strongly interacting $(0+1)$-dimensional quantum mechanical model at large N limit, whose effective action is Schwarzian one. As it will be shown, the model owns a Hubbard band in the spectral function by transferring the degree of freedoms from quasi particle peak to side band, a Hall mark of the strongly interacting systems. Such features are observed in the spectral functions of several fcc A$_3$C$_{60}$ system as well as in the transition metal Oxides. We compare with the density functional theory (DFT)~\cite{Hohenberg:1964zz} DOS results as Fig.4B in Ref.~\cite{Nomura:2015}. The interesting point is that our analytical results on the $0$ dimensional system quantitatively reproduces the density of state of the dynamical mean-field theory (DMFT)~\cite{Georges:1996zz} approach.

The paper is organized as follow.
In Sec.\ref{sec:Schwarzian_CFs}, we study the correlation functions from Schwarzian action at both zero and finite temperature. In Sec.\ref{sec:Schwarzian_LQ}, we study the zero temperature and thermal retarded Green's functions with loop correction from pNGBs, and local dynamical susceptibility of quantum liquid. The higher order local spin-spin correlation function beyond local susceptibility are also investigated. Generalization of AdS$_2$ spacetime as near IR horizon of RN black hole in AdS$_{d+1}$ spacetime is studied in Appendix.\ref{app:high-D-generalization}. In Sec.\ref{sec:hpt-corr}, we study higher point correlation functions, the thermal OTOCs functions in SK formalism.

\section{Correlation functions from Schwarzian}
\label{sec:Schwarzian_CFs}

In this section, based on the low energy effective action of the Schwarzian theory of time reparametrization from $2$-D gravity, we calculate the correlation functions, especially the $2$-point one~\cite{Witten:1998qj}.

\subsection{The action of the model}

The prototypical SYK model is described by the partition function $Z(J)=\int D\psi_i e^{-S}$, with an action
\beqa
S = \int d\tau \bigg(  \frac{1}{2}\sum_{i=1}^N  \chi_i \partial_\tau \chi_i - \frac{1}{4!} \sum_{i,j,k,l} J_{ijkl} \chi^i \chi^j \chi^k \chi^l  \bigg), \qquad
\eeqa
where $\chi^i$ are $N$ Majorana fermions, satisfying $\{\chi_i, \chi_j\}=\delta_{ij}$, interacting with random interactions involving $4$ fermions at a time. $J_{ijkl} $ is a Gaussian random infinite-range exchange interaction of all-to-all quartic coupling, which are mutually uncorrelated and satisfies the Gaussian's probability distribution function $P(J_{ijkl})\sim \exp{(- N^3 J_{ijkl}^2/12J^2 )}$, which leads to zero mean ${\mathbb E}[{J_{ijkl}}]=0$ and variance ${\mathbb E}[{J_{ijkl}^2}]=3!J^2/N^3$ with width of order $J/N^{3/2}$, respectively. The ${\mathbb E}[\cdots]$ denotes an average over disorder. The $J$ is the only one effective coupling after the disorder averaging for the random coupling $J_{ijkl}$. The random couplings $J_{ijkl}$ represents disorder, and does not correspond to a unitary quantum mechanics~\cite{Witten:2016iux,Kitaev:2017awl}. For euclidean time $\tau=it$, the model can be viewed alternatively as a $1$-dimensional statistical model of Majorana fermions.

\subsection{Effective action of gravity and soft mode}

In the linearized theory of the boundary action of $2$-D gravity, i.e., $1$-D effective action as
\beqa
S_{\text{eff}} = - C_g  \int dt \phi_r(t) \text{Sch}(f(t),t),  \label{Eq:S_eff}
\eeqa
$C_g$ is a constant depending on the bulk gravity parameter, $t$ is the boundary time coordinate, $f(t)$ is the field variable, $\phi_r(t)$ is the normalizable part of the dilaton, which is a constant on the cutoff boundary and plays a role of external coupling, while the divergent part that blows up at boundary is absorbed by a counter term, and can be identified as the source in NAdS$_2$/NCFT$_1$ description, Sch$(f,t)$ is the Schwarzian derivative defined as
\beqa
\text{Sch}(f,t) = \bigg(\frac{f^{\prime\prime}}{f^\prime}\bigg)^\prime - \frac{1}{2} \bigg( \frac{f^{\prime\prime}}{f^\prime}\bigg)^2 = \frac{f^{\prime\prime\prime}}{f^\prime} - \frac{3}{2} \bigg( \frac{f^{\prime\prime}}{f^\prime}\bigg)^2 ,  \qquad \label{Eq:Sch(f,t)}
\eeqa
where the prime $\prime$ denotes the derivative with respect to $t$. The zero modes is described by the Schwarzian action. The effective action has a global SL$(2)$ invariance, which is obvious that by noticing Sch$((af+b)/(cf+d),t)=$Sch$(f,t)$. By doing variation with respect to $f(t)$, the action becomes
\beqa
\delta S_{\text{eff}} \sim  - C_g \int \phi_r(t) dt \frac{[\text{Sch}(f(t),t)]'}{f'} \delta f,
\eeqa
and by using the property as
\beqa
\frac{[\text{Sch}(f,t)]'}{f'} = \bigg[\frac{1}{f'}\bigg(\frac{( f')'}{f'}\bigg)'\bigg]' ,
\eeqa
and that $\delta (f')^{-1} = - (f')^{-2}\delta f'$, one obtains the field equation of motion with respect to $t(s)$ turns out to be
\beqa
\bigg[ \frac{1}{f'}  \bigg(  \frac{(f'\phi_r)'}{f'}  \bigg) \bigg]' = 0 ,
\eeqa
which becomes $[\text{Sch}(f,t)]'/t'=0$ when $\phi_r$ is a constant. One of most simplest but non-trivial solution might be a non-constant functions with constant Schwarzian.

\subsubsection{Zero temperature soft mode propagator}

Consider a linear transformation $f(\tau)= \tau(t) $, then according to the composition rule of Schwarzian derivative as
\beqa
\text{Sch}( g(f) , t ) = f'^2 \text{Sch}(g(f),f) + \text{Sch}(f,t) ,  \label{Eq:composition-rule}
\eeqa
where $g(f)=g(f(t))$ and $f=f(t)$, one has $\text{Sch}(f,t) = \tau'^2 \text{Sch}(f,\tau) + \text{Sch}(\tau,t)$, where $\text{Sch}(f,\tau)=1/2$, when $\tau$ is a linear function of $t$, Sch$(t,s)$ is constant and satisfies the equation of motion of the Schwarzian action, i.e., Sch$(\tau,t)'/f'=0$. In the perturbative approach, one can set
\beqa
\tau(t) \equiv t + \epsilon k(t),  \label{Eq:ft-tau-T=0}
\eeqa
where $t=i {t}$ is imaginary time and $\epsilon \ll 1$ is the expansion parameter, which can be chosen as the bulk gravitational interaction coupling, i.e., $\epsilon = \kappa_N \sim \sqrt{G_N} $, which is proportional to $G_N^{1/2}$ in gravity or large $N^{-1/2}$ as in the SYK model. By expanding the Schwarzian action, one has
\beqa
\text{Sch}(f,t) =  \epsilon  k'''   + \epsilon^2\bigg( - \frac{3}{2} k''^2 - (k'' k')'  \bigg) + O (\epsilon^3) . \qquad
\eeqa
By dropping total derivative term, the leading order action at $\epsilon^2$ order reads as
\beqa
S_{\text{eff}} = \frac{3C}{2}   \epsilon^2 \int dt   k''^2  ,
\eeqa
where $C\propto \bar\phi_r$. By doing a Fourier transformation $k(t)=\sum_{n }k_n e^{ i n t}$,
the action becomes
\beqa
S_{\text{eff}} = \frac{3C}{2}  \int dt   n^4  k_n k_{-n}, \label{Seff:Sch-T=0}
\eeqa
where hence and forth, the Einstein summation notation convention is applied for repeated $n\in {\mathbb Z}$.
Thus, the soft mode propagator can be obtained
\beqa
&& \epsilon^2 \vev{k(t_1)k(t_2)} = \frac{1}{6\pi C} \sum_{n\ne 0} \frac{e^{in t}}{n^4} \nn\\
&& = \frac{1}{6\pi C}\bigg(  \frac{1}{24}(\abs{t}  - \pi )^4 - \frac{\pi^2}{12}(\abs{t} - \pi)^2 + \frac{7\pi^4}{360}   \bigg), \nn\\
&& 
\epsilon^2\vev{k(t_1)k'(t_2)} =  \frac{1}{36 \pi C} \text{sgn}(t)  ( \pi - \abs{t} )(2\pi - \abs{t}) \abs{t}, \nn\\
&& \epsilon^2\vev{k'(t_1)k'(t_2)} = \frac{1}{6\pi C} \bigg( - \frac{\pi^2}{6} + \frac{1}{2} (\pi - \abs{t})^2 \bigg) ,  \label{Eq:k12-T=0}
\eeqa
where $ t \equiv t_1 - t_2 $ and we have used that
\beqa
\text{Li}_4(z) + \text{Li}_4(\frac{1}{z}) &=& - \frac{1}{4!}(2\pi i)^4 B_4(1+\frac{i}{2\pi}\log{z}) \nn\\
&=&  -\frac{z^4}{24}+\frac{\pi  z^3}{6}-\frac{\pi ^2 z^2}{6}+\frac{\pi ^4}{45},  \qquad
\eeqa
where $z = e^{it}$. $\text{Li}_n(x)$ is the the polylogarithmic function defined by the series Li$_{n}(x)=\sum_{k=1}^n z^k/k^n$ for $\abs{z}<1$ and $B_n(x)$ is the Bernoulli polynomial.

\subsubsection{Thermal soft mode propagator}

Consider a thermal circle transformation $f(\tau)=\tan(\pi \tau/\beta) $ satisfying $f(\tau+\beta)=f(\tau)$ with period length $\beta=2\pi$, in other words, by imposing a mapping from
\beqa
f(t)=\tan\frac{\tau(t)}{2}, \quad \tau(t) = i\frac{2\pi}{\beta}t \equiv i t, \label{Eq:ft-tau-T!=0}
\eeqa
then the composition rule of Schwarzian derivative in Eq.(\ref{Eq:composition-rule}) leads to an action results as $\text{Sch}(f,t)=\tau'^2/2+\text{Sch}(\tau,t)$, when $\tau$ is a linear function of $t$, Sch$(t,s)$ is constant and satisfies the equation of motion of the Schwarzian action, i.e., Sch$(f,t)'/f'=0$. This can be traced back to the bulk equation of motion, which gives the dilaton solution.

In the perturbative approach as in Eq.(\ref{Eq:ft-tau-T=0}),
after expanding the effective Schwarzian action, higher order self-interaction terms for $k(\tau(t))$ are present, which is suppressed by factor of $\epsilon$. By expanding the Schwarzian action, one has
\beqa
\text{Sch}(f,t) 
&=& \frac{1}{2} + \epsilon ( k' + k'''  ) + \epsilon^2\bigg( \frac{1}{2} k'^2 - \frac{1}{2} k''^2 - (k'' k')'  \bigg) \nn\\
&+& O (\epsilon^3 ) .
\eeqa
By dropping total derivative term, the leading order action at $\epsilon^2$ order reads as
\beqa
S_{\text{eff}} = \frac{C}{2}   \epsilon^2 \int dt (  k''^2 - k'^2  ) ,
\eeqa
where $C\propto C_g \bar\phi_r$. By doing a Fourier transformation as $k(t)=\sum_{n}k_n e^{ i n t}$ where $t\in[0,2\pi]$, the action becomes
\beqa
S_{\text{eff}} = \frac{C}{2}  \epsilon^2 \int dt (  n^4 - n^2  ) k_n k_{-n}, \label{Seff:Sch-T!=0}
\eeqa
where for repeated $n$ the Einstein summation convention is applied as stated before.
Thus, the soft mode propagator can be obtained as
\beqa
&& \epsilon^2\vev{k(t_1)k(t_2)} = \frac{1}{2\pi C} \sum_{n\ne 0, \pm 1} \frac{e^{in t}}{n^2(n^2-1)} \nn\\
&&= \frac{1}{2\pi C}\bigg(  1 + \frac{\pi^2}{6} - \frac{(\abs{t}-\pi)^2}{2} + (\abs{t}-\pi) \sin{\abs{t}} + \frac{5}{2} \cos{\abs{t}}   \bigg), \nn\\
&& 
\epsilon^2\vev{k(t_1)k'(t_2)} =  \frac{\text{sgn}(t)}{2\pi C}\bigg( ( \pi - \abs{t} ) ( 1 - \cos{\abs{t}} )  - \frac{3}{2} \sin{\abs{t}}  \bigg), \nn\\
&& \epsilon^2\vev{k'(t_1)k'(t_2)} = \frac{1}{2\pi C} \bigg( 1 + \frac{1}{2}\cos{\abs{t}} - (\pi - \abs{t}) \sin{\abs{t}} \bigg) ,  \ii \nn\\
&& \label{Eq:k12-T!=0}
\eeqa
where $ t \equiv t_1 - t_2 $ and we have used that
\beqa
\text{Li}_2(z) + \text{Li}_2(\frac{1}{z}) &=& - \frac{1}{2}(2\pi i)^2 B_2(1+\frac{i}{2\pi}\log{z}) \nn\\
&=& \frac{\pi^2}{3} - \pi z + \frac{z^2}{2} , \nn
\eeqa
where $z = e^{it}$.

\subsection{Effective action of matter and two point correlation functions}

The $n$-point function of a matter field, e.g., a scalar $\Phi$ in NAdS$_2$, can be computed by coupling the matter field to the bulk gravity in AdS$_2$, and then rewriting the action by using $f(t)$, and rescaling by a factor $f'(t)^\Delta$ at the insertion of each operator.

For a massive scalar $\Phi$ in AdS$_2$ spacetime in Poincar\'{e} coordinate, since all gravitational configurations in $2$-dimentional spacetime can be described by the metric, and the effective action is~\cite{Maldacena:2016upp}
\beqa
S_{\chi} &=&  \frac{1}{2} \int d^2x \sqrt{-g} [ (\nabla \Phi )^2 + m^2 \Phi^2 ] \nn\\
&=& - N \int dt dt' \frac{\Phi_0(t) \Phi_0(t')}{\abs{t-t'}^{2\Delta}} + \cdots  , \quad
\eeqa
where $ N = {(\Delta-1/2)\Gamma[\Delta]}/{[\sqrt\pi \Gamma(\Delta - \frac{1}{2})]} $ and we have used the asymptotic behavior of $\Phi$ at boundary
\beqa
\Phi(t,z) =  z^{1 - \Delta} \Phi_0(t)  + \cdots  \quad z \to 0,
\eeqa
where $\Phi_0(t)$ can be viewed as a source for a scalar operator with conformal dimension $\Delta$, e.g., for free scalar in pure AdS$_2$/CFT$_1$ case,  $ \Delta  = {1}/{2} + \sqrt{{1}/{4} + m^2} \ge 1$. Consider that the trajectory of the boundary curve is $f(t)$, which can be transformed to the desired boundary conditions as
\beqa
\Phi_0(t) = [f'(t)]^{1-\Delta} \Phi_0(f(t)).
\eeqa
Then, the effective action can be re-parameterized as
\beqa
\bar{S}_{\text{eff}} = -N \int dt dt' \bigg( \frac{f'(t) f'(t')}{[f(t)-f(t')]^2} \bigg)^{\Delta} \Phi_0(t) \Phi_0(t') . \qquad \quad \label{Eq:S-chi_eff}
\eeqa
The two point function of the dual field ${\mathcal O}(t)$ to the source $\Phi_0$ can be read as
\beqa
G(t,t') \equiv \vev{ {\mathcal O}(t) {\mathcal O}(t')  } \sim \bigg( \frac{f'(t) f'(t')}{[f(t)-f(t')]^2} \bigg)^{\Delta} .  \qquad \label{Eq:2pt-OO}
\eeqa

\subsection{Thermal correlation functions}

Consider expand the boundary $t$ around the saddle of a thermal circle, according to Eqs.(\ref{Eq:ft-tau-T!=0}) and (\ref{Eq:ft-tau-T=0}) as
\beqa
f(t) = \tan\frac{t+\epsilon k(t)}{2}, \quad \text{or} \quad f(t) = e^{i t + i \epsilon k(t) }, \qquad \label{Eq:PNG_2}
\eeqa
where we have dropped a common factor $2\pi/\beta$, so to recover one has to rescale $t \to 2\pi/\beta t$. The two point function of the dual operators can be expanded as
\beqa
&& G(t_1,t_2)  = \frac{ 1 +  \epsilon {\mathcal C}_1(t_{12})  + \epsilon^2 {\mathcal C}_2(t_{12}) + O(\epsilon^3)  }{\big(2\sin\frac{t_{12}}{2}\big)^{2\Delta} } ,  \qquad \qquad \nn\\
&=& G_0(t_1,t_2)[1 +  \epsilon {\mathcal C}_1(t_{12})  + \epsilon^2 {\mathcal C}_2(t_{12}) + O(\epsilon^3)], \qquad  \label{Eq:G12_C12}
\eeqa
where ${\mathcal C}_n(t_{ij})\equiv{\mathcal C}_n(t_i,t_j)$. By neglecting the perturbation expansion term, $\epsilon\to 0$, the AdS$_2$ thermal two point functions is recovered as
\beqa
G_0(t_1,t_2)  = \frac{1}{\big(2\sin\frac{t_{12}}{2}\big)^{2\Delta} }  =  \bigg(\frac{\pi}{\beta\sin\frac{\pi t_{12}}{\beta}}\bigg)^{2\Delta}  ,  \qquad \label{Eq:G12_C12-tree}
\eeqa
where in the last equality, we have recovered the thermal factor $t_{12} \to (2\pi) /\beta t_{12}$.
The leading correction to the thermal two point-Green's function can be expressed more explicitly as
\beqa
&&  {\mathcal C}_1(t_{12}) = \Delta \bigg(   k'(t_1) + k'(t_2) - \frac{k(t_1) - k(t_2) }{\tan\frac{t_{12}}{2}} \bigg), \nn\\
&&  {\mathcal C}_2(t_{12}) =\frac{1}{2} \Delta ^2 \left(  k'(t_1)+k'(t_2)+\frac{k(t_2)-k(t_1)}{\tan\frac{t_{12}}{2}} \right)^2 \nn\\
&& +\frac{1}{4} \Delta  \bigg(  \frac{[k(t_1)-k(t_2)]^2}{(\sin\frac{t_{12}}{2})^2} -2 [k'(t_1)^2+k'(t_2)^2]\bigg), \qquad \quad
\label{Eq:C12}
\eeqa
where for the brief ness, we do not recover the thermal factor. It will be obvious in the following that the higher order ${\mathcal C}_n(t_1, t_2)$ for $n\ge3$ will not contributes to the leading order of all point functions.


\begin{figure}[ht!]
\centering
   \subfigure[ $G(t_1,t_2)$]
   {\includegraphics[scale=0.5]{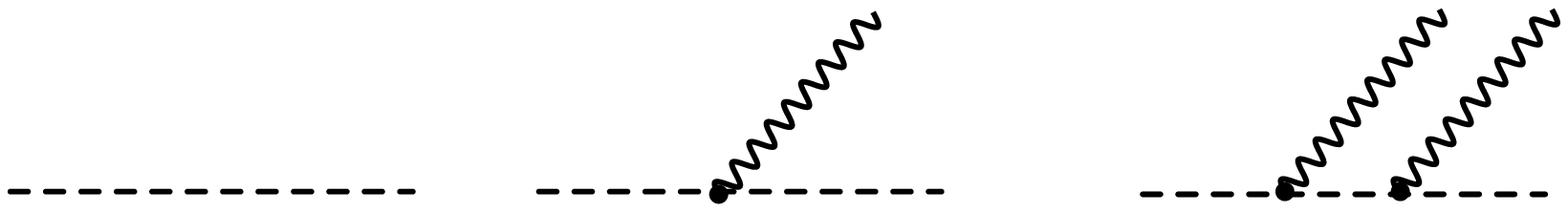}  }\qquad\qquad
   \subfigure[ $G_{2}(t_1,t_2)$]
   {\includegraphics[scale=0.5]{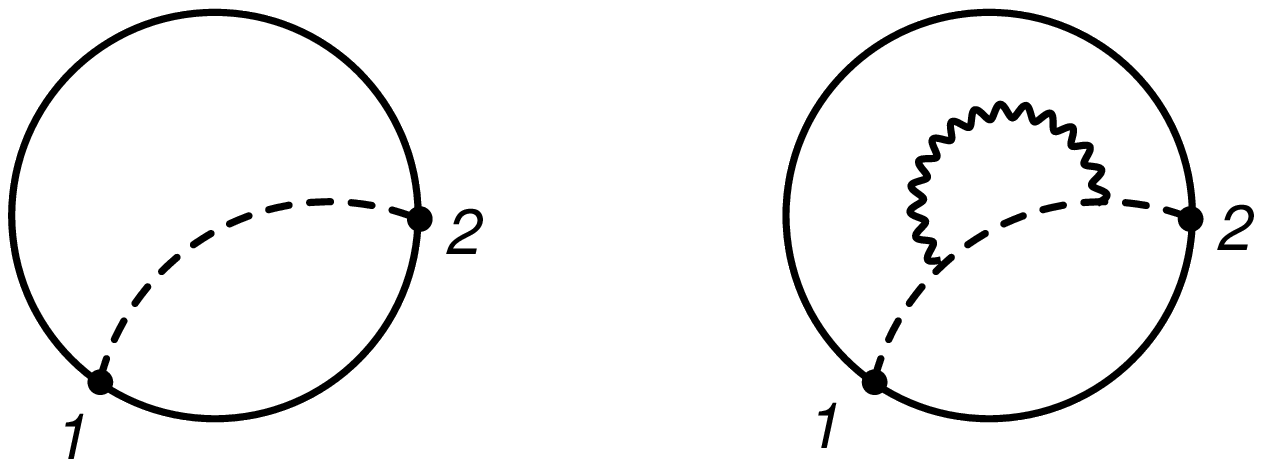}}
    \caption{ Feynman diagrams (a) Two point functions of scalars (dashed line) $G(t_1,t_2)$ and the corrections from gravitational soft mode (double wave lines) for ${\mathcal C}_1(t_{12})$ and ${\mathcal C}_2(t_{12})$ as shown in Eq.(\ref{Eq:G12_C12}). (b) Feynman diagrams for two point correlation functions $G_2(t_1,t_2)$ of scalar fields with loop corrections from soft modes as shown in Eq.(\ref{Eq:2pt-4pt-6pt-8pt-T!=0}).
    } \label{fig1}
\end{figure}


\begin{widetext}
The generating functional of connected correlators can be expanded as
\beqa
W &=& \log\vev{Z} = \log\vev{e^{-S_{\text{eff}}}}  = \log   \bigg[ \int {\mathcal D}\mu[f]  e^{ -\bar{S}_{0}  }  \bigg( 1 +  \epsilon^2 N \int \prod_{i=1}^2 dt_i  \Phi_0(t_i)  \frac{  {\mathcal C}_2(t_{12}) + \epsilon^2  {\mathcal C}_4(t_{12})  }{\big( 2 \sin\frac{t_{12}}{2} \big)^{2\Delta} }  \nn\\
&+& \frac{1}{2}  \epsilon^2  \int  \prod_{i=1}^4 dt_i \Phi_0(t_i) \frac{  :{\mathcal C}_1(t_{12}){\mathcal C}_1(t_{34}):  + \epsilon^2  ( : {\mathcal C}_2(t_{12}){\mathcal C}_2(t_{34}):  +  2 :{\mathcal C}_1(t_{12}){\mathcal C}_3(t_{34}) :  )  }{\big( 2 \sin\frac{t_{12}}{2} \big)^{2\Delta} \big( 2 \sin\frac{t_{34}}{2} \big)^{2\Delta}}  \nn\\
&+& \frac{1}{6}  \epsilon^4 \int  \prod_{i=1}^6 dt_i  \Phi_0(t_i)  \frac{   3:{\mathcal C}_1(t_{12}){\mathcal C}_1(t_{34}){\mathcal C}_2(t_{56}):  + \epsilon^2  ( :{\mathcal C}_2(t_{12}) {\mathcal C}_2(t_{34}) {\mathcal C}_2(t_{56})    : +  6 :{\mathcal C}_1 {\mathcal C}_2 {\mathcal C}_3    : + 3:{\mathcal C}_1 {\mathcal C}_1 {\mathcal C}_4    :   )   }{\big( 2 \sin\frac{t_{12}}{2} \big)^{2\Delta} \big( 2 \sin\frac{t_{34}}{2} \big)^{2\Delta} \big( 2 \sin\frac{t_{56}}{2} \big)^{2\Delta} }    \nn\\
&+& \frac{1}{24}  \epsilon^4 \int  \prod_{i=1}^8 dt_i \Phi_0(t_i) \frac{  :{\mathcal C}_1(t_{12}){\mathcal C}_1(t_{34}){\mathcal C}_1(t_{56}){\mathcal C}_1(t_{78}): + \epsilon^2(  6: {\mathcal C}_1(t_{12}){\mathcal C}_1(t_{34}){\mathcal C}_2(t_{56}){\mathcal C}_2(t_{78}): + 4 : {\mathcal C}_1{\mathcal C}_1{\mathcal C}_1{\mathcal C}_3:    )  }{\big( 2 \sin\frac{t_{12}}{2} \big)^{2\Delta} \big( 2 \sin\frac{t_{34}}{2} \big)^{2\Delta} \big( 2 \sin\frac{t_{56}}{2} \big)^{2\Delta} \big( 2 \sin\frac{t_{78}}{2} \big)^{2\Delta}} + \cdots \bigg) \bigg], \nn
\eeqa
where ${\mathcal C}_n(t_{ij})\equiv{\mathcal C}_n(t_i,t_j)$, ${\mathcal D} \mu[f]\equiv {\mathcal D} f(t)/f'(t)$ is an SL$(2,R)$ invariant measure~\cite{Stanford:2017thb}, $\bar{S}_{0} = - N \int dt_1 dt_2 { \Phi_0(t_1) \Phi_0(t_2) }/{\big( 2 \sin({t_{12}}/{2}) \big)^{2\Delta} } $ is the same as Eq.(\ref{Eq:S-chi_eff}) with $f(t)=\tan({t}/2)$ without soft mode $\epsilon k(t)$ correction,
and we have dropped the odd-leg source term considering that $\vev{{\mathcal C}_1(t_{12})}=0$ or $\vev{k(t)}=0$. $:\cdots:$ means the time ordering. From the functional $Z$, one can reads $2n$-point functions as
\beqa
&& \vev{ \prod_{i=1}^{2n} {\mathcal O}_{\Phi_0}(t_i)}  = \frac{1}{Z(\Phi_0)} \prod_{i=1}^{2n}\frac{\delta}{\delta \Phi_0(t_i)} Z(\Phi_0)|_{\Phi_0\to 0}. \qquad \quad  \label{Eq:2n-pt}
\eeqa
From the generating functional $W[\Phi_0]$, one can read all connected $2n$-points functions from irreducible Feynman diagrams. Thus, as the leading expansion, the two, four, six and eight-point functions, are respectively given by
\beqa
G_2 &=&  N   \frac{ 1 + \epsilon^2 \vev{ {\mathcal C}_2(t_{12}) }}{\big( 2 \sin\frac{t_{12}}{2} \big)^{2\Delta} }  , \quad G_4 =  N    \frac{ \epsilon^2 \vev{ :{\mathcal C}_1(t_{12}) {\mathcal C}_1(t_{34}) : } }{\big( 2 \sin\frac{t_{12}}{2} \big)^{2\Delta} \big( 2 \sin\frac{t_{34}}{2} \big)^{2\Delta}}  , \nn\\
G_6 &=&  N   \frac{ \epsilon^4 \vev{ : {\mathcal C}_1(t_{12}) {\mathcal C}_2(t_{34}) {\mathcal C}_1(t_{56})  : }   }{\big( 2 \sin\frac{t_{12}}{2} \big)^{2\Delta} \big( 2 \sin\frac{t_{34}}{2} \big)^{2\Delta} \big( 2 \sin\frac{t_{56}}{2} \big)^{2\Delta} }  , \quad G_8 =  N    \frac{ \epsilon^4\vev{ :{\mathcal C}_1(t_{12}){\mathcal C}_1(t_{34}){\mathcal C}_1(t_{56}){\mathcal C}_1(t_{78}): }  }{\big( 2 \sin\frac{t_{12}}{2} \big)^{2\Delta} \big( 2 \sin\frac{t_{34}}{2} \big)^{2\Delta} \big( 2 \sin\frac{t_{56}}{2} \big)^{2\Delta} \big( 2 \sin\frac{t_{78}}{2} \big)^{2\Delta}} .  \nn\\
G_{10} &=&  N    \frac{ \epsilon^6\vev{ :{\mathcal C}_1(t_{12}){\mathcal C}_1(t_{34}){\mathcal C}_2(t_{56}){\mathcal C}_1(t_{78}){\mathcal C}_1(t_{9,10}): }  }{\big( 2 \sin\frac{t_{12}}{2} \big)^{2\Delta} \big( 2 \sin\frac{t_{34}}{2} \big)^{2\Delta} \big( 2 \sin\frac{t_{56}}{2} \big)^{2\Delta} \big( 2 \sin\frac{t_{78}}{2} \big)^{2\Delta}\big( 2 \sin\frac{t_{9,10}}{2} \big)^{2\Delta}} ,
\label{Eq:2pt-4pt-6pt-8pt-T!=0}
\eeqa
where $G_2=G_2(t_1,t_2)$, $G_4=G_4(t_1,t_2,t_3,t_4)$, etc. $:\cdots:$ means the time ordering and $\vev{\cdots} \equiv Z_0^{-1}\int {\mathcal D} \mu[f] \cdots e^{-\bar{S}_0}$ and $Z_0=Z(\Phi_0)|_{\Phi_0\to0}$. The generalization to more higher order $2n$-point functions is straightforward.


\end{widetext}


\subsection{Zero temperature correlation functions}

Consider expand the boundary $t$ around the saddle of imaginary time $\tau$ as
\beqa
f(t) = t + \epsilon k(t),
\eeqa
The two point function of the dual operator can be expanded as
\beqa
G(t_1,t_2) = \frac{1 +  \epsilon {\mathcal B}_1(t_{12})  + \epsilon^2 {\mathcal B}_2(t_{12}) + O(\epsilon^3)}{t_{12}^{2\Delta} }   , \qquad \quad \label{Eq:B12}
\eeqa
where ${\mathcal B}_n(t_{12})\equiv{\mathcal B}_n(t_{1}-t_{2})$ with $n=1,2,\cdots$ and
\beqa
&& {\mathcal B}_1(t_{12}) = \Delta \bigg(   k'(t_1) + k'(t_2) - 2 \frac{k(t_1) - k(t_2) }{t_{12}} \bigg), \nn\\
&& {\mathcal B}_2(t_{12}) = \Delta  \left( \frac{(k(t_1)-k(t_2))^2}{t_{12}^2}- \frac{k'(t_1)^2+k'(t_2)^2}{2}\right) \nn\\
&& + 2\Delta ^2 \left( \frac{k'(t_1)+k'(t_2)}{2}- \frac{k(t_1)- k(t_2)}{t_{12}}\right)^2 , \label{Eq:C12}
\eeqa
where $t_{12}\equiv t_1-t_2$. In analogy to the finite temperature case, the two, four, six and eight-point functions, are respectively given by
\beqa
G_2 &=&  N   \frac{ 1 + \epsilon^2 \vev{ {\mathcal B}_2(t_{12}) }  }{  t_{12}^{2\Delta} }  , \nn\\
G_4 &=&  N   \frac{ \epsilon^2 \vev{ :{\mathcal B}_1(t_{12}) {\mathcal B}_1(t_{34}) : }    }{ t_{12}^{2\Delta} t_{34}^{2\Delta}}  , \nn\\
G_6 &=&  N   \frac{ \epsilon^4 \vev{ : {\mathcal B}_1(t_{12}) {\mathcal B}_1(t_{34}) {\mathcal B}_2(t_{56})  : }  }{t_{12}^{2\Delta} t_{34}^{2\Delta} t_{56}^{2\Delta} }  , \nn\\
G_8 &=&  N  \frac{ \epsilon^4 \vev{ :{\mathcal B}_1(t_{12}){\mathcal B}_1(t_{34}){\mathcal B}_1(t_{56}){\mathcal B}_1(t_{78}): } }{t_{12}^{2\Delta} t_{34}^{2\Delta} t_{56}^{2\Delta} t_{78}^{2\Delta}} , \qquad \quad \label{Eq:2pt-4pt-6pt-8pt-T=0}
\eeqa
where $G_2=G_2(t_1,t_2)$, $G_4(t_1,t_2,t_3,t_4) $, etc.

\subsection{Loop corrections to two point function}

\subsubsection{Finite temperature case}

By using the correlation function ${\mathcal C}_{2}(t_1,t_2)$ in Eqs.(\ref{Eq:C12}) and the soft mode propagators in Eq.(\ref{Eq:k12-T!=0}), one obtains the loop corrections to the two point function as
\beqa 
&& \epsilon^2 \vev{{\mathcal C}_2(t_{12})} \nn\\ 
&& = \frac{1}{2\pi C} \bigg[ \frac{\Delta}{4 \sin^2\frac{t_{12}}{2}}  \big[t_{12}^2-2 \pi  t_{12}+2 (\pi -t_{12}) \sin t_{12}  \nn\\
&& +  4\sin^2\frac{t_{12}}{2}    )\big]  +\frac{\Delta^2}{2} \bigg( \frac{t_{12} - 2 \pi}{\tan\frac{t_{12}}{2}} - 2\bigg) \bigg( \frac{t_{12}}{\tan\frac{t_{12}}{2}} -2\bigg)  \bigg], \qquad \quad\label{Eq:<C2>-t12}
\eeqa
which recovers Eq.(4.36) in Ref.~\cite{Maldacena:2016hyu}. The Feynman diagrams of the loop corrections from pNGBs are depicted in Fig.\ref{fig1}. In the large Lorentzian time with $t \to i \hat{t}$ and in the contour chosen in Eq.(\ref{OTOC-Schwinger-Contour}), one has the two point functions as
\beqa
\epsilon^2 \vev{{\mathcal C}_2(t_{12})}= \frac{1}{2\pi C} \bigg[ \bigg( 1 - \frac{\pi^2}{4} \bigg) \Delta  + 2 \Delta^2 \bigg) \bigg] \sim \frac{\pi}{\beta} \Delta^2. \qquad \quad
\eeqa
which is a constant and independent of $\hat{t}$.

\subsubsection{Zero temperature case}

By using the correlation function ${\mathcal B}_{2}(t_1,t_2)$ in Eqs.(\ref{Eq:B12}) and the soft mode propagators in Eq.(\ref{Eq:k12-T=0}), one obtains the correction to the two point function as
\beqa
\epsilon^2\vev{{\mathcal B}_2(t_{12})} 
= \frac{1}{18\pi C} \bigg[ \Delta  \left(\pi  t_{12}-\frac{t_{12}^2}{4}\right)-\pi  \Delta ^2 t_{12} \bigg]. \qquad \quad \label{Eq:B2}
\eeqa

In the large Lorentzian time with $t \to i \hat{t}$ and in the contour chosen in Eq.(\ref{OTOC-Schwinger-Contour}), the loop correction to the two point function turns out to be Lorentzian time independent as
\beqa
\epsilon^2\vev{{\mathcal B}_2(t_{12})} &=& \frac{1}{72} \pi  (4 \Delta -3).
\eeqa

\section{Quantum liquid with Schwarzian corrections}
\label{sec:Schwarzian_LQ}

In this section, we study the retarded Green's functions as well as local spin-spin correlation functions of quantum liquid with Schwarzian correlation in terms of ``Schwarzian liquid'', which can be related to the spectral functions and local dynamical susceptibility of strongly interacting quantum liquid including not only spin glass phase but also NFL phase. We also generalize the AdS$_2$ vacuum in $2$-D gravity to higher dimensional RN-AdS$_{d+1}$ vacuum in Einstein gravity with Maxwell action as in Appendix.\ref{app:high-D-generalization}.

\subsection{Quantum liquid from AdS$_2$}

\subsubsection{Global AdS$_2$ spacetime}

In $(1+1)$ dimensional spacetime, the spacetime metric of hyperbolic AdS$_2$ black hole in global coordinate is given by
\beqa
ds^2 &=& - \bigg( \frac{r^2}{r_0^2} - 1 \bigg) dt^2 + \bigg( \frac{r^2}{r_0^2}  - 1 \bigg) ^{-1} dr^2 ,
\eeqa
where the second expression is for global coordinate. The gauge field in two dimensional spacetime is
\beqa
A_t(r) =\mu (r - r_0),
\eeqa
which leads to a constant $U(1)$ field strength, since $F_{tr}=E=-\mu$. 
The temperature of the black hole turns out to be
\beqa
T = \frac{1}{2\pi r_0} .   \label{Eq:T-r0}
\eeqa
In Cartesian gauge, the metric
\beqa
ds^2= -dx^2 - dy^2 + dz^2
\eeqa
can be classified into three classes, according to the index $k=-1$ for embedding conic curves $ -x^2-y^2+z^2=k=-1$, which is obvious in the Lorentzian gauge as 
\beqa
&& ds^2 = - \cosh^2\rho d\tau^2 + d\rho^2,  \quad \rho\in(0,+\infty), \quad \tau \in(0,2\pi)  , \nn\\
&&  (x,y,z) = \cosh\rho( \sin\tau, \cos\tau,  \tanh\rho), ~ y\pm i x = e^{\pm i \tau}\cosh\rho  \qquad \nn\\
&& ds^2 = - e^{2\sigma} dt^2 + d\sigma^2 , \quad \sigma\in(-\infty,+\infty), \quad t\in(-\infty,+\infty),  \nn\\
&& (x,y,z) = (e^\sigma t,  \cosh\sigma - e^\sigma t^2/2,  \sinh\sigma  +  e^\sigma t^2/2 ), \nn\\
&& ds^2 = - \sinh^2\rho d\tau^2 + d\rho^2 , \quad \rho\in(0,+\infty), \quad \tau\in(0,2\pi) , \nn\\
&& (x,y,z) = \sinh\rho( \sinh\tau,   \cosh\tau, \coth\rho), ~ x\pm z = \pm e^{\pm\tau} \sinh\rho, \qquad \nn
\eeqa
where for hyperbolic case, the $(\rho,\tau)$ is just the usual bulk Rindler like coordinates at positive side of Rindler space with boundary at $\rho\to\infty$. The Poincar\'{e} time $t$ runs from $-\infty$ to $+\infty$, while the Rindler time $\tau$ is $2\pi$ periodic. In the Schwarzschild gauge,
\beqa
ds^2 &=& -  \bigg( \frac{r^2}{r_0^2} + 1 \bigg)  dt^2 +  \bigg( \frac{r^2}{r_0^2} + 1 \bigg)^{-1}   dr^2 = \frac{-dt^2 + dz^2}{[\sin{(z/r_0)}]^2} ,   \nn\\
 A_t &=& \mu r_0 \bigg( \cot\frac{z}{r_0} - 1 \bigg),  \qquad r = r_0 \cot\frac{z}{r_0} , \nn\\
ds^2 &=& - r^2 dt^2 +  \frac{1}{r^2}  dr^2 = \frac{-dt^2 + dz^2}{z^2} , \qquad \nn\\
 A_t &=& \mu \bigg( \frac{1}{z} - \frac{1}{z_0} \bigg),  \qquad  r = \frac{1}{z}, \nn\\
ds^2 &=& -  \bigg( \frac{r^2}{r_0^2} - 1 \bigg)  dt^2 +  \bigg( \frac{r^2}{r_0^2} - 1 \bigg)^{-1}   dr^2 = \frac{-dt^2 + dz^2}{[\sinh{(z/r_0)}]^2} , \nn\\
A_t &=& \mu r_0 \bigg( \coth\frac{z}{r_0} - 1 \bigg),  \qquad r = r_0 \coth\frac{z}{r_0} , \label{Eq:AdS2}  \label{Eq:AdS2-bd}
\eeqa
where $r\in(r_0,\infty)$ and $r_0$ is the Rindler horizon with $\ell$ being AdS$_2$ radius, the transformation between $r$ and $z$ relates the parabolic orbits and to elliptic/hyperbolic orbits. For elliptic orbit a a 2$D$ slice with normal along the spatial directionof $SL(2,R)$ group, it is not
At infinite boundary, the elliptic and hyperbolic type just reduces to be parabolic one.


\subsubsection{Poincar\'{e} AdS$_2$:zero temperature CFT$_1$}

In the AdS$_2$ spacetime in the energy coordinate $z$, the metric and the gauge field is linear in $1+1$-dimensional spacetime as
\beqa
ds^2 = - \frac{1}{4c_\mu } \frac{\ell^2}{z^2}(-dt^2 + dz^2), ~ A_t = \mu \bigg(  1 - \frac{z_\star}{z} \bigg) , \quad \quad
\eeqa
which is a AdS$_2$ spacetime in the Poincare coordinate, $c_\mu$ contains UV information from a $2$-D gravity. For the convenience, one may define an effective AdS$_2$ radius as
\beqa
\ell_L^2 \equiv -\frac{1}{4c_\mu }\ell^2.
\eeqa
In the momentum spacetime, the Klein-Gordon equation of a charge scalar in out-wave $e^{ -i\omega t + i k x}$ ($\partial_t \to - i \omega, \partial_x \to i k$) becomes
\beqa
\bigg[ \partial_z^2  + \bigg(\omega + q\mu \Big( 1 - \frac{z_\star}{z} \Big)  \bigg)^2  - \frac{m^2\ell^2}{z^2} \bigg] \phi(z,k) = 0, \qquad
\eeqa
which leads to the wave functions
\beqa
\phi(z) =   c_1 M_{i q \mu z_\star ,-\nu_q}(2 i z \bar\omega)+c_2 W_{i q \mu z_\star ,-\nu_q}(2 i z \bar\omega), \qquad
\eeqa
where $\bar\omega=\omega + q\mu$ and the conformal dimension is
\beqa
\nu_q = \sqrt{\frac{1}{4} -  \frac{m^2 \ell^2}{4c_\mu } - q^2 \mu^2 z_\star^2} . \quad
\eeqa
In the special case with $c_\mu =-1/4$, it just recovers the original one.

\begin{widetext}
In the near horizon limit, one obtains the asymptotic behavior of the boson wave function as
\beqa
\phi(z) & \overset{z\to \infty}{\sim} & 2^{i q \mu z_\star} \bigg(\frac{c_1 (-1)^{\frac{1}{2}-\nu_q + i q \mu z_\star} \Gamma (1-2 \nu_q) }{\Gamma \left(i q \mu z_\star -\nu_q+\frac{1}{2}\right)}  + c_2 \bigg) (i \bar\omega)^{i q \mu z_\star} e^{-i \bar\omega z} z^{i q \mu z_\star} + \frac{c_1 \Gamma (1-2 \nu_q) 2^{-i
   q \mu z_\star} (i \bar\omega)^{-i q \mu z_\star}}{\Gamma \left(-i q \mu z_\star -\nu_q+\frac{1}{2}\right)} e^{i \bar\omega z} z^{-i q \mu z_\star} . \qquad \quad
\eeqa
The out-going wave is $e^{-i\omega t + i \bar\omega z - i q\mu z_\star \log{z} }$, which implies that the in-falling boundary condition to be $c_1=0$.
On the other hand, in the infinite boundary $z\to 0$, one has
\beqa
\phi(z)  \overset{z\to 0}{\sim}   B(\omega) z^{\nu_q+\frac{1}{2}} + A(\omega) z^{-\nu_q+\frac{1}{2}},
\eeqa
where $A$ and $B$ are identified as source and response, respectively, and can be expressed more explicitly as
\beqa
 A(\omega) = \frac{2^{\frac{1}{2}-\nu_q} (i \bar\omega)^{\frac{1}{2}-\nu_q} \left[c_2 \Gamma (2 \nu_q)+c_1 \Gamma \left(-i q \mu z_\star + \nu_q+\frac{1}{2}\right)\right]}{\Gamma \left(-i q \mu z_\star +\nu_q+\frac{1}{2}\right)} , \quad B(\omega) = \frac{c_2 2^{\nu_q+\frac{1}{2}} (i \bar\omega)^{\nu_q+\frac{1}{2}} \Gamma (-2 \nu_q)}{\Gamma \left(-i q \mu z_\star -\nu_q+\frac{1}{2}\right)} .
\eeqa
The two point Green's function ban be read as
\beqa
G(\omega) 
= 4^{-\nu_q} (i \bar\omega)^{-2 \nu_q}  \frac{\Gamma (2 \nu_q)  }{\Gamma (-2 \nu_q)} \frac{\Gamma \left(-i q \mu z_\star -\nu_q+\frac{1}{2}\right)}{ \Gamma \left(-i q \mu z_\star + \nu_q+\frac{1}{2}\right)}. \qquad \quad
\eeqa
By doing an inverse Fourier transformation, one has
\beqa
\int_{-\infty}^{+\infty} d\omega e^{-i\omega t} \omega^a =  - e^{\frac{i \pi  a}{2}} \sin (\pi  a) \Gamma (a+1)   \left\{ \begin{aligned}
& \frac{\text{sgn}^{-1}(t)+1}{\abs{t} ^{a+1}} , \quad a>0  \\
& \frac{\text{sgn}(t)+1}{\abs{t} ^{a+1}},  \quad a<0. 
\end{aligned} \right. \quad
\int_{-\infty}^{+\infty} d\omega e^{-i\omega t} \abs{\omega}^a  =  - 2 \sin \left(\frac{\pi  a}{2}\right) \frac{\Gamma (a+1)}{\abs{t}^{a+1}} , \qquad \quad
\eeqa
where $e^{-i\pi a}=(-1)^a=(-i)^{2a}$ and $\text{sgn(t)}=t/\abs{t}$. Thus, in the coordinate spacetime, assuming $t\gg 0$, then one obtains the retarded Green's function in real coordinate spacetime, which just recovers the ansatz of the form the two point correlation function at strong coupling at zero temperature~\cite{Maldacena:2016hyu,Gross:2017vhb},
\beqa
G(t) 
&=& \sqrt{\frac{2}{\pi }} e^{-i \pi  \nu_q} \sin (2 \pi  \nu_q) \Gamma (1-2
   \nu_q)  \frac{\text{sgn}(t)}{\abs{t} ^{1 - 2 \nu_q} }  \sim  b \frac{\text{sgn}(t)}{\abs{t} ^{2\Delta_+}}. \qquad \quad
\eeqa
It is worthy of noticing that the result reproduces the SYK uniform saddle point solution, by making a match as below
\beqa
b &\equiv &\bigg ( \frac{1}{2\pi J^2} \big( 1 - \frac{2}{p} \big) \tan\frac{\pi}{p} \bigg)^{1/p} = \sqrt{\frac{2}{\pi }} e^{-i \pi  \nu_q} \sin (2 \pi  \nu_q) \Gamma (1-2
   \nu_q),
\eeqa
where $\Delta \equiv p^{-1} = {1}/{2} - \nu_q = \Delta_-$.
In particularly, in the case that $\nu_q= {1}/{4}$, ${m^2 \ell^2}/{(4c_\mu) } = q^2 \mu^2 z_\star^2$, the Green's function just recovers the two-point function of SYK model with conformal dimension $\Delta^{IR}_\pm=1/4$ due to an emergent conformal symmetry at low energies and large $N$ at zero temperature,
\beqa
G(t) = \vev{[{\mathcal O}_\Delta(t),{\mathcal O}_\Delta(0)]} \sim \frac{ \text{sgn}(t)}{\sqrt{\abs{t} }} , \quad t\gg 1,
\eeqa
where $\sim$ denotes a common factor $\sqrt{2} e^{-\frac{i \pi }{4}}$ is dropped. It is useful to use the Fourier transforms for symmetric and antisymmetric function as
\beqa
\int_{-\infty}^{+\infty} e^{i\omega t} \frac{1}{\abs{t}^{2\Delta}} \big\{ \text{sgn}(t), 1 \big\} = 2  \Gamma(1-2\Delta) \abs{\omega}^{2\Delta-1} \big\{ i  \cos{(\pi \Delta)} \text{sgn}(\omega), \sin{(\pi \Delta)}  \big\}.
\eeqa

\subsubsection{Global AdS$_2$:finite temperature CFT$_1$}

Consider the AdS$_2$ metric in global coordinates as in (\ref{Eq:AdS2}) for hyperbolic case in the vacuum $b=1$ as
\beqa
ds^2 = -\frac{1}{4c_\mu } \frac{-dt^2+dz^2}{z_0^2 \sinh^2{(z/z_0)}} , \quad  A_t(z) =  \mu z_0 \bigg( 1 - \coth{\frac{z}{z_0}} \bigg),
\eeqa
where $z\in(0,\infty)$.
It is worthy of noticing that it can be transformed into
\beqa
 ds^2 = -\frac{1}{4c_\mu } \frac{-dt^2+dz^2}{ \sinh^2{z}} , \quad A_t(z) =  \mu  ( 1 - \coth{z} ),
\eeqa
by making a replacement
\beqa
t \to t z_0, \quad z \to z z_0, \quad \mu \to \frac{\mu}{z_0} , \label{Eq:t-z_z0}
\eeqa
For the simplicity, let's consider the $z_0=1$ case at the beginning, we can obtain a general results by making an inverse rescaling
\beqa
\omega \to \omega z_0, \quad z \to \frac{z}{z_0}, \quad \mu \to \mu z_0, \quad z_0 = \frac{1}{2\pi T} ,  \quad \label{Eq:z0}
\eeqa
the parameter $z_0$ is related to the temperature, according to Eq.(\ref{Eq:T-r0}).
The Klein-Gordan equation are
\beqa
\phi^{\prime\prime}(z) + \bigg[ \omega^2 + \bigg( \frac{1}{4} - \nu_1^2 \bigg) \frac{1}{\sinh{z}} \bigg] \phi(z) = 0 , \quad \nu_1 = \sqrt{ \frac{1}{4} - \frac{m^2}{4c_\mu } } .
\eeqa
For neutral scalar case ($\mu=0$), the wave functions are
\beqa
\phi(z) &=&  i^{-2 \nu_1} \tanh ^{\frac{1}{2}-\nu_1}(z) \left(-\text{sech}^2(z)\right)^{-\frac{i \omega }{2}} \bigg[  c_2 \tanh ^{-\nu_1}(z) \, _2F_1\left(\frac{1}{4} (-2 \nu_1-2 i \omega
   +1),\frac{1}{4} (-2 \nu_1-2 i \omega +3);1-\nu_1;\tanh ^2(z)\right)  \nn\\
   && +c_1 i^{2 \nu_1} \tanh ^{ \nu_1}(z) \, _2F_1\left(\frac{1}{4} (2 \nu_1-2 i
   \omega +1),\frac{1}{4} (2 \nu_1-2 i \omega +3);\nu_1+1;\tanh ^2(z)\right)\bigg].
\eeqa
For the charged scalar case, the Klein-Gordan equation are
\beqa
\phi^{\prime\prime}(z) + \bigg[ \omega^2 + \bigg( \frac{1}{4} - \nu_1^2 \bigg) \frac{1}{\sinh^2{z}} + 2 q\mu\bar\omega (1 - \coth{z}) \bigg] \phi(z) = 0 , \quad
\nu_1 = \sqrt{ \frac{1}{4} - \frac{m^2}{4c_\mu } -q^2 \mu^2} .  \label{Eq:nu_1-qmu}
\eeqa
The wave functions are
\beqa
\phi(z) &=& -i \cosh (z) \tanh ^{\frac{1}{2}-\nu_1}(z) (\tanh (z)+1)^{-\frac{1}{2} i (2 \mu  q+\omega +i)} (\tanh (z)-1)^{\nu_1+i \mu  q+\frac{i \omega }{2}} \nn\\
&& \bigg[ c_1 \left(\frac{\tanh (z)}{\tanh (z)-1}\right)^{-\nu_1} \,
   _2F_1\left(-i q \mu -\nu_1+\frac{1}{2},-i q \mu -\nu_1-i \omega +\frac{1}{2};1-2 \nu_1;\frac{2 \tanh (z)}{\tanh (z)-1}\right) \nn\\
&& +c_2 (-2)^{2 \nu_1} \left(\frac{\tanh (z)}{\tanh (z)-1}\right)^{\nu_1} \,
   _2F_1\left(-i q \mu +\nu_1+\frac{1}{2},-i q \mu +\nu_1-i \omega +\frac{1}{2};1 + 2 \nu_1;\frac{2 \tanh (z)}{\tanh (z)-1}\right)\bigg] .
\eeqa
The wave function can also be re-expressed as
\beqa
\phi(\rho) &=&  -\frac{i \rho ^{\frac{1}{2}-\nu_1} (\rho +1)^{-\frac{1}{2} i (2 \mu  q+\omega +i)} (\rho -1)^{\nu_1+i \mu  q+\frac{i \omega }{2}}
   }{\sqrt{1-\rho ^2}} \bigg[ c_1 \, _2F_1\left(-i q \mu -\nu_1+\frac{1}{2},-i q \mu -\nu_1-i \omega
   +\frac{1}{2};1-2 \nu_1;\frac{2 \rho }{\rho -1}\right) \nn\\
   && +c_2 (-2)^{2 \nu_1} \left(\frac{\rho }{\rho -1}\right)^{2 \nu_1} \,
   _2F_1\left(-i q \mu +\nu_1+\frac{1}{2},-i q \mu +\nu_1-i \omega +\frac{1}{2};1 + 2 \nu_1;\frac{2 \rho }{\rho -1}\right)  \bigg],
\eeqa
where $\rho\equiv \tanh{z}$.
In the near horizon limit ($\rho\to 1$), the wave function can be re-expressed as
\beqa
\psi(\rho) \overset{\rho \to 1}{\sim} a(\omega) (1-\rho)^{\frac{i\omega}{2}} + b(\omega) (1-\rho)^{-\frac{i\omega}{2}},
\eeqa
where $\sim $ means that we have dropped a common factor $-i 2^{\nu_1-\frac{1}{2}} (-1)^{\nu_1+i \mu  q+\frac{i \omega }{2}}$ in front of the wave function and the coefficients are
\beqa
a(\omega) &=& 2^{-\frac{i \omega }{2}} \Gamma (-i \omega ) \left(\frac{c_1 \Gamma (1-2 \nu_1)}{\Gamma \left(i q \mu -\nu_1+\frac{1}{2}\right) \Gamma \left(-i q \mu -\nu_1-i \omega +\frac{1}{2}\right)}+\frac{c_2 \Gamma (2
   \nu_1+1)}{\Gamma \left(i q \mu +\nu_1+\frac{1}{2}\right) \Gamma \left(-i q \mu +\nu_1-i \omega +\frac{1}{2}\right)}\right), \nn\\
b(\omega) &=& 2^{\frac{i \omega }{2}} \Gamma (i \omega ) \left(\frac{c_1 \Gamma (1-2 \nu_1)}{\Gamma
   \left(-i q \mu -\nu_1+\frac{1}{2}\right) \Gamma \left(i q \mu -\nu_1+i \omega +\frac{1}{2}\right)}+\frac{c_2 \Gamma (1 + 2 \nu_1)}{\Gamma \left(-i q \mu +\nu_1+\frac{1}{2}\right) \Gamma \left(i q \mu + \nu_1 +i \omega +\frac{1}{2}\right)}\right).
\eeqa
Since $e^{-i\omega t  - i \frac{1}{2}\log{(1-\rho)} }$ is the infalling wave, will impose the in-falling wave condition that $a(\omega)=0$, from which the relation between $c_2$ and $c_1$ can be determined.
In the UV limit, one has
\beqa
\phi(z) \overset{z\to 0}{\sim}   c_1 \rho^{\frac{1}{2}-\nu_1}+ c_2 4^{\nu_1} e^{4 i \pi  \nu_1} \rho^{\frac{1}{2}+\nu_1}   \equiv   A(\omega) \rho^{\Delta_-} + B(\omega)  \rho^{\Delta_+} ,
\eeqa
where for $\sim$, we have dropped a common factor $ -i (-1)^{\nu_1+i \mu  q+\frac{i \omega }{2}} $. The conformal dimension  is defined as $\Delta_\pm = {1}/{2} \pm \nu_1$. Thus, the retarded Green's functions are
\beqa
G^R(\omega) = \frac{B(\omega)}{A(\omega)} = 4^{\nu_1} e^{4i\pi \nu_1} \frac{c_2}{c_1} = -4^{\nu_1} e^{4i\pi \nu_1} \frac{\Gamma (1-2 \nu_1) \Gamma \left(i q \mu +\nu_1+\frac{1}{2}\right) \Gamma \left(-i q \mu +\nu_1-i \omega +\frac{1}{2}\right)}{\Gamma (1 + 2 \nu_1) \Gamma \left(i q \mu -\nu_1+\frac{1}{2}\right)
   \Gamma \left(-i q \mu -\nu_1-i \omega +\frac{1}{2}\right)}.
\eeqa
By using the rescaling relation in Eq.(\ref{Eq:z0}), one obtains the Green's function as
\beqa
G^R(\omega) &=& z_0^{2\Delta_-} \frac{B(\omega)}{A(\omega)} = 4^{\nu_1} z_0^{ 2\nu_1 - 1 } e^{ 4i \pi \nu_1 } \frac{c_2}{c_1} \nn\\
&=& - 2 (\pi T)^{1-2\nu_1} e^{4i \pi \nu_1} \frac{\Gamma (1-2 \nu_1) \Gamma \left(i  \frac{q\mu}{2\pi T} +\nu_1+\frac{1}{2}\right) \Gamma \left( +\nu_1-i \frac{\omega+ q\mu}{2\pi T} +\frac{1}{2}\right)}{\Gamma (1 + 2 \nu_1) \Gamma \left(i  \frac{q \mu}{2\pi T} -\nu_1+\frac{1}{2}\right) \Gamma \left( -\nu_1-i \frac{\omega + q\mu }{2\pi T} +\frac{1}{2}\right)} , \label{Eq:GR-T!=0}
\eeqa
where the pre-factors $z_0$ are due to the rescaling of the coordinates in Eq.(\ref{Eq:t-z_z0}).
For neutral case, i.e., $\mu=0$ and $q=0$, one has
\beqa
G^R(\omega) 
 &=& -4^{\nu_1} e^{4i\pi \nu_1} \frac{\Gamma [ 2(1-\Delta_+) ]}{\Gamma [ 2(1+\Delta_+ ) ]}  \frac{\Gamma \left( \Delta_+ \right)}{\Gamma \left(  1 - \Delta_+ \right) } \frac{  \Gamma \left( \Delta_+ -i \frac{\omega}{2\pi T} \right)}{
   \Gamma \left( 1 - \Delta_+ -i \frac{\omega  }{2\pi T} \right)},
\eeqa
where $\Delta_+ =  {1}/{2} + \nu_1$. The equation shows that the dimension $\Delta_+$ sets the quasi-normal mode frequencies as $i\omega_n \beta=2\pi(\Delta_- + n )$.  Therefore,  by doing an inverse Fourier transformation and according to the integral identity as in Eq.(\ref{Eq:integ1-chi}), one obtains the retarded Green's function in real coordinate spacetime which just recovers the ansatz of the form the two point correlation function at strong coupling at finite temperature~\cite{Maldacena:2016hyu,Gross:2017vhb},
\beqa
G^R(t ) = \vev{{\mathcal O}_\Delta(t){\mathcal O}_\Delta(0)} = e^{2i\pi (2\Delta_+ - 1   )} \frac{2\Delta_+ - 1}{\Gamma [ 2(1+\Delta_+ ) ]}  \frac{\Gamma \left( \Delta_+ \right)}{\Gamma \left(  1 - \Delta_+ \right) }   \frac{\pi}{\beta} \frac{\text{sgn}(t)}{[\sinh\frac{\pi t}{\beta}]^{2\Delta_+}}  \sim  b  \frac{\text{sgn}(t)}{[\frac{\beta}{\pi}\sinh\frac{\pi t}{\beta}]^{2\Delta_+}} ,  \label{Eq:GR-Liouville}
\eeqa
where $\text{sgn}(t)\equiv t/\abs{t}$ is a step function. At this step, by making a comparison with that of SYK model as in Eq.(\ref{Eq:SYK_two-pt}), one has
\beqa
&&  2 J^2 b^{\frac{1}{\Delta_+}} \pi = (1-2\Delta_+)\tan{\pi \Delta_+} , \quad b = e^{2i\pi (2\Delta_+ - 1   )} \frac{2\Delta_+ - 1}{\Gamma [ 2(1+\Delta_+ ) ]}  \frac{\Gamma \left( \Delta_+ \right)}{\Gamma \left(  1 - \Delta_+ \right) }   \bigg(\frac{\pi}{\beta}\bigg)^{2\Delta_+ -1} .
\eeqa

The finite temperature retarded Green's function can be expanded as
\beqa
G^R(t)   \sim   \text{sgn}(t) \bigg( \frac{\pi}{\beta} \frac{1}{\sinh\frac{\pi t}{\beta}} \bigg)^{2 \Delta}  \overset{t \to 0}{\propto} \frac{1}{t^{2 \Delta}} - \frac{ \Delta\pi^2}{3\beta^2t^{2( \Delta-1)}}  + \frac{ \Delta(1+5 \Delta)\pi^4}{90\beta^4t^{2( \Delta - 2)}} + \cdots . \qquad \quad
\eeqa
In frequency space, it can be re-expressed as $G^R(\omega)=-i G(-i\omega+\epsilon)$. As expected, at low temperature, i.e., in the large $\beta$ limit, the retarded Green's function recovers the zero temperature one as $G^R(t)  \sim t^{-2\Delta}$.


The retarded Green's function obtained above describes a general class of strongly interacting system such as NFL with $\Delta=1/4$~\cite{Parcollet:1999,Sachdev:2015efa,Cai:2016jxd,Bi:2017yvx,Wu:2018vua} and spin fluid or quantum spin glass with $\Delta=1/2$, or random/disordered paramagnet~\cite{Sachdev:1992fk,Grempel:1998,Grempel:1999,Grempel:2001}, which describes the quantum fluctuations near a critical quantum Heisenberg spin glass. The retarded Green's functions of CFT$_1$ or NCFT$_1$ in AdS$_2$ and NAdS$_2$ spacetime, can be generalized to be those in higher dimensional spacetime, as explored in Appendix.\ref{app:high-D-generalization}.

\subsection{Schwarzian retarded Green's functions}

The two point correlation function of quantum liquid without Schwarzian correction in real time is
\beqa
G(t) = \frac{1}{i^{2\Delta}} \frac{1}{[2 \frac{\beta}{2\pi} \sinh\frac{t}{2} \frac{2\pi}{\beta} ]^{2\Delta}}  \overset{\beta\to \infty}{=}  \frac{1}{i^{2\Delta}} \frac{1}{ t^{2\Delta}} = G(t), \qquad \quad\label{Eq:GTt-Gt}
\eeqa
from which, one obtains the retarded Green's function, or the susceptibility of quantum liquid as defined in Eq.(\ref{Eq:GRA-Gpm}) as
\beqa
G^{R}(\omega) &=& -i \int_{-\infty}^{+\infty} dt \theta(t) e^{i\omega t} G(t) = -i \bigg(\frac{2\pi}{\beta}\bigg)^{2\Delta}  \frac{e^{-i \pi  \Delta } \Gamma (1-2 \Delta ) \Gamma (\Delta -i \omega \frac{\beta}{2\pi} )}{\Gamma (1-\Delta -i \omega \frac{\beta}{2\pi} )}, \qquad \quad \label{Eq:GR-susceptibility}
\eeqa
where $0<\Delta<1/2$, $\beta>0$, $\text{Im}~\omega> 0$ and we have used the integral in Eq.(\ref{Eq:I0}). One can restore the temperature by multiplying each $\omega$ with factor $\beta/(2\pi)$. The temperature dependent factor in front origins from thermal correlation function in Eq.(\ref{Eq:GTt-Gt}), so that it recovers the quantum correlation function in the zero temperature limit. For zero temperature case, one has
\beqa
 G^R(\omega)
 &=& (-1)^{-2\Delta}\omega^{2\Delta-1}
 \Gamma(1-2\Delta), \qquad \quad \label{Eq:GR-susceptibility-T=0}
\eeqa
where $\Delta<1/2$, $\text{Im}~\omega> 0$ and $(-1)^{-2\Delta}= e^{2i\pi \Delta } $.

The pNGBs loop corrections to the imaginary time thermal two point functions in Eq.(\ref{Eq:<C2>-t12}) can be re-expressed as real time one, by replacing $t_{12}$ with real time $i t$ as
\beqa 
&& \epsilon^2 \vev{{\mathcal C}_2(t_{12})}
= \frac{1}{2\pi C} \bigg( \Delta  [ ( 2\Delta +1)+ \frac{\Delta}{2}    \left(t^2+2 i \pi  t\right)] + \frac{\Delta  (2 \Delta +1) t (t+2 i \pi )}{4 \sinh^2\left(\frac{t}{2}\right) } -i \Delta  (2 \Delta +1) (\pi -i t) \frac{\coth \left(\frac{t}{2}\right) }{\sinh \left(\frac{t}{2}\right) } \bigg),
\eeqa
where we will assigned every $t$ with a factor multiplying factor $2\pi/\beta$. It can also be separated as two parts, one has even symmetry for time, while the other has odd symmetry as
\beqa
 \epsilon^2 \vev{{\mathcal C}_2(t_{12})}_e
&=& \frac{1}{2\pi C} \bigg(  \Delta  [ ( 2\Delta +1)+ \frac{\Delta}{2}   t^2 ] + \frac{\Delta  (2 \Delta +1) t^2}{4 \sinh^2\left(\frac{t}{2}\right) } - \Delta  (2 \Delta +1)  \frac{t \cosh \left(\frac{t}{2}\right) }{\sinh \left(\frac{t}{2}\right) } \bigg),  \nn\\
\epsilon^2 \vev{{\mathcal C}_2(t_{12})}_o
&=& \frac{1}{2\pi C} \bigg( \Delta^2  t + \frac{\Delta  (2 \Delta +1)  t}{2 \sinh^2\left(\frac{t}{2}\right) } - \Delta  (2 \Delta +1)  \frac{\cosh \left(\frac{t}{2}\right) }{\sinh \left(\frac{t}{2}\right) } \bigg) \pi i  .
\eeqa
By doing Fourier transformation, the second and third part of the odd sector will be vanishing unless $-1<\Delta<0$, thus the non-vanishing part within $0<\Delta <1/2$, comes form the even part, which turns out to be
\beqa
\bar{G}^R(\omega) &=& G^R(\omega)[ 1 +  \epsilon^2 \vev{{\mathcal C}_2(\omega)}_e  +  \epsilon^2 \vev{{\mathcal C}_2(\omega)}_o ], \qquad
\eeqa
where
\beqa
 \epsilon^2 \vev{{\mathcal C}_2(\omega)}_e &=&  \frac{1}{2 \pi  C} \bigg( \Delta  (2 \Delta +1) + \frac{\Delta^2}{2} [  ( \psi ^{(0)}_{\Delta -i \omega  } )^2 + \psi ^{(1)}_{\Delta -i \omega  }   ] - \frac{\Delta^2 + \omega^2 }{2} [  ( \psi ^{(0)}_{\Delta+1 -i \omega  } )^2 + \psi ^{(1)}_{\Delta+1 -i \omega }  ] - (1+2\Delta)(  1 - i \omega \psi ^{(0)}_{\Delta -i \omega  }  )  \bigg),\nn\\
\epsilon^2 \vev{{\mathcal C}_2(\omega)}_o &=&  \frac{ 1}{2 \pi C}\big( - i \pi \Delta ^2 \psi ^{(0)}_{\Delta -i \omega} \big), \qquad\quad
\eeqa
where we have used the integrals as in Eqs.(\ref{Eq:I-odd_chi}) and (\ref{Eq:I-even_chi}).

\subsubsection{Finite temperature case}

Therefore, for finite temperature case, the pNGBs loop corrected two point thermal retarded Green's function becomes

\beqa
 \bar{G}_T^R(\omega)&& = G_T^R(\omega)\bigg[ 1 + \frac{1}{2 \pi  C} \bigg( \Delta  (2 \Delta +1) + \frac{\Delta^2}{2} \Big(  \big( \psi ^{(0)}_{\Delta -i \omega\frac{\beta}{2\pi}  } \big)^2 + \psi ^{(1)}_{\Delta -i \omega\frac{\beta}{2\pi}  }   \Big)  - \frac{\Delta^2 + \omega^2 \frac{\beta ^2}{4 \pi ^2}}{2} \Big(  \big( \psi ^{(0)}_{\Delta+1 -i \omega\frac{\beta}{2\pi}  } \big)^2 + \psi ^{(1)}_{\Delta+1 -i \omega\frac{\beta}{2\pi} }   \Big)  \nn\\
 && -(1+2\Delta)\Big(  1 - i \omega\frac{\beta}{2\pi} \psi ^{(0)}_{\Delta -i \omega\frac{\beta}{2\pi}  }  \Big)   - i \pi \Delta ^2 \psi ^{(0)}_{\Delta -i \omega\frac{\beta}{2\pi}} \bigg) \bigg] ,
 \qquad \quad  0<\Delta <1/2, \quad \beta>0, \quad \text{Im}\omega> 0 . \label{Eq:GR_loop}
\eeqa
where $G^R(\omega)$ is defined in Eq.(\ref{Eq:GR-susceptibility}), and we have used that the iteration relation
\beqa
I^{(0)}_{\Delta+1-i \omega \frac{\beta}{2\pi} } = \frac{ \frac{\beta ^2}{4 \pi ^2} \omega ^2+ \Delta ^2}{2 \frac{\beta ^2}{4 \pi ^2} \left(2 \Delta ^2+\Delta \right)} I^{(0)}_{\Delta-i \omega\frac{\beta}{2\pi} } ,
\eeqa
where $I^{(0)}_{\Delta-i \omega }$ is defined in Eq.(\ref{Eq:I0}).


\subsubsection{Zero temperature case}


For zero temperature case, one can do the Fourier transformation upon Eq.(\ref{Eq:B2}), which leads to
\beqa
\bar{G}^R(\omega) &=& G^R(\omega)\bigg[1 + \frac{ \Delta (\Delta -1) (2 \Delta -1)}{18\pi C} \bigg( -\frac{i \pi  }{\omega } + \frac{1}{2 \omega ^2} \bigg) \bigg],
\eeqa
where $\Delta <1/2$ and $G^R(t) = (-1)^{2\Delta}\omega^{2\Delta-1}\Gamma(1-2\Delta)$ is given as in Eq.(\ref{Eq:GR-susceptibility-T=0}). The dynamical local susceptibility of Shcwarzian liquid becomes
\beqa
\bar\chi_{\text{loc}}^{(2)}(\omega) &=& \int_{-\infty}^{+\infty} dt \theta(t) e^{i\omega t} \bar{G}_2(t) \bar{G}_2(-t)\nn\\
&=& \chi_{\text{loc}}^{(2)}(\omega)\bigg( 1 + \frac{(2 \Delta -1) (4 \Delta -1) \Delta  \left(9 C+\pi
   \Delta  (\Delta -1)^2\right)}{162 \pi  C^2 \omega ^2} + \frac{(\Delta -1)  (4 \Delta -3) (2 \Delta -1) (4 \Delta -1) \Delta ^2}{648 \pi ^2 C^2 \omega ^4}\bigg), \quad
\eeqa
where $\Delta<{1}/{4}$, and $\chi_{\text{loc}}^{(2)}(\omega)=(-i \omega )^{4 \Delta -1} \Gamma (1-4 \Delta )$ as given in Eq.(\ref{Eq:S_Delta-omega-T=0}). After obtains loops correction from the Schwarzian effective action, the local suscpetibility becomes more singular at zero frequency limit $\omega=0$. While these terms is vanishing when $\Delta=1/2,1/4$, the physical consequence of which can be observed at finite temperature.

\end{widetext}

\subsection{Retarded Green's function}

It turns out that the loop correction from pNGBs to the thermal correlation functions, or the retarded Green's functions in Eq.(\ref{Eq:GR_loop}), leads to a dynamically generated high energy Hubbard band in spectral function, which corresponds to the destruction of quasi-particle states in the spectral function/DOS of quantum liquid, as shown in Fig.\ref{fig2}, Fig.\ref{fig3}, Fig.\ref{fig4}, Fig.\ref{fig5} and Fig.\ref{fig6}.

\subsubsection{$\Delta=1/4$: NFL}

\begin{figure}[ht!]
\centering
   \subfigure[ Evolution of $\bar{G}^R(\omega)$ with coupling $(2\pi C)^{-1}$ ]
   { \includegraphics[scale=0.48]{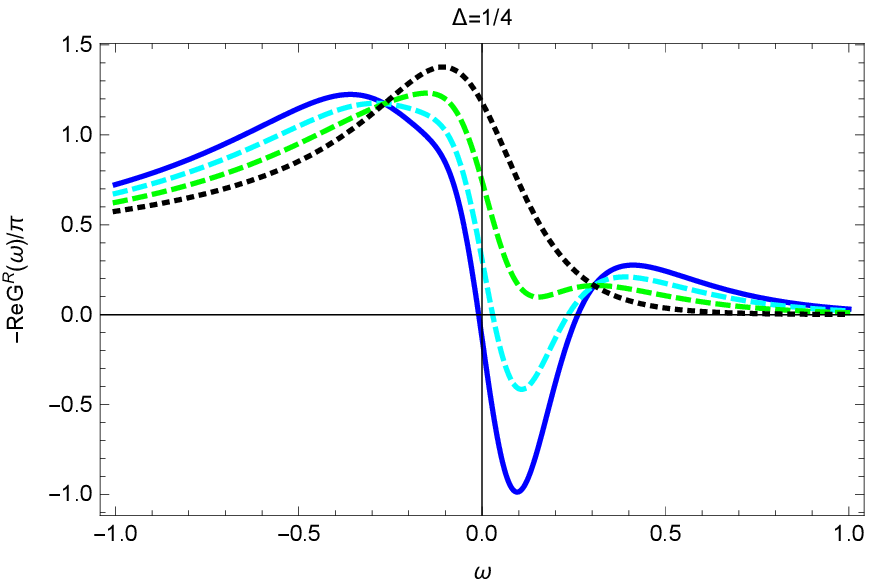} \quad \includegraphics[scale=0.48]{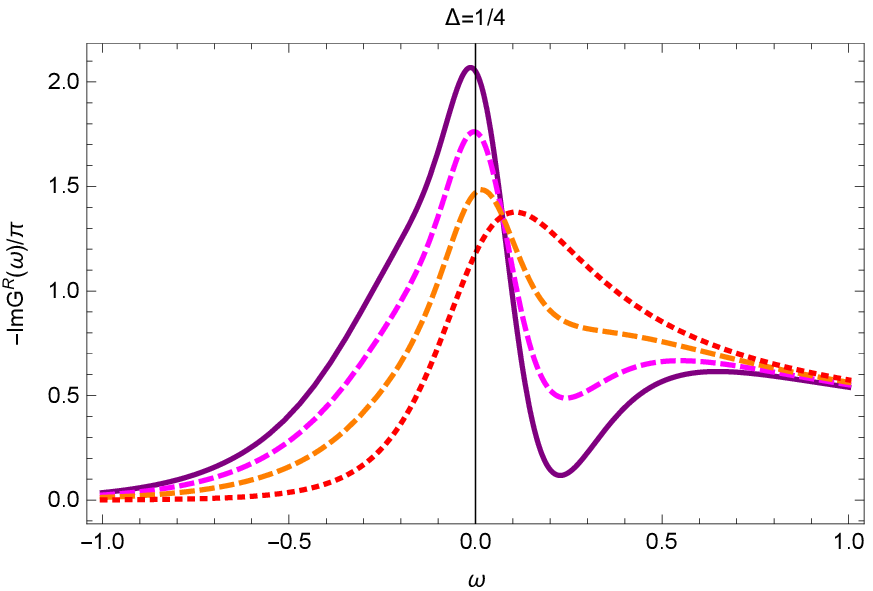}  }
   \subfigure[ Evolution of $\bar{G}^R(\omega)$ with temperature $\beta^{-1}$ ]
   { \includegraphics[scale=0.48]{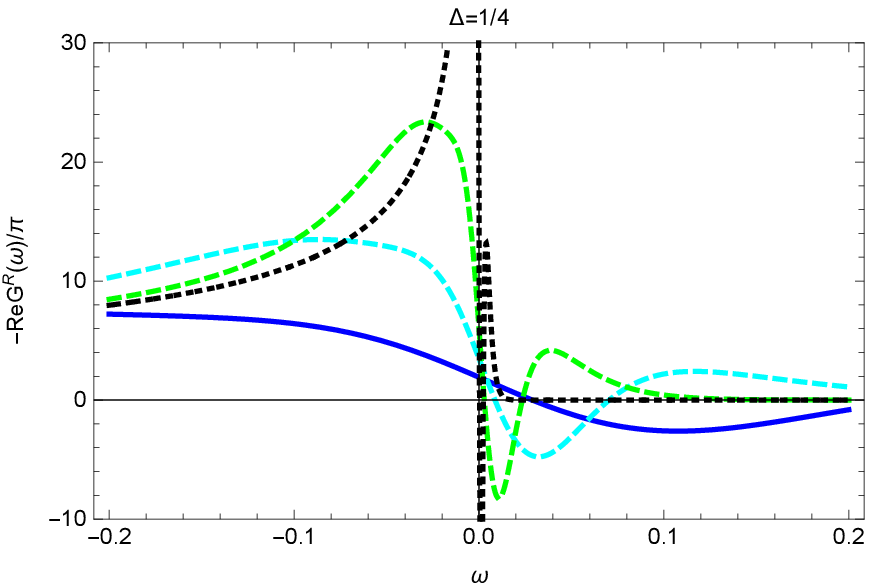} \quad
   \includegraphics[scale=0.48]{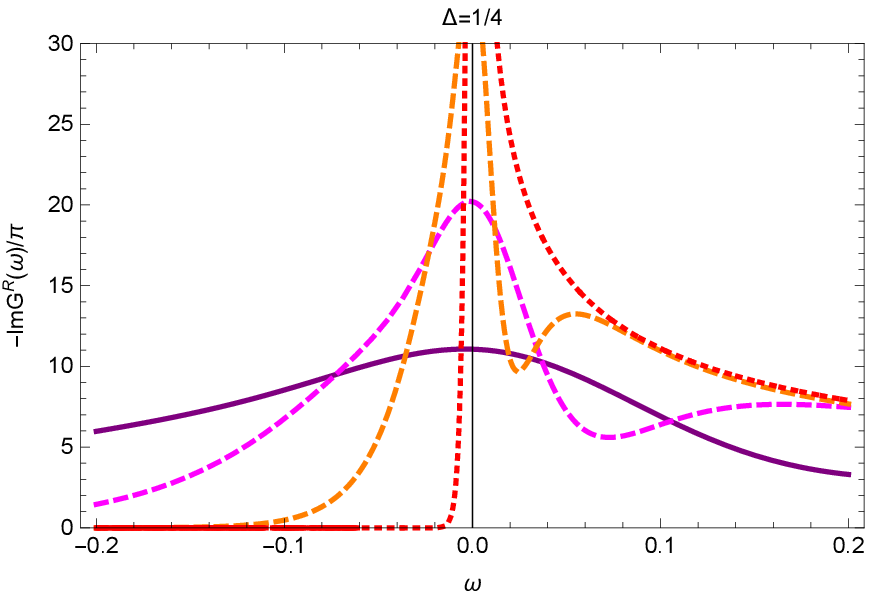}  }
    \caption{Dynamical susceptibility or retarded Green's functions of Schwarzian NFL with $\Delta=1/4$: $\bar\chi(\omega) = -G^R(\omega)/\pi $ given in Eq.(\ref{Eq:GR_loop}): (a) Evolution with coupling strength $(2\pi C)^{-1}$ : $C=1/3\pi$ (blue/purple solid line), $C=1/2\pi$ (cyan/magenta dashed line); $C=1/\pi$ (green/orange dashed line) and $C=+\infty$ (black/red dotted line). (b) Evolution with temperature $T=\beta^{-1}$: In front of $\bar\chi(\omega)$, we have multiplying a temperature depending factor $\pi/\beta$. For different $\beta$ : $\beta=2\pi$ (blue/purple solid line), $\beta=20\pi/3$ (cyan/magenta dashed line), $\beta=20\pi$ (green/orange dashed line); $\beta=200\pi$ (black/red dotted line). We have chosen input parameters as $\beta=2\pi$.
    } \label{fig2}
\end{figure}

In Fig.\ref{fig2}(a), we show the coupling strength evolution of retarded Greens function of NFL with Schwarzian correction in terms of ``Schwarzian NFL'' with $\Delta=1/4$ by increasing strength ($\sim C^{-1}$) of pNGBs loop corrections to matter two point correlation functions, due to Eq.(\ref{Eq:<C2>-t12}). The $C=\infty$ case (dotted red curves) corresponds to conventional NFL with fragile quasi-particle picture. With the increasing of the coupling strength (or decreasing of $C$) up to $C^{-1}=3\pi$ (purple solid line), the DOS accumulates more in the $\omega=0$ region as the metallic phase with Fermi liquid behavior, meanwhile it develops a ``slope-dig-ramp'' shoulder structure, i.e., a Hubbard band at $\omega\approx 0.22$, which is dynamically generated DOS at finite frequency. The Hubbard band is a smoking gun indicating the presence of a bad metal phase. Among the intermediate range, there is a temperature dependent crossover between the Fermi liquid regime and bad metal regime in the strongly correlating regime, in which the quasi-particle picture is still fragile or even broken down. This signature of NFL phase with Schwarzian correction, i.e., a DOS with Hubbard band in strongly correlated region, is significant different from the conventional NFL phase, where there is no Hubbard band structure present at all. In particular, this signature has been observed in experiments in strongly correlated system, which can be calculated by DFT approach. While to fit the experimental data in the quantum liquid with Schwarizain correction, one needs only three input parameters: the temperature $\beta$, the conformal dimension $\Delta$ and coupling strength of low energy effective Schwarzian action $\sim C_g^{-1}$ that comprise the UV information of various $2$-D gravity. Moreover, our exact analytical results quantitatively reproduce the DOS obtained from DMFT approach with a state-of the art numerical calculation from first principles of many-body theory~\cite{Nomura:2015}.

In Fig.\ref{fig2}(b), we also show the temperature evolution of retarded Green's functions of Schwarzian NFL with $\Delta=1/4$ by decreasing temperature (or by increasing $\beta$) untill $T=1/(200\pi)$, which approximately corresponds to zero temperature case (dotted lines). The decreasing of the temperature from $T=1/(2\pi)$ to $T=1/(20\pi)$, the dig of DOS moves from $\omega\approx 0.22$ to more lower frequency region at $\omega\approx 0.02$, and so does the location of Hubbard band, which indicates that the dynamics is due to the pNGBs from spontaneous and explicit symmetry breaking. Meanwhile, the DOS accumulates rapidly and results in a peak at $\omega=0$, which implies that the quantum liquid becomes more metallic like in zero temperature limit.


\begin{figure}[ht!]
\centering
   \subfigure[ -Re$G^R(\omega)$ ]
   {\includegraphics[scale=0.46]{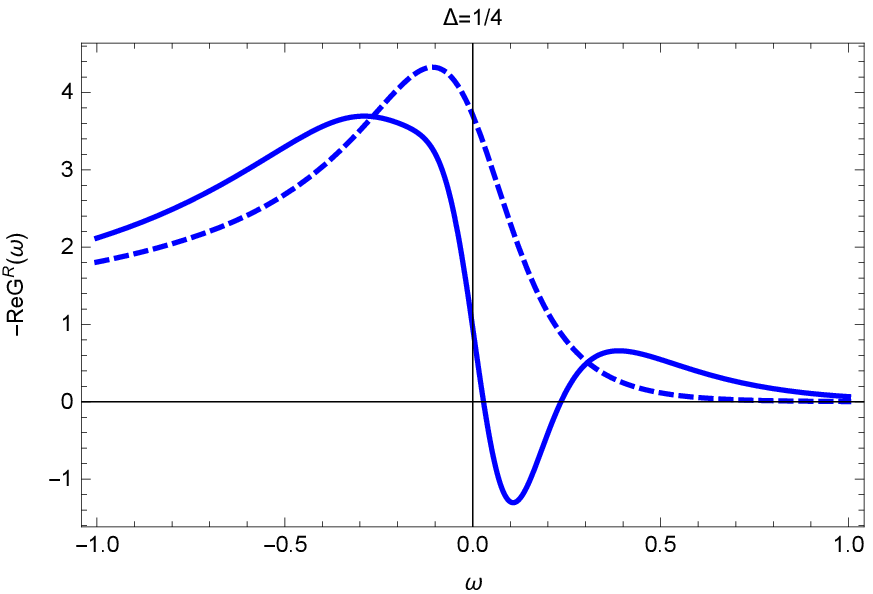}  } \quad
   \subfigure[ -Im$G^R(\omega)\sim$ DOS ]
   {\includegraphics[scale=0.46]{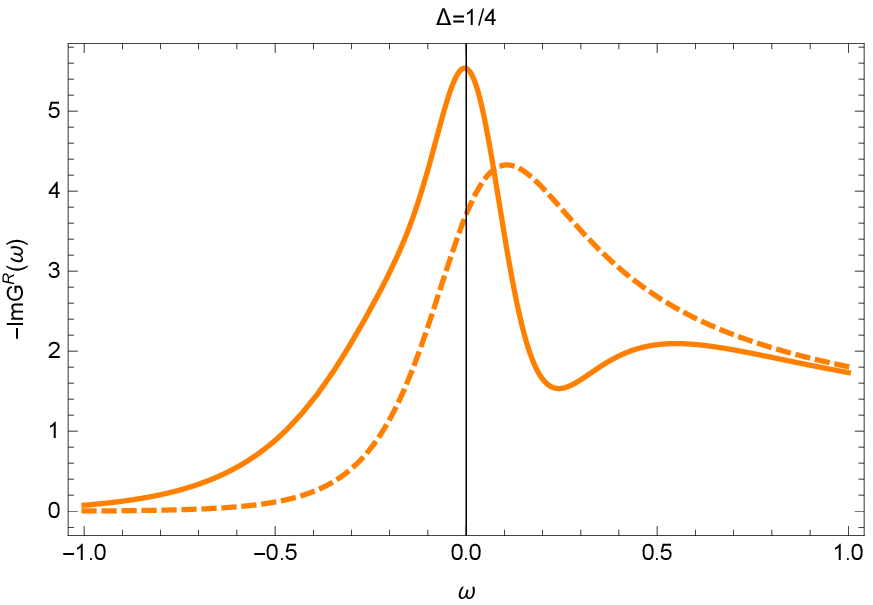}  }
   \subfigure[ Re$\chi_{\text{loc}}^{(2)}(\omega)$ ]
   {\includegraphics[scale=0.46]{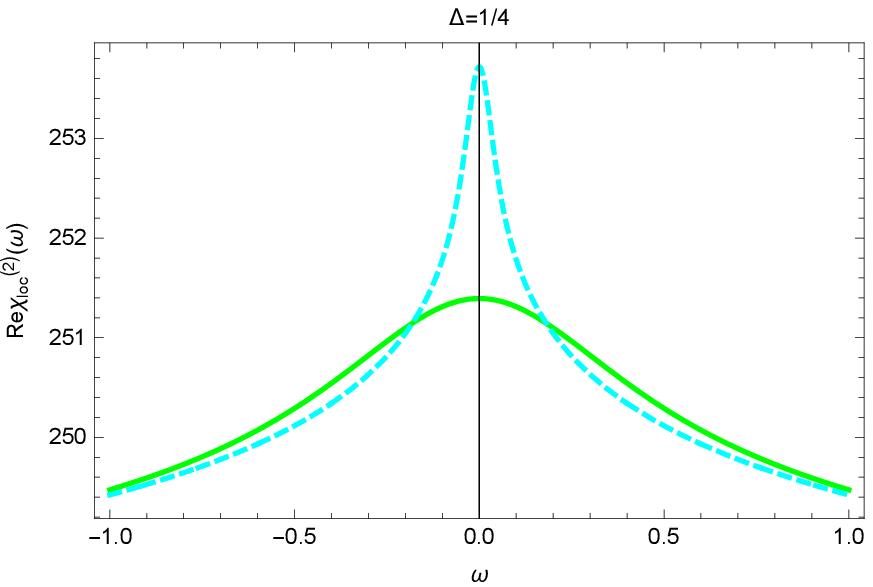}  } \quad
   \subfigure[ Im$\chi_{\text{loc}}^{(2)}(\omega)$ ]
   {\includegraphics[scale=0.46]{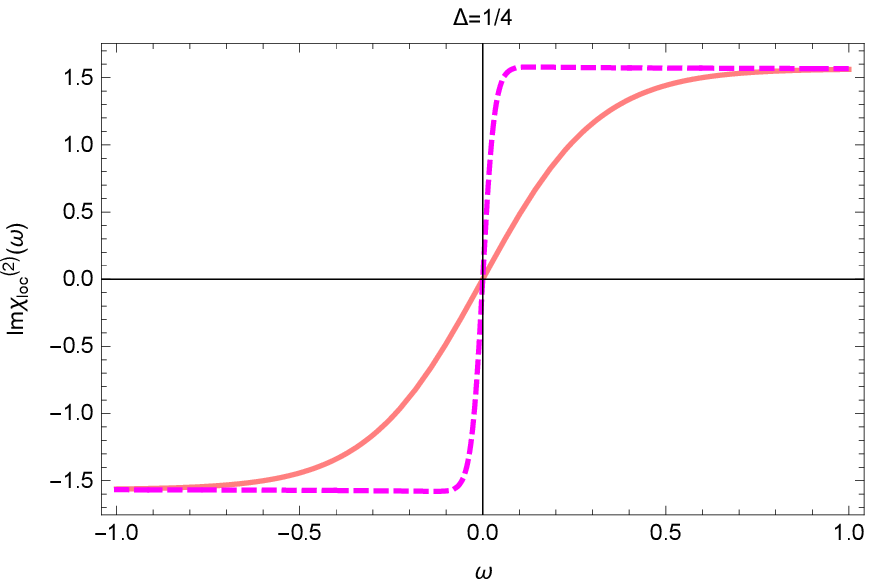}  }
    \caption{Dynamical susceptibility or thermal retarded Green's functions $\chi(\omega) = G^R(\omega)$ given in Eq.(\ref{Eq:GR_loop}) for quantum liquid with (solid line) or without (dashed line) Schwarzian correction. For $\Delta=1/4$ cases:  (a) -Re~$G^R(\omega)$ (blue line); (b) -Im~$G^R(\omega)$ (orange line). We have chosen $\beta=2\pi$ and $C=1/(2\pi)$. Local dynamical spin-spin correlation functions of quantum liquid $\chi_{\text{loc}}^{(2)}(\omega) $ as given in Eq.(\ref{Eq:S_Delta-omega}) in high temperature case with $\beta=2\pi$ (solid green/pink line) and low temperature case with $\beta=20\pi$ (dashed cyan/magenta line): (c) Re~$\chi_{\text{loc}}^{(2)}(\omega)$, (d) Im~$\chi_{\text{loc}}^{(2)}(\omega) \sim \tanh(\omega\beta/2)$. To avoid singularity of $\chi_{\text{loc}}^{(2)}(\omega)$ at $\Delta=1/4$, we have chosen $\Delta=1/4-\epsilon$ with $\epsilon = 10^{-3}$.
    } \label{fig3}
\end{figure}

In Fig.\ref{fig3}(a-b), we show the retarded Green's function at finite temperature with $\beta=2\pi$ and coupling $C=1/(2\pi)$ for $\Delta=1/4$ case. It is worthy of noticing that the real part of $\chi^{(2)}(\omega)$ owns a peak at $\omega=0$ and decays with the increasing of $\abs{\omega}$, and is expected to be a delta function $\delta(\omega)$ at $\omega=0$ in the zero temperature limit as shown in Fig.\ref{fig3}(c). While the imaginary part of local dynamical susceptibility, i.e., $\chi''(\omega)\propto $Im$\chi^{(2)}(\omega)$ behaviors like a smoothness function $\sim\tanh{\omega}$ as given in Eq.(\ref{Eq:Im-chi2_loc}) and shown in Fig.\ref{fig3}(d), which is expected to be a step function jumping at $\omega=0$ in the zero temperature limit.

\emph{Boson and fermions}--
In this paper, we mainly focus on the bosonic retarded Green's function of quantum liquid with Schwarzian correction, a similar procedure might be imposed to fermion's case, which leads to NFL underlying fundamental Dirac or Weyl fermions~\cite{Sachdev:2015efa,Cai:2016jxd,Bi:2017yvx,Wu:2018vua}. For $\Delta=1/4$ case, one just recovers the fractionalized Fermi liquid of lattice Anderson model~\cite{Sachdev:2010um}. To obtain thermal fermionic retarded Green's function of quantum liquid, one needs to solve the wave functions of Dirac fermions in $(1+1)$-dimensional spacetime in global AdS$_2$ coordinate. The exact solutions to the $2$-D Dirac fermion wave functions are shown in Appendix.~\ref{app:Fermion_AdS2}. 

\subsubsection{$\Delta=1/3$: quantum liquid}

In this section, we study the spectral functions of a specific quantum liquid with Schwarzian correction with a conformal dimension $\Delta=1/3$. This is an intriguing phase between Schwarzian NFL phase ($\Delta=1/4$) and Schwarzian spin glass ($\Delta=1/2$) as will be discussed in more detail in the following section.

\begin{figure}[ht!]
\centering
   \subfigure[ -Re$G^R(\omega)$ ]
   {\includegraphics[scale=0.46]{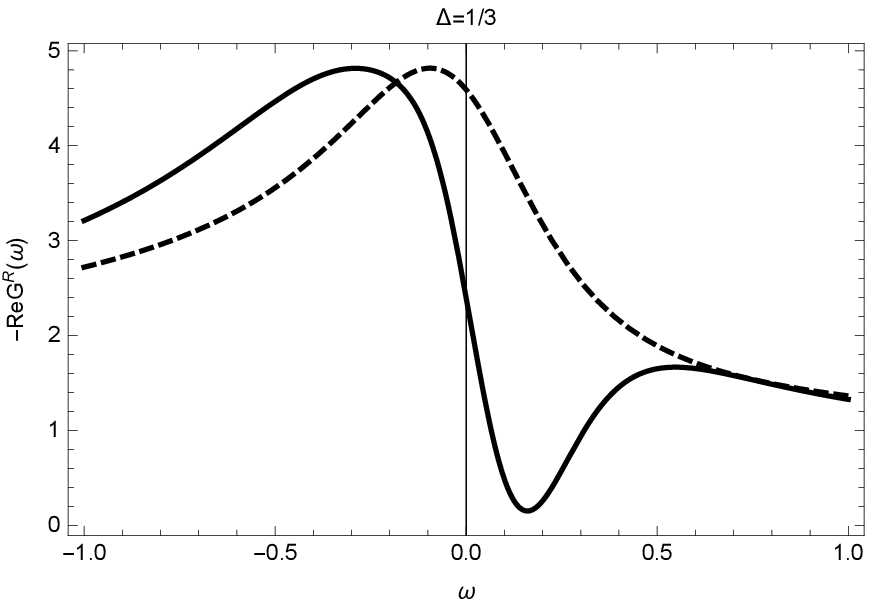}  } \quad
   \subfigure[ -Im$G^R(\omega)\sim$ DOS ]
   {\includegraphics[scale=0.46]{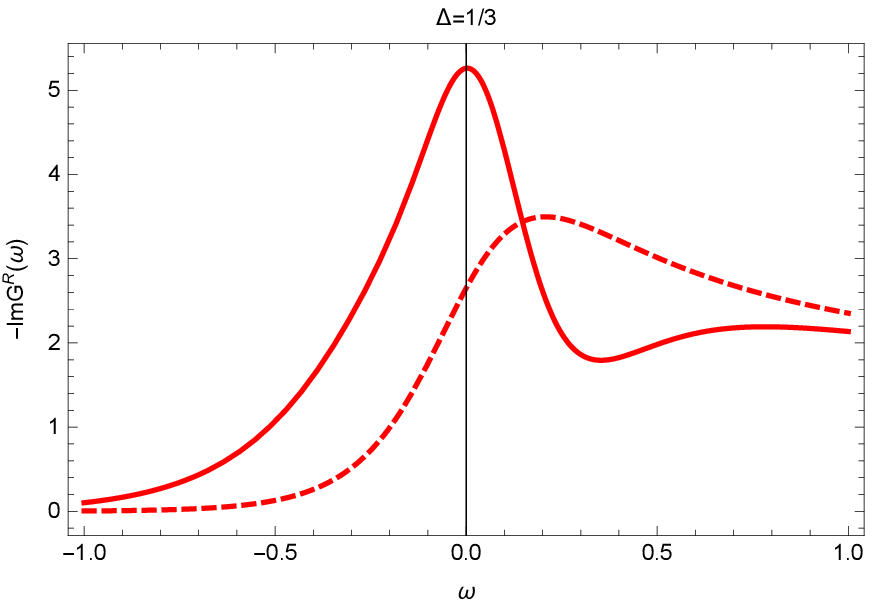}  }
   \subfigure[ Re$\chi_{\text{loc}}^{(2)}(\omega)$ ]
   {\includegraphics[scale=0.47]{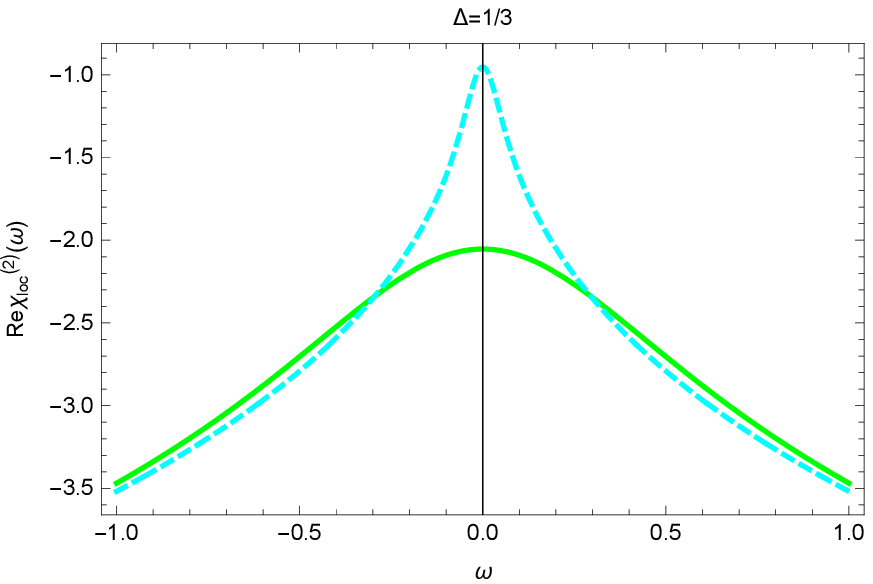}  } \quad
   \subfigure[ Im$\chi_{\text{loc}}^{(2)}(\omega)$ ]
   {\includegraphics[scale=0.47]{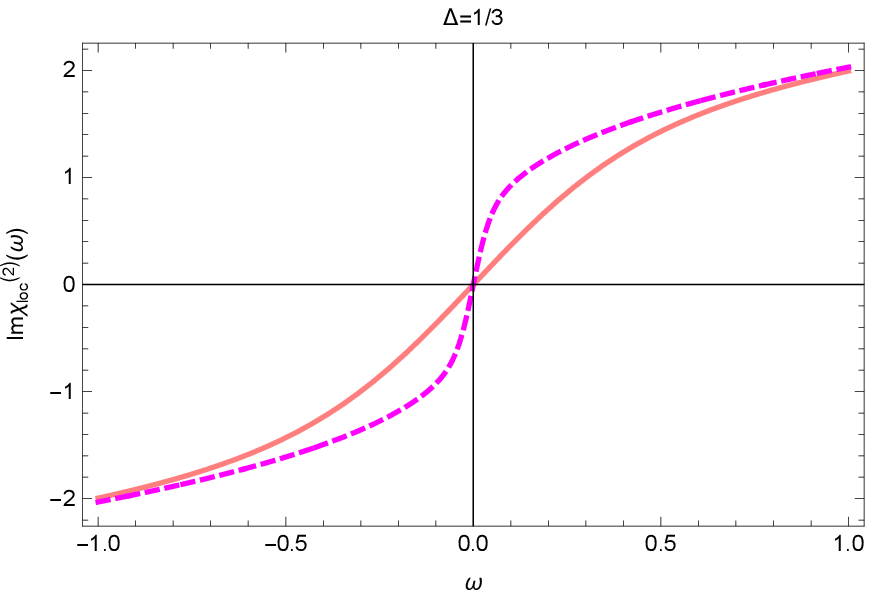}  }
    \caption{Susceptibility or thermal retarded Green's functions $\chi(\omega) = G^R(\omega) $ of quantum liquid with (solid line) or without (dashed line) Schwarzian correction as given in Eq.(\ref{Eq:GR_loop}). For $\Delta=1/3$ case: (a) -Re~$G^R(\omega)$ (black line); (b) -Im~$G^R(\omega)$ (red line).
    We have chosen $\beta=20\pi$ and $C=1/(2\pi)$.
    Local dynamical spin-spin correlation functions of quantum liquid $\chi_{\text{loc}}^{(2)}(\omega) $ as given in Eq.(\ref{Eq:S_Delta-omega}): (c) Re~$\chi_{\text{loc}}^{(2)}(\omega)$, (d) Im~$\chi_{\text{loc}}^{(2)}(\omega)$ at high temperature with $\beta=2\pi$ (solid green/pink line) or at low temperature with $\beta=20\pi$ (dashed cyan/magenta line)
     } \label{fig4}
\end{figure}

In Fig.\ref{fig4}.(a-b), we show the retarded Green's functions of quantum liquid with or without Schwarzian correction for $\Delta=1/3$ case, and we also plot the corresponding local dynamical susceptibility in Fig.\ref{fig4}.(c-d), which characters the local spin-spin correlation of disordered state.

\subsubsection{$\Delta=1/2$: spin glass}

\begin{figure}[ht!]
\centering
   \subfigure[ Evolution of $\bar{G}^R(\omega)$ with coupling $(2\pi C)^{-1}$ ]
   {\includegraphics[scale=0.48]{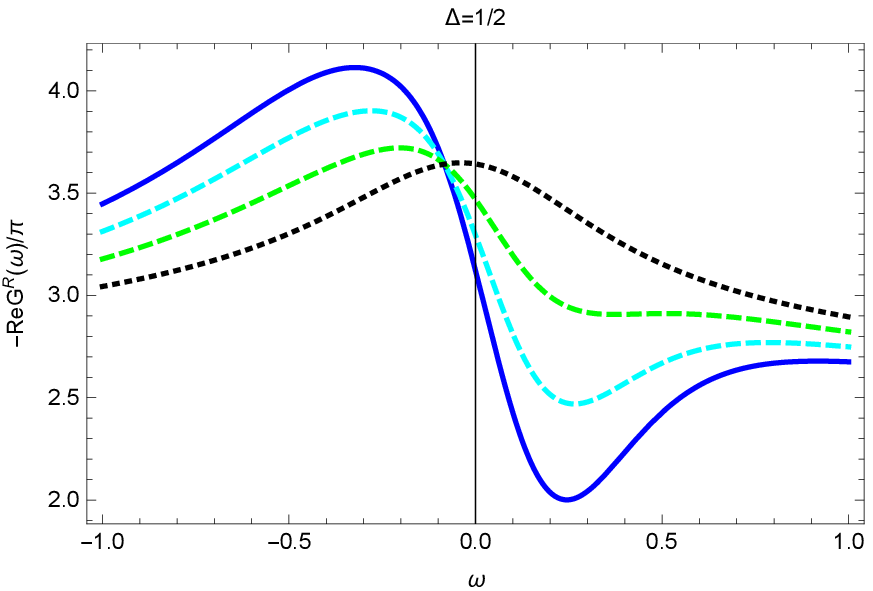} \quad \includegraphics[scale=0.48]{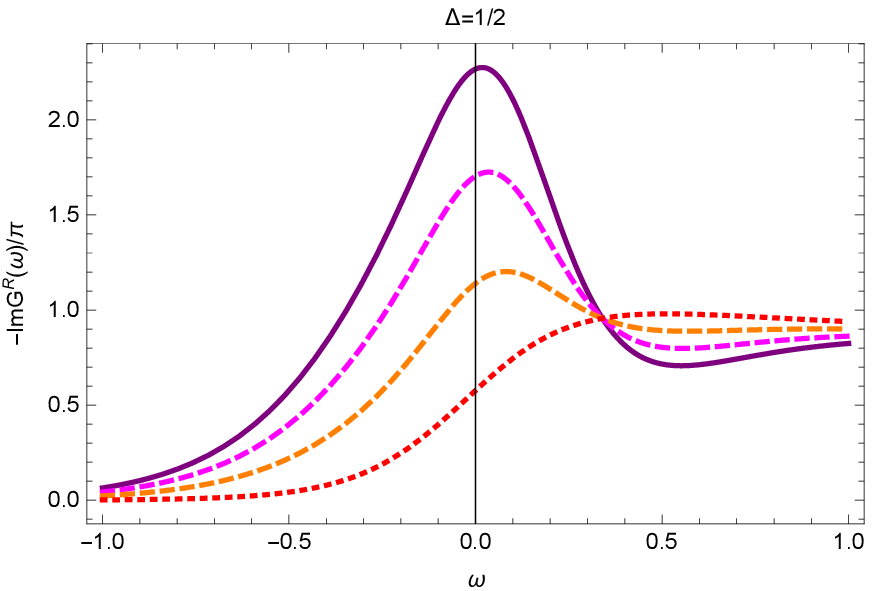}    }
   \subfigure[ Evolution of $\bar{G}^R(\omega)$ with temperature $\beta^{-1}$ ]
   {\includegraphics[scale=0.48]{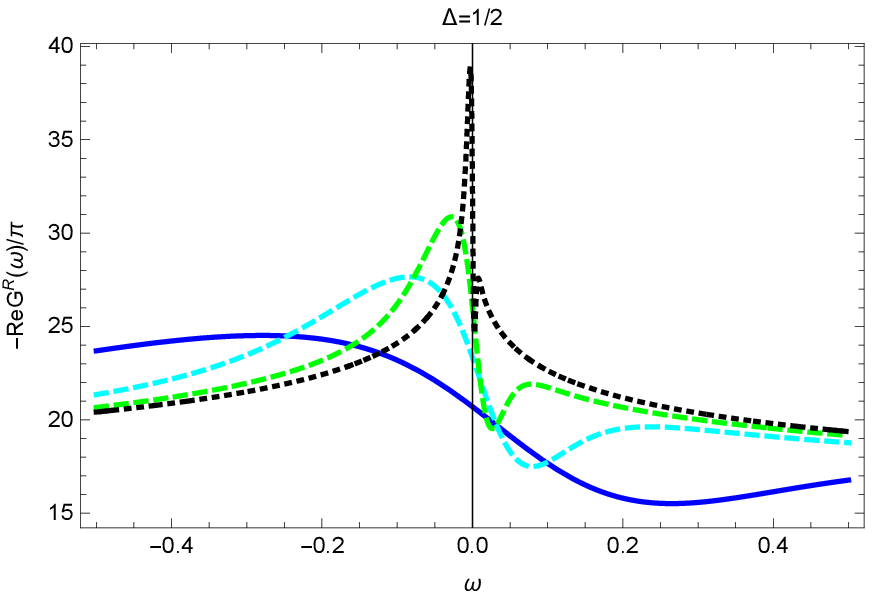} \quad \includegraphics[scale=0.48]{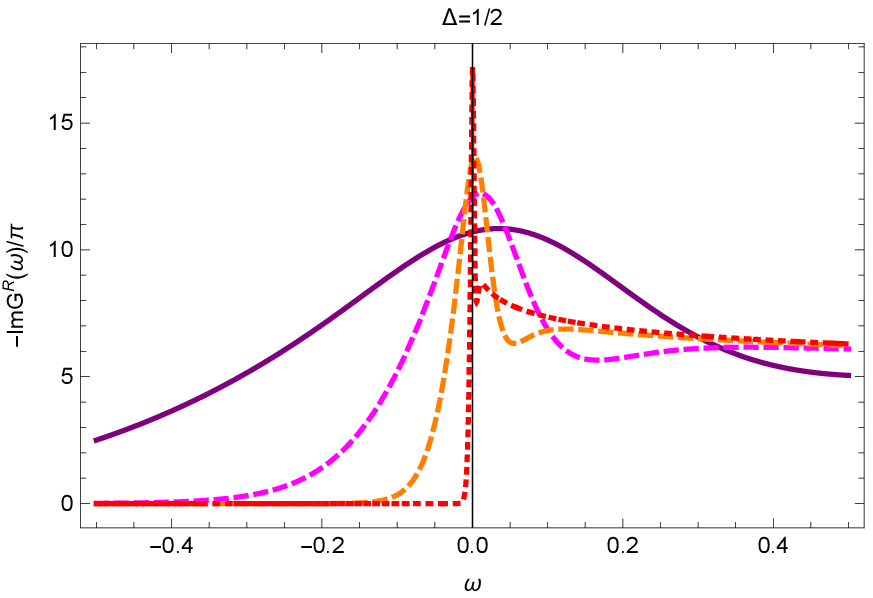}    }
    \caption{Dynamical susceptibility or retarded Green's functions of Schwarzian spin glass with $\Delta=1/2$: $\chi(\omega) = -\bar{G}^R(\omega)/\pi $ is given in Eq.(\ref{Eq:GR_loop}). (a) Evolution with different coupling strength $(2\pi C)^{-1}$: $C=1/3\pi$ (blue/purple solid lines), $C=1/2\pi$ (cyan/magenta dashed line); $C=1/\pi$ (green/orange dashed line) and $C=+\infty$ (black/red dotted line).
    (b) Evolution with different temperature $T$: In front of Eq.(\ref{Eq:GR_loop}), we have multiplying a temperature depending factor $\pi/\beta$. For different $T=\beta^{-1}$ : $\beta=2\pi$ (blue/purple solid lines), $\beta=20\pi/3$ (cyan/magenta dashed line); $\beta=20\pi$ (green/orange dashed line); $\beta=200\pi$ (black/red dotted line).
    We have chosen input parameters as $\beta=2\pi$.
    } \label{fig5}
\end{figure}

The Fig.\ref{fig5}(a) shows the evolution of retarded Green's function with respect to the coupling coefficient $ C^{-1}$ (where $C\sim C_g \bar\phi_r$ ), for spin glass (with $\Delta=1/2$) with Schwarzian correction, in terms of ``Schwarzian spin glass''. The coupling coefficient $ C^{-1}$ (where $C\sim C_g \bar\phi_r$ ) characterizes the coupling strength of pNGBs loop corrections to matter two point correlation functions, according to Eq.(\ref{Eq:<C2>-t12}). The $C=\infty$ case (dotted red curves) corresponds to conventional spin glass, and the local susceptibility becomes exact step function $\theta(\omega)$ in the limit $\epsilon \to 0$. With the increasing of the coupling strength (or decreasing of $C$) up to $C^{-1}=3\pi$ (solid purple line), the DOS accumulates more in the $\omega=0$ region and develops a small dig at the $\omega\approx 0.5$.

In Fig.\ref{fig5}(b), we also show the evolution of retarded Green's functions of Schwarzian spin glass with respect to the temperature. By decreasing temperature (or by increasing $\beta$) from $T=1/(2\pi)$ down to $T=1/(20\pi)$ as well as $T=1/(200\pi)$ (black/red dotted line). As expected, the DOS spread out among the frequency space at finite temperature, but there is still a peak at $\omega=0$, and a plateau in the $\omega>0$ region at low temperature limit as $T \to 0$.


\subsubsection{Large $p$ or small $\Delta$ behavior}

\begin{figure}[ht!]
\centering
   \subfigure[ $\Delta=1/8$ ]
   {\includegraphics[scale=0.48]{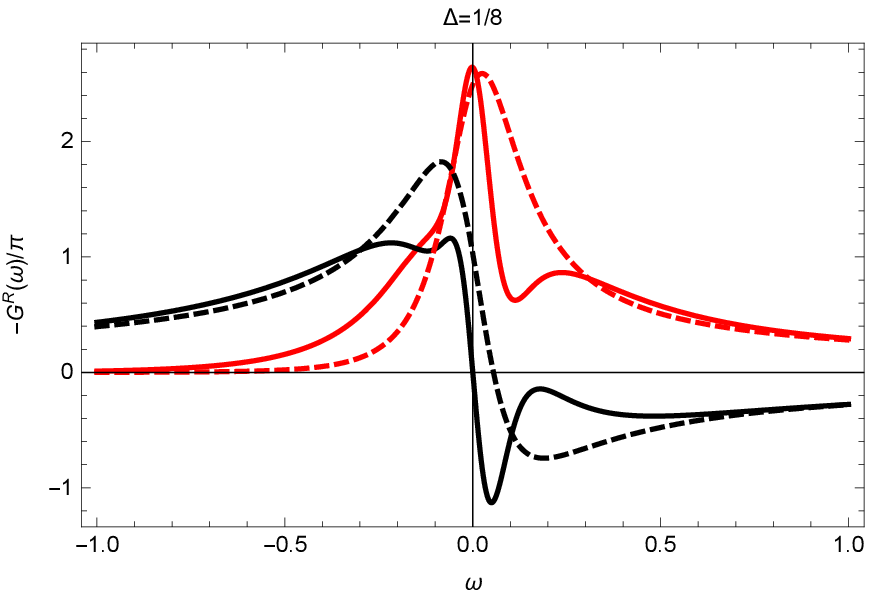}  }\quad
   \subfigure[ $\Delta=1/16$ ]
   {\includegraphics[scale=0.48]{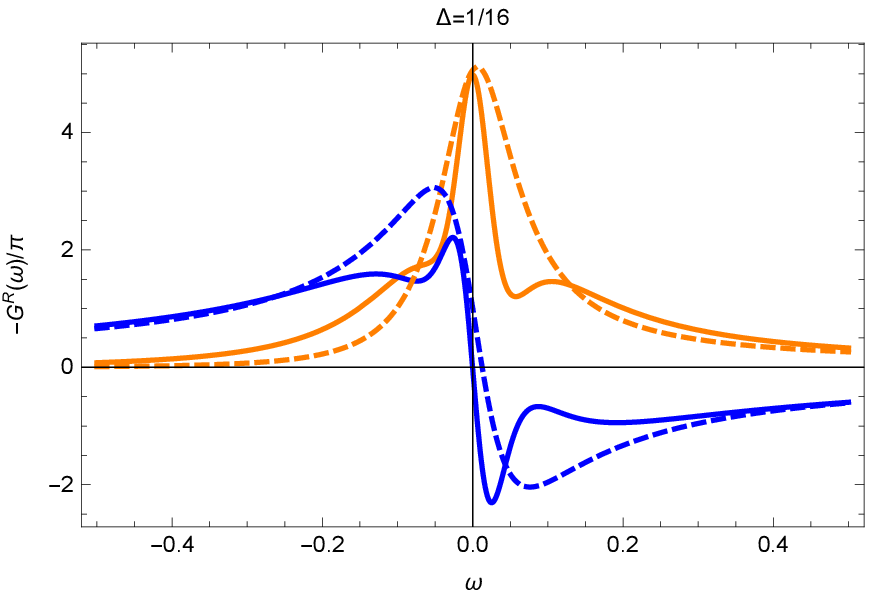}  }\\
   \subfigure[ $\Delta=1/32$ ]
   {\includegraphics[scale=0.48]{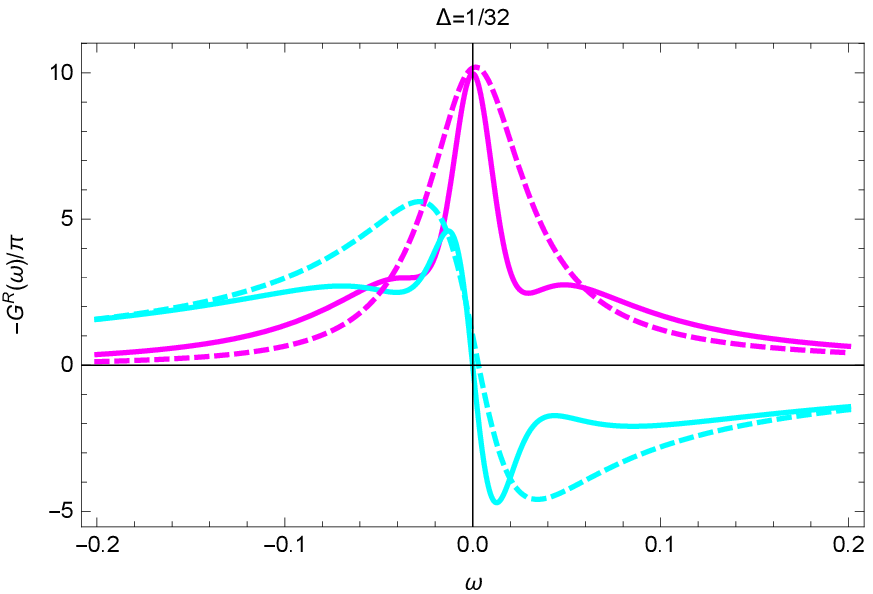}  }\quad
   \subfigure[ $\Delta=1/64$ ]
   {\includegraphics[scale=0.48]{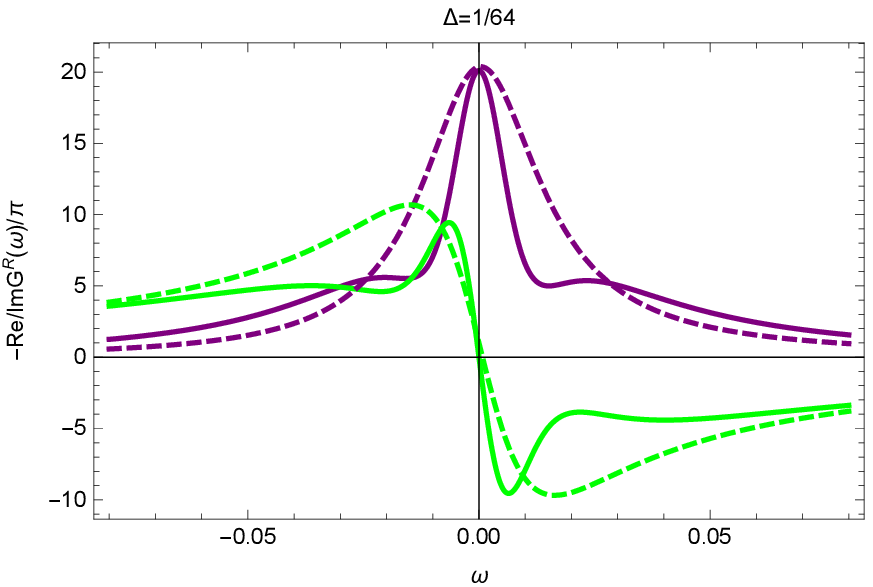}  }\\
   \subfigure[ Re$\chi_{\text{loc}}$ ]
   {\includegraphics[scale=0.48]{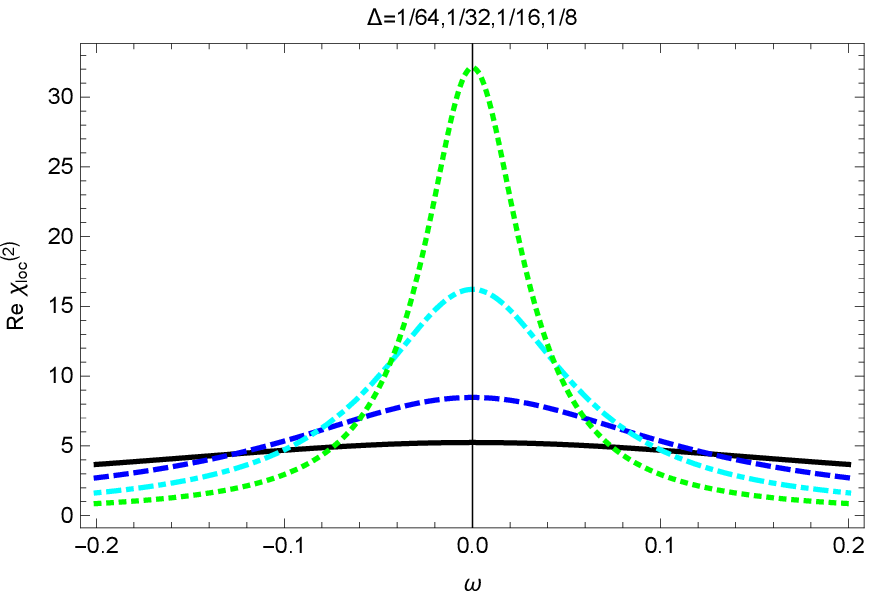}  }\quad
   \subfigure[ Im$\chi_{\text{loc}}$ ]
   {\includegraphics[scale=0.48]{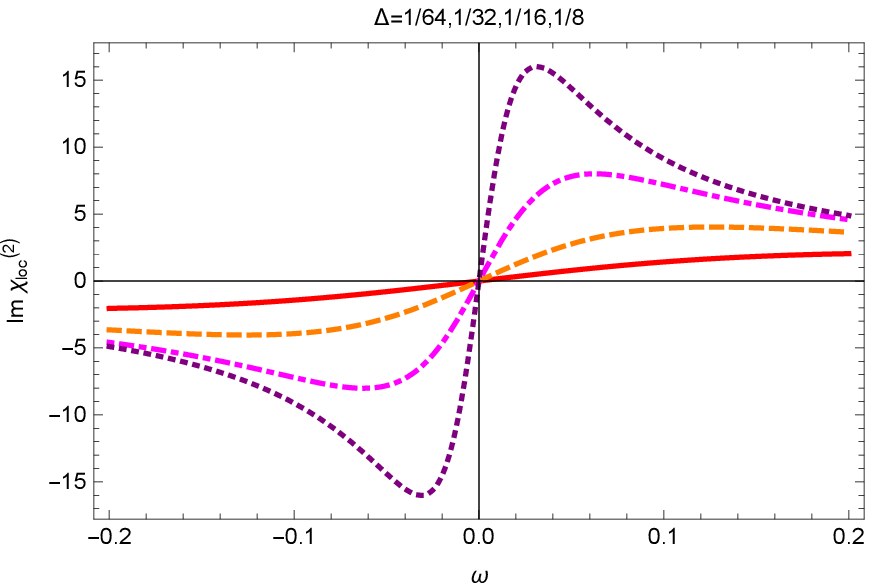}  }
    \caption{Retarded Green's functions $G^R(\omega) $ of quantum liquid with (solid line) or without (dashed line) Schwarzian correction as given in Eq.(\ref{Eq:GR_loop}). Re/Im $G^R(\omega)$ for
    (a) $\Delta=1/8$ (black or red line); (b) $\Delta=1/16$ (blue or orange line) ; (c) $\Delta=1/32$ (cyan or magenta line) ; (d) $\Delta=1/64$ (green or purple line). For all six cases, there are clean signatures of Hubbard band in DOS. Local dynamical susceptibility $\chi^{(2)}(\omega)$ with different $p\equiv 1/\Delta$ ($p=8,16,32,64$ corresponds to solid, dashed, dot-dashed and dotted lines, respectively): (e) Re$\chi^{(2)}(\omega)$; (f) Im$\chi^{(2)}(\omega)$.
    We have chosen input parameters as $\beta=2\pi$ and $C=1/(2\pi)$. With the increase of $p$, the density spectral function becomes more central localized at low frequency region, i.e., $\omega\sim 0$.
    } \label{fig6}
\end{figure}

In this section, we consider the physical consequence when $\Delta$ becomes smaller as shown in Fig.\ref{fig6}. This is equivalent to increasing the number of interacting particles, i.e., $p \equiv 1/\Delta$~\cite{Maldacena:2016hyu}).
As stated before in the introduction section, the $p$-fermion interacting vertex with $p\ge 4$ is only UV relevant in $(0+1)$-dimensional spacetime. Without loss of generality, we chose some specific value for conformal dimension as $\Delta=1/8,/16,1/32,1/64$, respectively. For larger $p$ or smaller $\Delta$, the spectral functions Im$G^R(\omega)$ becomes more sharper at $\omega=0$. Consequently, the life time of the quasi-particle becomes longer as shown in Fig.\ref{fig6}. Conversely, for smaller $p$ or larger $\Delta$, the life time becomes shorter and the DOS shows non-quasi-particle behavior at low frequency.

With the increasing of $p=\Delta^{-1}$, or the decreasing of conformal dimension $\Delta$ from $\Delta=1/8$ to $\Delta=1/64$, the spectral functions Im$G^R(\omega)$ become more and more centralized at $\omega=0$. It shows more metallic behavior at low frequency region, meanwhile the depth of Hubbard band increases towards low frequency region, and so does the location of Hubbard band.

\subsection{High order local spin-spin correlation}

By using the tree level retarded (real time) Green's function in Eq.(\ref{Eq:GR-Liouville}), it is straightforward to calculate the local spin-spin correlation function, namely the dynamical local spin susceptibility $\chi_{\text{loc}}(\omega)$ as defined in Eq.(\ref{Eq:chi_loc})
\beqa
 \chi_{\text{loc}}^{(2)}(\omega) = \int_{-\infty}^{+\infty} \theta(t) dt e^{i\omega t} G_2(t)G_2(-t) . \qquad \label{Eq:chi_T=0}
\eeqa
By using retarded Green's function defined in Eq.(\ref{Eq:GR-Liouville}), and according to Eq.(\ref{Eq:integ1-chi}), we are able to calculate the local spin susceptibility at zero temperature as
\beqa
 \chi_{\text{loc}}^{(2)}(\omega)   &=& \int_{-\infty}^{+\infty} dt ~\theta(t)  \frac{e^{i\omega t}}{t^{4\Delta}} = (-i\omega)^{-1+4\Delta}   \Gamma (1-4 \Delta ) \nn\\
 && = e^{-i\frac{\pi}{2}(4\Delta-1)} \omega^{-1+4\Delta} \Gamma (1-4 \Delta ),  \qquad\quad \label{Eq:S_Delta-omega-T=0}
\eeqa
where the conformal dimension is limited as $0<\Delta < {1}/{4}$ and the frequency must be in the upper complex plane $\text{Im}\omega>0 $. We have also used $(-i)=e^{-i\frac{\pi}{2}}$ in the last equality of above equation.
For finite temperature case, by making a rescaling $t\to (2\pi/\beta)t$ or $\omega\to \beta/(2\pi) \omega$, the dynamical local spin susceptibility at finite temperature becomes
\beqa
 \chi_{\text{loc}}^{ (2)T }(\omega) \equiv   \int_{-\infty}^{+\infty} \theta(t) dt e^{i\omega t} G_2\Big(\frac{2\pi}{\beta}t\Big)G_2\Big(-\frac{2\pi}{\beta}t\Big) . \qquad \qquad\label{Eq:chi_T!=0}
\eeqa
By using Eq.(\ref{Eq:GR-Liouville}), one has
\beqa
&& \chi_{\text{loc}}^{(2)T}(\omega)   =   \bigg( \frac{\beta}{2\pi} \bigg)^{1-4\Delta} \frac{\Gamma (1-4 \Delta ) \Gamma \Big(2 \Delta -i \frac{\beta}{2\pi} \omega \Big)}{\Gamma \Big(1 - 2 \Delta -i \frac{\beta}{2\pi} \omega\Big)}. \qquad \nn\\
&& \qquad 0<\Delta < \frac{1}{4}, \quad \beta>0, \quad \text{Im}\omega>0. \label{Eq:S_Delta-omega}
\eeqa
Based upon which, the higher order local spin susceptibility with respect to the frequency becomes
\beqa
&& \chi_{\text{loc}}^{ (3)T  }(\omega) = \frac{2\pi}{\beta} \partial_\omega \chi_{\text{loc}}^{(2)T}(\omega) , \nn\\
&& \chi_{\text{loc}}^{ (4)T  }(\omega) = \Big( \frac{2\pi}{\beta} \Big)^2 \partial_\omega^2 \chi_{\text{loc}}^{(2)T}(\omega). \qquad \quad
\eeqa
By imposing Eq.(\ref{Eq:S_Delta-omega}), one obtains leading higher order local spin susceptibility as
\beqa
 \{ \chi_{\text{loc}}^{(3)T   }, \chi_{\text{loc}}^{(4)T } , \chi_{\text{loc}}^{(5)T } \} &=& \chi_{\text{loc}}^{(2)T } \{ -\psi^{(0)}_{x}, \quad  \big(\psi^{(0)}_{x}\big)^2 + \psi^{(1)}_{x} , \nn\\
&& - [   \big(\psi^{(0)}_{x}\big)^3 + 3 \psi^{(0)}_{x}\psi^{(1)}_{x}  +\psi^{(2)}_{x} ] \}, \nn
\eeqa
where $x\equiv 2\Delta-i\omega {\beta}/{(2\pi)}$, the prime is with respect to the frequency and we have used the definition of functions defined in Eq.(\ref{Eq:psi_n-new}).

\subsubsection{Static local susceptibility}

The leading order low frequency behavior of local spin susceptibility is a constant, i.e., $\chi(\omega) = \text{const}. + O(\omega) $, in which, the constant term is inverse proportional to the temperature as
\beqa
\chi_{\text{loc}}^{(2)}(0) 
= \bigg( \frac{\beta}{\pi}\bigg)^{1-4 \Delta } \frac{ \Gamma \left(\frac{1}{2}-2 \Delta \right) \Gamma (2 \Delta )}{2 \sqrt\pi} .
\eeqa
While it turns out that for the special $\Delta=1/4$ case, the imaginary sector of $\chi(0)$ is divergent
\beqa
\chi_{\text{loc}}^{(2)}(0)  \propto
\left\{ \begin{aligned}
& -\frac{\pi}{\beta}, \quad \Delta = {1}/{2} \\
& -\frac{1}{4\Delta-1} + \log\frac{2\beta}{\pi}, \quad \Delta = {1}/{4}. 
\end{aligned} \right. \qquad
\eeqa
Consider first derivative of $\chi_{\text{loc}}''(\omega)$ with respect to frequency $\omega$, one obtains that the static local spin susceptibility $\chi_{\text{loc}}^{(3)}(0)$ is inversely proportional to the square of temperature, i.e., $\beta^{2}$ as
\beqa
\chi_{\text{loc}}^{(3)}(0) = \frac{i}{2}\sqrt\pi \bigg(\frac{\beta}{\pi}\bigg)^{1-4\Delta} \ii \cot{(2\pi\Delta)} \Gamma(\frac{1}{2}-2\Delta)\Gamma(2\Delta) . \qquad\quad
\eeqa
from which, it turns out that for the special $\Delta=1/2$ case, the imaginary part of $\chi_{\text{loc}}^{(3)}(0)$ is divergent
\beqa
\chi_{\text{loc}}^{(3)}(0)  \propto
\left\{ \begin{aligned}
& \frac{\pi}{\beta}\bigg( -\frac{1}{2\Delta-1} + 2 + 2\log\frac{\beta}{2\pi} \bigg), \quad \Delta = {1}/{2}  \\
& \frac{\pi^2}{2}, \quad \Delta = {1}/{4}. 
\end{aligned} \right.
\eeqa
While at the $2$-nd order derivative of $\chi_{\text{loc}}''(\omega)$ with respect to $\omega$, the static local spin susceptibility $\chi''''(0)$ for both $\Delta=1/2$ and $\Delta=1/4$ case,  becomes convergent and is inversely proportional to the cubic of temperature, i.e., $\beta^3$ as
\beqa
\chi_{\text{loc}}^{(4)}(0)  \propto
\left\{ \begin{aligned}
& \frac{2}{3}\frac{\pi^3}{\beta} , \quad \Delta = {1}/{2}  \\
& 14\zeta(3),  \quad \Delta = {1}/{4}. 
\end{aligned} \right.
\eeqa
where $\zeta (3)\approx 1.20206$ is the Riemann zeta function.

The effective bath for the local spin is given by the local spin-spin correlation function itself, which have nontrivial low frequency behavior, which appears only as a subdominant correction to the leading low frequency behavior $\chi^{(n)}(0)\sim \text{const}$ given $\beta$.


\subsubsection{Marginal NFL}

For $\Delta=1/4$ case,
\beqa
G^R(t)   \sim   \sqrt\frac{\pi}{\beta\sinh\frac{\pi t}{\beta}} ,
\eeqa
the result just recovers the retarded Greens'f function of fractionalized Fermi liquid phase of the lattice Anderson model~\cite{Sachdev:2010um}, which can be obtained in an analogy procedure for fermion case, by solving $2$-D Dirac equation as shown in Appendix.~\ref{app:Fermion_AdS2}. More generally, it is a special case of NFL in a doped Mott insulator~\cite{Parcollet:1999,Sachdev:2015efa,Cai:2016jxd,Bi:2017yvx,Wu:2018vua}.


For the $\Delta=1/4$ case, the  low frequency behavior of local spin susceptibility $\chi_{\text{loc}}^{(2)T}(\omega)$ is given by Eq.(\ref{Eq:chi-omega-Delta=1/4}).
\beqa
\chi_{\text{loc}}^{(2)T}(\omega) 
&=& \log \left(\frac{\beta }{2 \pi }\right)+\frac{1}{1-4 \Delta }-\gamma_E  -\psi ^{(0)}\left(\frac{1}{2} - i \frac{  \beta  \omega }{2 \pi }\right),   \nn\\
&& \label{Eq:chi-omega-Delta=1/4}
\eeqa
where we have used that $\psi(-{1}/{2})  = \psi({3}/{2}) = 2-\gamma_E - 2\log (2)$ and $\psi({1}/{2}) =-\gamma_E - 2\log (2) $. Thus, one obtains the universal form for low frequency behavior of the dynamical local spin-spin correlation susceptibility~\cite{Parcollet:1999}
\beqa
 \text{Im} \chi_{\text{loc}}^{(2) T}(\omega) = -\text{Im} \psi \bigg(\frac{1}{2}- i \frac{\omega \beta}{2\pi} \bigg) = \frac{\pi}{2} \tanh\frac{\omega}{2T} . \qquad \label{Eq:Im-chi2_loc}
\eeqa
which is simply a smoothed-out version of the step function at zero temperature ($\text{Im} \chi_{\text{loc}}^{(2)}(\omega) =  \pi \tanh(\pi \omega)$) or in large frequency limit, i.e., $T\to 0$, or $T\ll \omega$.
The local dynamical susceptibility implies that~\cite{Virosztek:1990}
\beqa
\chi_{\text{loc}}''(\omega)  \propto
\left\{ \begin{aligned}
& \frac{\omega}{2T}, \quad \omega \ll T \\
& \text{sgn}(\omega), \quad \omega \gg T 
\end{aligned} \right.
\eeqa
which is precisely of the form for spin and charge fluctuations in the phenomenological ``marginal non-Fermi liquid'' (mNFL) description of High-T$_c$ cuprates in the strange metal region. The marginal critical point can be viewed as a concrete realization of the \emph{bosonic fluctuation spectrum} needed to support a mNFL. In this case
\beqa
\chi_{\text{loc}}^{\prime}(\omega) \equiv \int d\omega \frac{\chi_{\text{loc}}''(\omega)}{\omega}  
&&\sim  \log\frac{1}{\abs{\omega}} , \quad \omega \gg T,
\eeqa
which just recovers the spin-polarization correlation function of ``marginal NFL''~\cite{Virosztek:1990,Chou:1995,Parcollet:1999}.

For higher order local spin-spin correlation functions,  at finite temperature case, one has
\beqa
\chi_{\text{loc}}^{(3)T } &=& \psi ^{(1)}\left(\frac{1}{2} -  i\frac{ \beta  \omega }{2 \pi }\right) , \nn\\
 \chi_{\text{loc}}^{(4)T } &=& -\psi ^{(2)}\left( \frac{1}{2} - i\frac{ \beta  \omega }{2 \pi }\right) .
\eeqa
Thus, one obtains
\beqa
 \text{Re}  \chi_{\text{loc}}^{(3)T }(\omega) &=&  
 \frac{\pi^2}{2}   \text{sech}^2( \frac{\beta}{2}\omega ), \nn\\
  \text{Im} \chi_{\text{loc}}^{(4)T }(\omega) 
  &=& \pi^3 {\tanh (  \frac{\beta}{2}\omega ) }{ \text{sech}^2( \frac{\beta}{2}\omega ) },  \qquad
\eeqa
where in the last equality, we have used reflection principle in Eq.(\ref{Eq:psi-reflection-1}) as
\beqa
\text{Im} \psi \bigg(\frac{1}{2}+i \frac{\beta}{2\pi} \omega \bigg) 
&=& \frac{\pi}{2} \tanh(\frac{\beta}{2}\omega) , \nn\\
\text{Re} \psi^{(1)} \bigg(\frac{1}{2}+i \frac{\beta}{2\pi}\omega \bigg) &=& \frac{\pi ^2}{2}  \text{sech}^2(  \frac{\beta}{2}\omega ),  \label{Eq:psi-reflection-1} \\
\text{Im} \psi^{(2)} \bigg(\frac{1}{2}+i \frac{\beta}{2\pi}\omega \bigg) &=&  \pi ^3 \tanh (  \frac{\beta}{2}\omega ) \text{sech}^2(  \frac{\beta}{2}\omega ). \nn
\eeqa

\subsubsection{Spin glass}

For $\Delta=1/2$ case,
\beqa
G^R(t)   \sim   \frac{\pi}{\beta\sinh\frac{\pi t}{\beta}},
\eeqa
the result also recovers the retarded Greens's function of spin glass~\cite{Sachdev:1992fk,Grempel:1998,Grempel:1999,Grempel:2001}, which describes the quantum fluctuations near a critical quantum Heisenberg spin glass.
The  low frequency behavior of local spin susceptibility $\chi_{\text{loc}}''(\omega)$ is given by Eq.(\ref{Eq:chi-omega-Delta=1/2}) as
\beqa
\chi_{\text{loc}}^{(2)T}(\omega)
&& = \frac{\pi }{\beta }+i \omega  \bigg[ 1 - \gamma_E +\frac{1}{2-4 \Delta } + \log \left(\frac{\beta }{2 \pi }\right) \nn\\
&& - \psi ^{(0)}\left(-\frac{i \beta  \omega }{2 \pi }\right) \bigg],
\eeqa
from which, one obtains
\beqa
\text{Re}\chi_{\text{loc}}^{(2)}(\omega) = \frac{\pi}{\beta} + \omega \text{Im}~\psi \bigg(-\frac{i \beta  \omega }{2 \pi }\bigg) = -\frac{\omega}{2}\coth\frac{\beta\omega}{2} . \qquad \quad
\eeqa
where in the last equality, we have used reflection principle in Eq.(\ref{Eq:psi-reflection-1}).
\beqa
\text{Im} \psi \bigg(i \frac{\beta}{2\pi} \omega \bigg) 
&=& \frac{\pi }{\beta  \omega }+\frac{1}{2} \pi  \coth \left(\frac{\beta  \omega }{2}\right), \nn\\
\text{Re} \psi^{(1)} \bigg(i \frac{\beta}{2\pi}\omega \bigg) &=& -\frac{2 \pi ^2}{\beta ^2 \omega ^2}-\frac{1}{2} \pi ^2 \text{csch}^2\left(\frac{\beta  \omega }{2}\right),  \label{Eq:psi-reflection-2}\\
\text{Im} \psi^{(2)} \bigg(i \frac{\beta}{2\pi}\omega \bigg) &=&  -\frac{8 \pi ^3}{\beta ^3 \omega ^3}-\pi ^3 \coth \left(\frac{\beta  \omega }{2}\right) \text{csch}^2\left(\frac{\beta  \omega }{2}\right). \nn
\eeqa
The higher order susceptibility is
\beqa
\chi_{\text{loc}}^{(3)}(\omega) &=& \frac{2\pi}{\beta} \bigg[ \frac{1 }{2(1-2 \Delta) } +   \log \left(\frac{\beta }{2 \pi }\right) + 1 -\gamma_E\nn\\
&&  -  \psi ^{(0)}\left(-\frac{i \beta  \omega }{2 \pi }\right) \bigg] + i \omega  \psi ^{(1)}\left(-\frac{i \beta  \omega }{2 \pi }\right) , \nn\\
 \chi_{\text{loc}}^{(4)}(\omega) &=&  \frac{4 \pi  }{\beta }\psi ^{(1)}\left(-\frac{i \beta  \omega }{2 \pi }\right) -i \omega  \psi ^{(2)}\left(-\frac{i \beta  \omega }{2 \pi }\right)  . \quad
 \label{Eq:chi-omega-Delta=1/2}
\eeqa
In this case, one obtains
\beqa
\text{Im} \chi_{\text{loc}}^{(3)}(\omega) &=& -\frac{2\pi}{\beta} \text{Im} \psi ^{(0)}\left(-\frac{i \beta  \omega }{2 \pi }\right) + \omega  \text{Re} \psi ^{(1)}\left(-\frac{i \beta  \omega }{2 \pi }\right) \nn\\
&=& \frac{\pi^2}{2\beta}   \frac{ \sinh (\beta  \omega )-\beta  \omega }{\sinh^2 \frac{\beta  \omega }{2}} , \nn\\
\text{Re} \chi_{\text{loc}}^{(4)}(\omega) &=& \omega \text{Im} \psi ^{(2)}\left(-\frac{i \beta  \omega }{2 \pi }\right) + \frac{4\pi}{\beta}\text{Re}\psi ^{(1)}\left(-\frac{i \beta  \omega }{2 \pi }\right) \nn\\
&=& \frac{\pi ^3}{\beta} \frac{ \beta  \omega  \coth \left(\frac{\beta  \omega }{2}\right) -2 }{\sinh^2\frac{\beta  \omega }{2}},
\eeqa
where we have used the refection principle in Eq.(\ref{Eq:psi-reflection-2}).


\section{High point correlation functions}
\label{sec:hpt-corr}

In this section, based on the low energy effective action of the Schwarzian theory of time reparametrization from $2$-D gravity, we calculate the high point functions, especially the $4$-point correlations function~\cite{Muck:1998rr,Aharony:1999ti}. The physical consequence of the four point functions can be detected in the system of quantum chaos~\cite{Shenker:2013pqa,Maldacena:2015waa,Gu:2016oyy,Cotler:2016fpe,Chen:2017dbb,Haehl:2017pak,Blake:2017ris}, which can be characterized by an exponential growth of the thermal out-of-time-order correlating (OTOC)~\cite{Larkin:1969,Roberts:2014ifa,Maldacena:2015waa,Tsuji:2016jbo,Caputa:2017rkm,Blake:2017ris} four point function with a scrambling time $\hat{t}$.

\subsection{Four-point function}

\subsubsection{Finite temperature case}

Suppose one has two operators $\V$ and $\W$ with the same conformal weight $\Delta$, which are dual to two bulk scalar field $\Phi$. The connected four point function is given by
\beqa
F^{(4)} & \equiv &\frac{\vev{\V(t_1)\V(t_2)\W(t_3)\W(t_4)}  }{\vev{\V(t_1)\V(t_2)}\vev{\W(t_3)\W(t_4)}} - 1 , \nn\\
&=& \epsilon^2 \vev{{\mathcal C}_1(t_{12}){\mathcal C}_1(t_{34})}.
\eeqa
By using the correlation function ${\mathcal C}_{1}(t_1,t_2)$ in Eqs.(\ref{Eq:C12}) and the soft mode propagators in Eq.(\ref{Eq:k12-T!=0}), the correlation to the four-point function turns out to be $G^{(4)}(t_1,t_2,t_3,t_4) = F^{(4)}(t_1,t_2,t_3,t_4)/[2\sin{(t_{12}/2)}]^{2\Delta}[2\sin{(t_{34}/2)}]^{2\Delta}$.

According to the relative ordering of the time, there are several possibility, one is $t_1>t_2>t_3>t_4$, in this case one obtains
\beqa
F^{(4)}_{VVWW} =  \frac{2\Delta^2}{\pi C} \bigg( 1 - \frac{t_{12}}{\tan\frac{t_{12}}{2}}\bigg) \bigg( 1 - \frac{t_{34}}{\tan\frac{t_{34}}{2}}\bigg), \qquad \label{Eq:G4_VWVW}
\eeqa 
which can be viewed as arising from energy fluctuations. After recovering the thermal factor $t\to (2\pi/\beta)t$, one just recovers the connected four point function in Eq.(3.131) in Ref.~\cite{Maldacena:2016hyu}. Each two point function ${\mathcal C}_1(t_{12})$ generates an energy fluctuation, which affects each other. This result does not depend on the relative distance between the pair of points. In the double limit of $t_{12} \to 0$ and $t_{34}\to0$, one has
\beqa
F^{(4)}_{\V\V\W\W} = \frac{\Delta^2}{72C\pi} t_{12}^2 t_{34}^2 .
\eeqa
The other results is obtained with the time order $t_1>t_3>t_2>t_4$, in this case, one has
\beqa
F^{(4)}_{\V\W\V\W} 
&=& F^{(4)}_{\V\V\W\W}  \nn\\
&+& \frac{ \Delta^2}{C}\bigg(   \frac{t_{23}}{\tan\frac{t_{12}}{2}\tan\frac{t_{34}}{2}}  - 2 \frac{\sin\frac{t_{23}}{2}\cos\frac{t_{14}}{2}}{\sin\frac{t_{12}}{2}\sin\frac{t_{34}}{2}}  \bigg). \qquad \quad  \label{Eq:G4_VVWW}
\eeqa
which depends on the overall separation of the two pair. In the absence of cross distance $t_{23}$, i.e., when $t_2=t_3$, the result $F^{(4)}_{\V\W\V\W}$ just recovers $F^{(4)}_{\V\V\W\W} $.

\begin{figure}[ht!]
\centering
   {\includegraphics[scale=0.5]{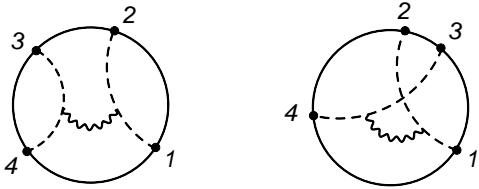}  }
    \caption{Feynman diagrams for four point functions $G_4(t_1,t_2,t_3,t_4)$ of scalar fields with loop corrections from soft modes, as shown in Eq.(\ref{Eq:2pt-4pt-6pt-8pt-T!=0}).
    } \label{fig7}
\end{figure}

\subsubsection{OTOCs}

A simple diagnostic of quantum chaos is consider a square of the commutator by taking an expectation value in some thermal state, by considering a quantity, i.e., the commutator of operators separated in time as~\cite{Roberts:2014ifa,Roberts:2014ifa,Kitaev:2015}
\beqa
&& C(t) = - \vev{[ \V(t), \W(0) ]^2}_\beta = C_1(t) - C_2(t), \\
&& C_1(t) = \vev{ \V(t) \W(0) \W(0) \V(t) } + \vev{ \W(0) \V(t) \V(t) \W(0) }, \nn\\
&& C_2(t) = \vev{ \V(t) \W(0) \V(t) \W(0) } + \vev{ \W(0) \V(t) \W(0) \V(t) }, \nn
\eeqa
where $\W(t)$ and $\V(t)$ are two different operators dual to the source $\Phi_0(t)$, and $\vev{\cdots}_\beta=Z^{-1}\text{Tr}[e^{-\beta H}\cdots]$, where the subscript $\beta$ is introduced to denote the thermal expectation value at temperature $T=\beta^{-1}$. The behavior of $C(t)$ in a chaotic system is
By expanding it, there are four point functions in $C$, two of them consists of $C_1(t)$ in terms of Lorentzian time ordered correlators (TOCs), i.e., $G^{(4)}_{\V\V\W\W}=\vev{ \V(t)\V(t)\W(0)\W(0) }$, while the other two of them are OTOCs of the form $G^{(4)}_{\V\W\V\W}=\vev{ \V(t)\W(0)\V(t)\W(0) }$, which can be used to diagnose chaos. The Feynman diagrams of four point TOCs and OTOCs functions are shown in Fig.\ref{fig7}.

\begin{figure}[ht!]
\centering
   {\includegraphics[scale=0.4]{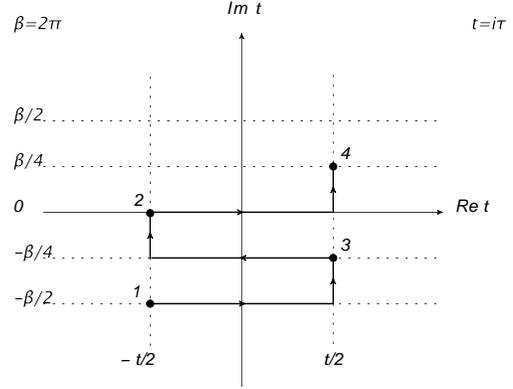}  }\qquad
    \caption{Schwinger-Keldysh four contour for OTOCs four-point, with time chosen as in Eq.(\ref{OTOC-Schwinger-Contour}).
    } \label{fig8}
\end{figure}


With the OTOCs four point function $G^{(4)}_{\V\W\V\W}$, by making the parameterization with SK four-contour~\cite{Schwinger:1960qe,Keldysh:1964ud}, which is depicted in Fig.\ref{fig8}
\beqa
(\hat{t}_i)^{\text{OTOCs}}
&=& \bigg( -\frac{\hat{t}}{2}-i \frac{\beta}{2},  \frac{\hat{t}}{2}  -  i\frac{\beta}{4}, -\frac{\hat{t}}{2}, \frac{\hat{t}}{2} + i \frac{\beta}{4} \bigg)
, \qquad\quad \label{OTOC-Schwinger-Contour}
\eeqa
where $i=1,3,2,4$ and $\beta=2\pi$. $\hat{t}$ is the separation of the early $V$ operator and the later $W$ operators. The contour goes from some initial time $\hat{t}_1$ within Euclidean domain, along the imaginary time axis to some time $\hat{t}_2$, then turns to the Euclidean domain time $\hat{t}_3$ again, and again runs along the imaginary time axis to $\hat{t}_4$.

\begin{figure}[ht!]
\centering
   \subfigure[ -Re$G^{(4)}_{\V\W\V\W}/\pi$ ]
   {\includegraphics[scale=0.49]{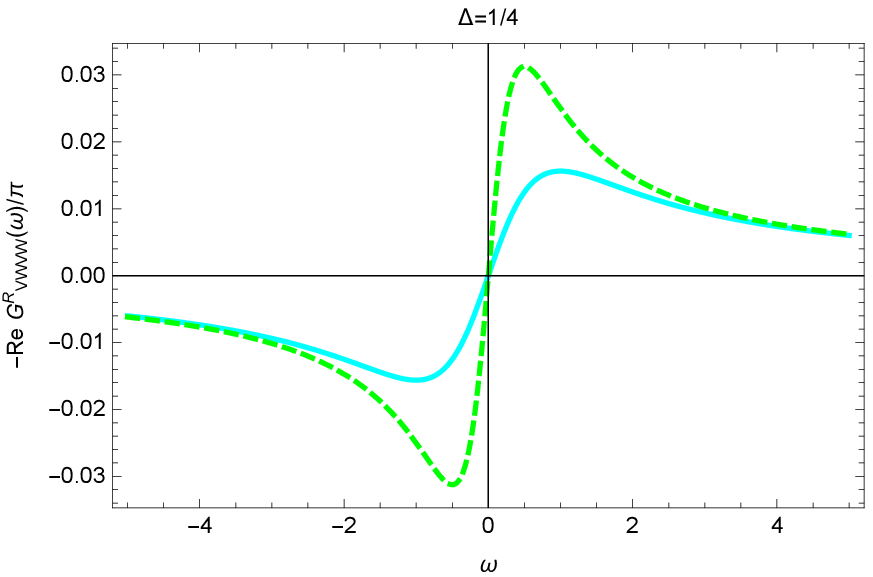}  }
   \subfigure[ -Im$G^{(4)}_{\V\W\V\W}/\pi$ ]
   {\includegraphics[scale=0.49]{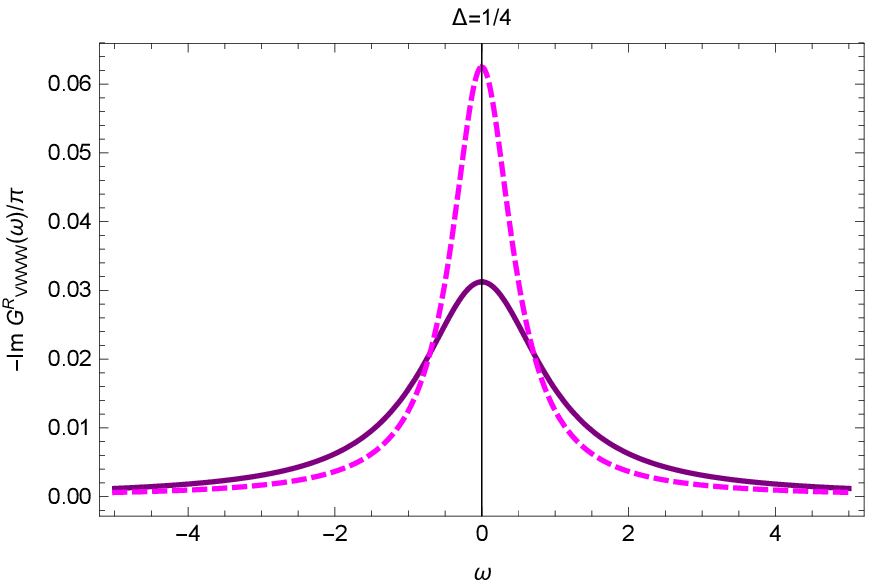}  }
    \caption{The four point OTOCs in frequency space with Lyapnov exponent as in Eq.(\ref{Eq:OTOCs-beta}) $-G^{(4)}_{\V\W\V\W}(\omega,\beta)/\pi$ of quantum liquid with Schwarzian correction: maximal chaotic behavior with $\lambda_L=1$ ($\beta=2\pi$) (cyan/purple thick lines) or non-maximal chaotic behavior with $\lambda_L=1/2$ ($\beta=4\pi$) (green/magnet dashed lines). We have chosen a set of input parameters as $\Delta=1/4$, $C=1/(2\pi)$.
    } \label{fig9}
\end{figure}

In the $\hat{t}\gg \beta$ limit, one obtains the TOCs and OTOCs four point functions, in the SK contour as depicted in Fig.\ref{fig8}
\beqa
 G^{(4)}_{\V\V\W\W} &=& \frac{1}{4^{2\Delta}} \frac{ 2 \Delta ^2}{\pi  C}, \quad \nn\\
 G^{(4)}_{\V\W\V\W} &=& - \frac{1}{4^{2\Delta}}\frac{\Delta^2}{ C} \cosh{\hat{t}} \sim \frac{1}{4^{2\Delta}} \beta \frac{\Delta^2}{C} e^{\lambda_L\hat{t}}, \label{Eq:G4_VVWW-G4_VWVW-SC} \qquad \quad
\eeqa
where $G^{(4)}_{\V\V\W\W}=G^{(4)}_{\V\V\W\W}(\hat{t}_1,\hat{t}_2,\hat{t}_3,\hat{t}_4)$ etc., and for the last equality of OTOCs, we have transferred the Euclidean time $\tau(t)$ in Eq.(\ref{Eq:ft-tau-T=0}) to Minkowski time, i.e., $t_i\to i\hat{t}$ and recover the temperature by rescaling $t_i \to 2\pi t_i /\beta  $,  while the $\beta$ in front comes from recovering of the thermal factor, i.e., by multiplying a factor $1\to \beta/2\pi$. The behavior of thermal OTOCs at later time shows a exponential expansion with a Lyapnov exponent $\lambda_L=2\pi/\beta$, which indicates the growth rate of chaos in thermal quantum systems with large number of degrees of freedom, and is bounded in a universal system~\cite{Maldacena:2015waa}, as $\lambda\le \lambda_L = 2\pi/\beta$. The Eq.(\ref{Eq:G4_VVWW-G4_VWVW-SC}) is valid under the condition that $t_r \ll t \ll t_s$, where $t_r$ is relaxation time, $t_s$ is the scrambling time and $t_s\sim \lambda_L^{-1}\log{C}$ when $C(t)$ becomes of $O(1)$ under time long time evolution.
By selecting the exponential increasing mode and doing Fourier transformation, in the choice of SK contour as in Eq.(\ref{OTOC-Schwinger-Contour}), or depicted in Fig.\ref{fig8}. The chaotic mode of OTOCs in frequency space becomes
\beqa
G^{(4)}_{\V\W\V\W}(\omega) &=& -i \frac{\beta}{2\pi} \int_{-\infty}^{+\infty} dt e^{i\omega t}\theta(t) G^{(4)}_{\V\W\V\W}(t) \nn\\
&=& -\frac{   \Delta ^2}{4^{2 \Delta}} \frac{1}{   2 C (  \omega -   i \lambda_L )}, ~ \text{Im}\omega > \lambda_L, \qquad \quad
\label{Eq:OTOCs-beta}
\eeqa
while the normal mode of TOC becomes
\beqa
G^{(4)}_{\V\V\W\W}(\omega) = -\frac{1}{4^{2\Delta}}\frac{\Delta ^2}{C \omega }, \quad \text{Im}\omega >0,
\eeqa
which is singular at $\omega=0$. The chaotic behavior of four-point OTOCs functions in frequency space is shown in Fig.\ref{fig9}. For maximal chaotic behavior with $\lambda_L=1$ ($\beta=2\pi$), it results in a non-zero frequency bump in the in low frequency region at large Lorentizian time. While at low temperature limit for ($\beta=4\pi$), the peak of the bulk moves more closer to low frequency range, or equivalently, a much more larger Lorentizian time to saturate the chaos, which corresponds to the non-maximal chaotic behavior with $\lambda_L=1/2$. In the zero temperature limit, $\lambda_L\to 0$ ($\beta\to\infty$), as expected, the peak of the bump moves to the $\omega=0$, and the mass spectrum of pNGB becomes NGB like.


\subsubsection{Zero temperature case}

By using the correlation function ${\mathcal B}_{1}(t_1,t_2)$ in Eqs.(\ref{Eq:B12}) and the soft mode propagators in Eq.(\ref{Eq:k12-T!=0}), the correlation to the connected four-point function turns out to be $G^{(4)}(t_1,t_2,t_3,t_4) = \epsilon^2 \vev{{\mathcal B}_1(t_{12}){\mathcal B}_1(t_{34})}/[t_{12}t_{34}]^{2\Delta}$. For the normal ordering of the time case, i.e., $t_1>t_2>t_3>t_4$, the correlation function turns out to be vanishing
\beqa
G^{(4)}_{\V\V\W\W} = 0, \label{Eq:G4_VWVW_T=0}
\eeqa
which means although the two point function ${\mathcal B}_1(t_{12})$ generates an energy fluctuation, they do not affects each other.  While for the crossing time ordering case, i.e., $t_1>t_3>t_2>t_4$, the corresponding four-point function turns out to be non-vanishing.
\beqa
G^{(4)}_{\V\W\V\W} &=& \frac{1}{t_{12}^{2\Delta}t_{34}^{2\Delta}}\frac{\Delta ^2 }{9 C } \frac{t_{23} \left(t_{23}^2-3 t_{13} t_{24}\right)}{t_{12} t_{34}} , \label{Eq:G4_VVWW_T=0}
\eeqa
which is proportional to the overall separation of the two pair $t_{23}$. In the absence of cross distance $t_{23}$, i.e., when $t_2=t_3$, the result $G^{(4)}_{\V\W\V\W}$ just recovers $G^{(4)}_{\V\V\W\W} =0 $. In the case $t_3 \to t_1$ and $t_2\to t_4$, one has $G^{(4)}_{\V\W\V\W}\sim t_{23}^2/t_{14}^2\sim 1$.


\subsection{Six-point function}

\begin{figure}[ht!]
\centering
   {\includegraphics[scale=0.5]{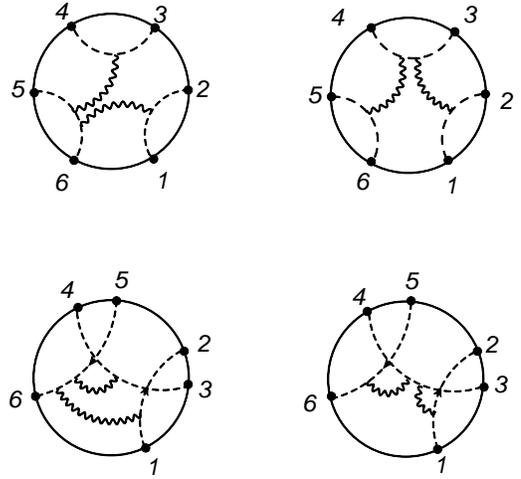}}
    \caption{
    Typical Feynman diagrams for six point correlation functions $G_6(t_1,\cdots,t_6)$ of scalar fields with loop corrections from soft modes as shown in Eq.(\ref{Eq:2pt-4pt-6pt-8pt-T!=0}), in TOCs and OTOCs, respectively.
    } \label{fig10}
\end{figure}

It is straightforward to calculate the higher point functions as in Eq.(\ref{Eq:2n-pt}) such as the six-point functions obtained in Eq.(\ref{Eq:2pt-4pt-6pt-8pt-T!=0}). The typical Feynman diagrams of six point TOCs and OTOCs functions are depicted in Fig.\ref{fig10}, respectively.

\begin{figure}[ht!]
\centering
   {\includegraphics[scale=0.4]{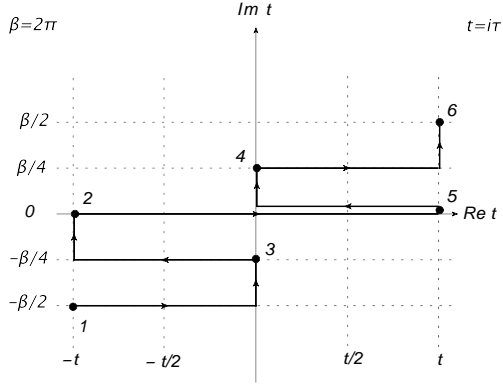}  }\qquad
    \caption{Schwinger-Keldysh six contour for six-point OTOCs, with time chosen as in Eq.(\ref{OTOC-Schwinger-Contour-6}). One may entail an infinitesimal Im$~{t_{52}}=\delta>0$ to entail that $t_2$ is earlier than $t_5$ along imaginary time line, and in the end set it to be zero, which does not affect the results.
    } \label{fig11}
\end{figure}

For the convenience of viewing physical consequence of six-point functions, one may generalize the SK four-contour in Eq.(\ref{OTOC-Schwinger-Contour}) to be six-contour as shown in Fig.\ref{fig11}, by increasing the real time with equal pace and imaginary time separately with
for OTOCs as
\beqa
&& (\hat{t}_i)^{\text{OTOC}} = \bigg( -\hat{t}- \frac{\beta}{2} i , -\hat{t} -  \frac{\beta}{4} i, 0, 0 + \epsilon i,  \hat{t} + \frac{\beta}{4} i  , \hat{t} + \frac{\beta}{2} i  \bigg) ,  \nn\\
&& \label{OTOC-Schwinger-Contour-6}
\eeqa
where the time order is $i=1,3,2,5,4,6$.

For example, in the SK six contour chosen in Fig.\ref{fig11}, the six-point correlation function can be expressed more elegantly as$G_6(t_1,\cdots,t_6)=F_6/4^{3\Delta}$, where $F_6$ are
\beqa
F_{\V\W\V\X\W\X} &=&  F_{\V\W\V\W\X\X} = \bigg( 1-\frac{1}{2} \pi  \cosh\hat{t} \bigg) F_{\V\V\W\W\X\X}, \nn\\
F_{\V\V\W\W\X\X} &=&  F_{\V\V\W\X\W\X} = \frac{\left(8 \Delta + 4 - \pi ^2\right) \Delta ^3}{4 \pi ^2 C^2},
\eeqa
where
\beqa
F_{\V\W\V\X\W\X} &=& \epsilon^3 \langle{{\mathcal C}_1(t_{1,2}){\mathcal C}_1(t_{3,4}){\mathcal C}_2(t_{5,6})\theta(t_{32})\theta(t_{54})\rangle}, \nn\\
F_{\V\W\V\W\X\X} &=& \epsilon^3 \langle{{\mathcal C}_1(t_{1,2}){\mathcal C}_1(t_{3,4}){\mathcal C}_2(t_{5,6})\theta(t_{32})\rangle}, \nn\\
F_{\V\V\W\X\W\X} &=& \epsilon^3 \langle{{\mathcal C}_1(t_{1,2}){\mathcal C}_1(t_{3,4}){\mathcal C}_2(t_{5,6})\theta(t_{54})\rangle}, \nn\\
F_{\V\V\W\W\X\X} &=& \epsilon^3 \vev{{\mathcal C}_1(t_{12}){\mathcal C}_1(t_{34}){\mathcal C}_2(t_{56})}.
\eeqa
In a similar manner, one can redefine $\tilde{F}_{\V\V\W\W\X\X}=F_{\V\V\W\W\X\X}(t_{3,4}\leftrightarrow t_{5,6})$ as
\beqa
\tilde{F}_{\V\V\W\W\X\X} &=& \epsilon^3 \vev{{\mathcal C}_1(t_{12}){\mathcal C}_2(t_{34}){\mathcal C}_1(t_{56})},
\eeqa
where $t$ is imaginary time, and $\theta(t_{ij})$ is the step function.
It turns out that, in the chosen SK six contour in Fig.(\ref{fig11}), one obtains
\beqa
&& \tilde{F}_{\V\W\V\X\W\X}-\tilde{F}_{\V\W\V\W\X\X}-\tilde{F}_{\V\V\W\X\W\X}+\tilde{F}_{\V\V\W\W\X\X}  \nn\\
&&=  \frac{\Delta^4 }{C^2} \cosh^2(\hat{t}) \sim \frac{\Delta^4}{4  C^2} e^{2\lambda_L\hat{t}} ,
\eeqa
with the OTOCs time as $\hat{t}_i$ with $i=(1,3,2,5,4,6)$ and $\lambda_L=2\pi/\beta$. The results just recover the $3$-OTCs of $6$-point functions in Ref.~\cite{Haehl:2017pak}.


\subsection{Eight-point function}

\begin{figure}[ht!]
\centering
   {\includegraphics[scale=0.5]{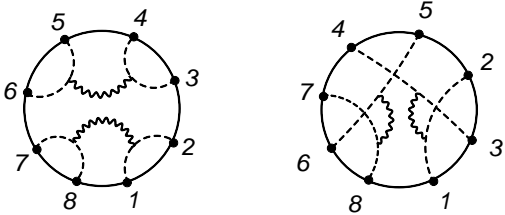}}
    \caption{
    Typical Feynman diagrams for eight point functions $G_8(t_1,\cdots,t_8)$ of scalar fields with loop corrections from soft modes as shown in Eq.(\ref{Eq:2pt-4pt-6pt-8pt-T!=0}), in TOCs and OTOCs, respectively.
    } \label{fig12}
\end{figure}

According to Eq.(\ref{Eq:2n-pt}), the eight point functions are obtained in Eq.(\ref{Eq:2pt-4pt-6pt-8pt-T!=0}). The typical Feynman diagrams of eight point TOCs and OTOCs functions are depicted in Fig.\ref{fig12}, respectively.

\begin{figure}[ht!]
\centering
   {\includegraphics[scale=0.4]{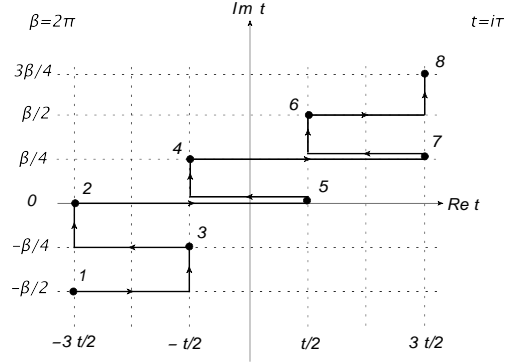}  }\qquad
    \caption{Schwinger-Keldysh eight contour for eight-point OTOCs, with time chosen as in Eq.(\ref{OTOC-Schwinger-Contour-8}). One may entail infinitesimals Im$~{t_{52}}=\delta>0$ and Im$~{t_{74}}=\delta'>0$ to entail that $t_{2,4}$ is earlier than $t_{5,7}$, respectively, along the imaginary time axis.
    } \label{fig13}
\end{figure}

For the convenience of viewing physical consequence of eight-point OTOCs functions, one may generalize the SK four-contour in Eq.(\ref{OTOC-Schwinger-Contour}) to be eight-contour as shown in Fig.\ref{fig13}, by increasing the real time with equal pace and imaginary time separately, as
\beqa
 && (\hat{t}_i)^{\text{OTOC}}
 = \bigg( -\frac{3}{2}\hat{t} -\frac{\beta}{2}i , -\frac{1}{2}\hat{t} - \frac{\beta}{4} i;  -\frac{3}{2}\hat{t}  , \frac{1}{2}\hat{t} +\epsilon i ; \nn\\
 && -\frac{1}{2}\hat{t} + \frac{\beta}{4} i ,  \frac{3}{2}\hat{t} + \frac{\beta}{4}  i + \epsilon' i;  \frac{1}{2}\hat{t}  + \frac{\beta}{2} i,  \frac{3}{2}\hat{t} + \frac{3\beta}{4} i  \bigg),  \label{OTOC-Schwinger-Contour-8}
\eeqa
where the time order is $i=(1,3,2,5,4,7,6,8)$.

In a similar manner as in calculating the six-point function, in the chosen SK eight-contour as in Fig.\ref{fig13} with the OTOCs time as $\hat{t}_i$ with $i=(1,3,2,5,4,7,6,8)$, the thermal eight-point functions turns out to be $G_{8}\sim F_{8}/4^{4\Delta}$, where $F_8$ are
\beqa
&& F_{\V\V\W\W\X\X\Y\Y} = F_{\V\V\W\X\W\X\Y\Y} = \frac{4 \Delta ^4}{\pi^2 C^2}, \nn\\
&& F_{\V\W\V\Z\W\X\Z\X} = F_{\V\W\V\W\Z\X\Z\X} \nn\\
&&= -F_{\V\W\V\Z\W\Z\X\X} = -F_{\V\V\W\Z\W\X\Z\X}  \nn\\
&&= -F_{\V\W\V\W\Z\Z\X\X} = - F_{\V\V\W\W\Z\X\Z\X} \nn\\
&&= \frac{4\Delta ^4 }{\pi ^2 C^2} \bigg(1 - \frac{\pi}{2}  \cosh{\hat{t}}\bigg)^2 \sim  \frac{ \Delta ^4 e^{2 \lambda_L \hat{t} }}{ 4 C^2}, \qquad
\eeqa
where
\beqa
F_{\V\V\W\W\X\X\Y\Y} &=& \epsilon^3 \langle{{\mathcal C}_1^4(t_{1,8})\rangle}, \nn\\
F_{\V\V\W\X\W\X\Y\Y} &=& \epsilon^3 \langle{{\mathcal C}_1^4(t_{1,8})\theta(t_{54})\rangle}, \nn\\
F_{\V\W\V\Z\W\X\Z\X} &=& \epsilon^3 \langle{{\mathcal C}_1^4(t_{1,8})\theta(t_{32})\theta(t_{54})\theta(t_{76})\rangle}, \nn\\
F_{\V\W\V\W\Z\X\Z\X} &=& \epsilon^3 \langle{{\mathcal C}_1^4(t_{1,8})\theta(t_{32})\theta(t_{76})\rangle}, \nn\\
F_{\V\W\V\Z\W\Z\X\X} &=& \epsilon^3 \langle{{\mathcal C}_1^4(t_{1,8})\theta(t_{32})\theta(t_{54})\rangle}, \nn\\
F_{\V\V\W\Z\W\X\Z\X} &=& \epsilon^3 \langle{{\mathcal C}_1^4(t_{1,8})\theta(t_{54})\theta(t_{76})\rangle}, \nn\\
F_{\V\W\V\W\Z\Z\X\X} &=& \epsilon^3 \langle{{\mathcal C}_1^4(t_{1,8})\theta(t_{32}))\rangle}, \nn\\
F_{\V\V\W\W\Z\X\Z\X} &=& \epsilon^3 \langle{{\mathcal C}_1^4(t_{1,8})\theta(t_{76})\rangle},
\eeqa
where we have introduced the new notation ${\mathcal C}_1^4(t_{1,8})\equiv{\mathcal C}_1(t_{1,2}){\mathcal C}_1(t_{3,4}){\mathcal C}_1(t_{5,6}){\mathcal C}_1(t_{7,8})$. It is easy to check that it satisfy the relation
\beqa
&& F_{\V\W\V\Z\W\Z\X\X} - F_{\V\W\V\W\Z\Z\X\X}  - F_{\V\V\W\X\W\X\Y\Y} \nn\\
&& + F_{\V\V\W\W\X\X\Y\Y} =0 \nn\\
&& F_{\V\V\W\Z\W\X\Z\X} - F_{\V\V\W\X\W\X\Y\Y}  - F_{\V\V\W\W\Z\X\Z\X} \nn\\
&& + F_{\V\V\W\W\X\X\Y\Y} =0, \nn\\
&& F_{\V\W\V\W\Z\X\Z\X} - F_{\V\W\V\W\Z\Z\X\X}  - F_{\V\V\W\W\Z\X\Z\X} \nn\\
&& + F_{\V\V\W\W\X\X\Y\Y}  = \frac{\Delta ^4 \cosh ^2({\hat{t}})}{C^2} \sim \frac{\Delta ^4 e^{2\lambda_L \hat{t}}}{4C^2}.
\eeqa
For the higher point OTCs, one would expect that the thermal system will approach the chaos much faster with time less than $\tau_L \equiv 1/\lambda_L$.

\section{Discussions and Conclusion}
\label{sec:dis-con}

The SYK model is an intriguing quantum mechanical model displaying both a spontaneous and explicit breaking of an emergent reparametrization symmetry Diff$_1$. The breaking patten of this symmetry determines many feature of the low energy dynamical property of the model and some are expected to be universal in strongly interacting IR fixed point at large N limit.

\emph{Features of SYK like model}--
The most fabulous features of the SYK model is the solvability in the strongly interacting IR fixed point at large N limit. The mass spectrum of the SYK model is obtained by solving Schwinger-Dyson equation and the spectrum of two-point and four-point function, as well as more higher-point functions are computed~\cite{Polchinski:2016xgd,Maldacena:2016hyu,Gross:2017aos}.

The other interesting features of the model is that in the strong coupling limit ($\beta J\gg 1$), the four point function saturates the maximal chaotic bound since it is dominated by the universal sector of gravity~\cite{Maldacena:2015waa}，which is characteristic of a gravity theory with black hole solutions~\cite{Shenker:2013pqa}. The saturation means it achieves the maximally allowed chaos quantified by the Lyapunov exponent $\lambda=2\pi\beta$, the growing rate of a thermal four-point OTOCs functions, as defined on the Keldysh contour~\cite{Kitaev:2015,Jensen:2016pah}, $\vev{ V_i(0) W_j(t) V_i(0) W_j(t)}_\beta\sim e^{\lambda t}/N$, which is true at a time range between the dissipation time and the scrambling time, i.e., $t\in(\lambda^{-1},\lambda^{-1}\log{N})$. The exponential growing manner reflects a underlying chaotic dynamics.

Another novel feature of the model is the emergent conformal symmetry, i.e., the time reparametrizations diffeomorphism symmetry Diff$_1$, or Virasoro symmetry, at low energy and its spontaneous and explicit breaking~\cite{Maldacena:2016hyu,Maldacena:2016upp}.

\emph{Spontaneous breaking of Diff$_1$}--
In the SYK model, the emergent Diff$_1$ symmetry is spontaneously broken down to SL$(2,R)$ symmetry~\cite{Witten:1987ty,Alekseev:1988ce}, which is kept in the Schwarzian action (the Lagrangian) at finite frequency. From the gravity viewpoint, the Diff$_1$ symmetry is an approximate asymptotic boundary symmetry of the perfect AdS$_2$ at IR conformal fixed point ($\omega=0$ or $J = \infty$), and is spontaneously broken down to a one dimensional global conformal group SO$(2,1)\sim$SL$(2,R)$ symmetry, or large diffeomorphism \text{Diff} owned by the AdS$_2$ symmetry.

\emph{Explicit breaking of Diff$_1$}--
In the SYK model, the emergent Diff$_1$ symmetry is also explicitly broken, since the symmetry is not kept by the Lagrangian any more as one slightly moves away from the IR conformal fixed point, where the kinetic term $\partial_\tau$ becomes relevant at low frequency or strong coupling region ($\omega \ll 1$ or $J\gg 1$). From the gravity viewpoint, the bulk spacetime is slightly deviated from AdS$_2$ vacuum to near-AdS$_2$ (NAdS$_2$) by taking account of the backreaction due to arbitrary tiny energy excitation.

\emph{Diff$_1$ symmetry breaking pattern}--
The pattern of spontaneous breaking of the Diff$_1$ results in an infinite number of zero mode, namely, the NGBs characterized by the coset Diff$_1/SL(2,R)$. As the Diff$_1$ symmetry is explicitly broken, the leading order dynamical correction is described by a $SL(2,R)$ invariant Schwarzian derivative of the reparameterization $f(t)$ in terms of effective Schwarzian action in Eq.(\ref{Eq:Sch}) as described in Appendix.~\ref{app:SYK}, which determines many aspects of the theory. As will be seen, the dynamics of the Schwarzian correction to the quantum correlations of SYK model is characterized by Schwarizian action with an SL$(2,R)$ unbroken symmetry. As the Diff$_1$ symmetry is explicitly breaking due to a small but non-vanishing derivation $\epsilon \sim \kappa \sim J^{-1}$ (this is equivalent to a small $\omega\ll1$ or the presence of a relevant kinetic term $\partial_\tau$), associated with an infinitesimal fluctuation field $k(t)$ in a dynamical reparameterization function $f = t + \epsilon k(t) $, as in Eqs.(\ref{Eq:ft-tau-T=0}) and (\ref{Eq:ft-tau-T!=0}) for zero and finite temperature cases, respectively. To be brief, the Diff$_1$ symmetry is parameterized by $f$ and is explicitly broken by the small fluctuation $\epsilon k(t)$, which is parameterized by the coset Diff$_1/$SL$(2,R)$.

\emph{Quantum and thermal correlations}--
For zero temperature case as in Eq.(\ref{Seff:Sch-T=0}), the zero modes of the fluctuation field, i.e., NGBs, leads to a zero action in the IR conformal fixed point ($J=\infty$ or $\epsilon=0$), while the soft modes of the fluctuation field, i.e., pNGBs, leads to a non-vanishing action when the classic solution is deviated away from the conformal limit (a small but finite $\epsilon \sim J^{-1}$). For finite temperature case as in Eq.(\ref{Seff:Sch-T!=0}), the first exciting state of pNGBs, i.e., the $n=\pm 1$ soft modes, also leads to a zero action. In both cases, the $2$-point function is singular and needs to be regularized. Consequently, the $2$-point and $4$-point correlation functions of matter field, obtains loop corrections from the pNGBs as the reminiscent effect of the broken of time reparamterization invariance Diff$_1$.

In conclusion, we study the retarded Green's function of quantum liquid with Schwarzian correction, which can be depicted by a $(0+1)$-dimensional strongly interacting quantum mechanical/statistics model dual to a general $(1+1)$-dimensional classical dilaton gravity model. Based upon the two point correlation functions of matter, which get loop corrections from pNGBs in coset Diff$_1/$SL$(2,R)$, we obtain the bosonic retarded Green's functions as well as local dynamical susceptibility, i.e., the $2$-nd order local spin-spin correlation functions for quantum liquid. We also calculate the four point as well as higher point thermal OTOCs functions in SK formalism, which cultivate the quantum chaos at large real time.

To manifest our results, we show the spectral functions of not only Schwarzian spin-glass described with conformal dimension $\Delta=1/4$ ($p=4$) but also Schwarzian NFL with $\Delta=1/2$ ($p=2$), as well as a specific quantum liquid phase with $\Delta=1/3$ ($p=3$). Large $p$-body behavior of quantum liquid with Schwarzian correction are studied too. Moreover we make comparison with the leading order retarded Green's functions, which just recovers the results of quantum liquid from AdS$_2$/CFT$_1$ approach.

In the infrared (IR) conformal fixed point with zero frequency ($\omega = 0$) where the Diff$_1$ symmetry is emergent, the spectral functions owns Fermi liquid~\cite{Landau} peak in DOS at $\omega=0$ and leads to typical metallic behavior. The symmetry is spontaneously broken to SL$(2,R)$ and leads to zero modes on the boundary in terms of "boundary graviton"~\cite{Maldacena:2016upp}, which are the Fourier modes of the Diff$_1$ symmetry. At finite frequency ($\omega\ne 0$), the Diff$_1$ symmetry is explicit broken. As its physical consequence, the system develops a feature which is interpreted as bad metalic behavior with a high energy Hubbard band dynamically generated. In the intermediate region, there is a temperature dependent crossover between Fermi liquid phase and bad metal phase in the strongly correlation regime, in which the quasi-particle picture is fragile or even broken down.

We make generalization of $4$-point correlation to higher point correlation functions. As non-inclusive demos, we show concise analytic results on $6$-point as well as $8$-point thermal OTOCs functions in SK contour, which exhibit exponential growth until progressively a longer timescale and thus sensitive to more fine grained quantum chaos. We also obtain analytic expression for $3$-order and $4$-th order local spin-spin correlation functions.

The quantum liquid with Schwarzian correction can be related not only to quantum spin glass or disordered metals without quasi-particles scenery depicted by SYK like model, but also NFL phase~\cite{Virosztek:1990,Parcollet:1997ysb,Parcollet:1999,Sachdev:2010um,Sachdev:2015efa,Cai:2016jxd}. We study the matter retarded Green's function by taking account of the loop corrections from pNGBs to the matter two point correlation function, and unexpectedly find a Hubbard band or dynamically generated DOS in the spectral functions, which is due to the spontaneous and explicit breaking of time reparameterization symmetry and is a distinct signature of quantum liquid with Schwarzian correction, comparing with the conventional strongly interacting quantum liquid. The existence of pNGBs mode in the quantum liquid with Schwarzian correction also provides a dynamical mechanism for explaining the commonly observations of bad metal in strongly correlated system.



\vspace{0.2in}  \centerline{\bf{Acknowledgements}} \vspace{0.2in}

We thank valuable discussion with Gleb Arutyunov, Roberto Emparan, Ara Go, Sunly Khimphun, Ki-Seok Kim, Bum-Hoon Lee, Yue-Zhou Li, Andrew Lucas, Zhan-Feng Mai, Eun-Gook Moon, Ioannis Papadimitriou, Varun Sethi, Run-Qiu Yang, Hossein Yavartanoo, Junggi Yoon, Kentaroh Yoshida and Yun-Long Zhang. This work is supported by Mid-career Researcher Program through the National Research Foundation (NRF) of Korea under grant No.~NRF-2016R1A2B3007687. YS is also supported in part by Basic Science Research Program through NRF under grant No.~NRF-2016R1D1A1B03931443. We also would like to thank the hospitality of 2017 APCTP focus workshop program ''Geometry and Holography of Quantum critical point`` during their visiting at the POSTECH, Pohang.

\appendix

\section{SYK model and Schwarzian}
\label{app:SYK}

\emph{SYK model introduction}--In path integral, the prototypical SYK model is described by the partition function an action $Z(J)=\int D\psi_i e^{-S}$, with the action as
\beqa
S = \int d\tau \bigg(  \frac{1}{2}\sum_{i=1}^N  \chi_i \partial_\tau \chi_i - \frac{1}{4!} \sum_{i,j,k,l} J_{ijkl} \chi^i \chi^j \chi^k \chi^l  \bigg), \qquad
\eeqa
where $\chi^i$ are $N$ Majorana fermions, satisfying $\{\chi_i, \chi_j\}=\delta_{ij}$, interacting with random interactions involving $4$ fermions at a time. $J_{ijkl} $ is a Gaussian random infinite-range exchange interaction of all-to-all quartic coupling, which are mutually uncorrelated and satisfies the Gaussian's probability distribution function $P(J_{ijkl})\sim \exp{(- N^3 J_{ijkl}^2/12J^2 )}$, which leads to zero mean ${\mathbb E}[{J_{ijkl}}]=0$ and variance ${\mathbb E}[{J_{ijkl}^2}]=3!J^2/N^3$ with width of order $J/N^{3/2}$, respectively. The ${\mathbb E}[\cdots]$ denotes an average over disorder. The $J$ is the only one effective coupling after the disorder averaging for the random coupling $J_{ijkl}$. The random couplings $J_{ijkl}$ represents disorder, and does not correspond to a unitary quantum mechanics~\cite{Witten:2016iux,Kitaev:2017awl}. For euclidean time $\tau=it$, the model can be viewed alternatively as a $1$-dimensional statistical model of Majorana fermions. For finite temperature case, the quantum mechanical model can be alternatively depicted in a quantum statistics. By using a Hubbard-Stratonovich transformation, it is possible to rewrite the original partition function of SYK model as a functional integral of the form~\cite{Kitaev:2015,Grempel:1998,Grempel:1999,Grempel:2001} as
\beqa
 && Z = e^{-\beta F} = \int {\mathcal D}{G}{\mathcal D}\Sigma \exp{ (- N \bar{S} ) } , \quad \nn\\
 && \bar{S} = -\bigg[ \log\text{Pf}(\partial_\tau - \Sigma) - \frac{1}{2} \int d\tau_1 d\tau_2 \bigg( \Sigma {G}  - \frac{J^2}{4} {G}^p  \bigg) \bigg] ,  \qquad ~ \label{Eq:Sb_SYK}
\eeqa
where $\bar{S}=S/N$ is a disorder-averaged non-local effective action by doing Gaussian integral over the disorder and integrating out fermions after introducing a bilocal field $G(\tau,\tau')$ and a Lagrange multiplier field $\Sigma(\tau,\tau')$. Pf denotes the Pfaffian, and the first term of the action can also be re-expressed as $\log[\text{det}(\partial_\tau - \Sigma)]/2$, $\tau,\tau^\prime$ are Matsubara times and $p=4$ denotes the number of Majorana fermion in the vertex.

At large $N$ limit, i.e, a model with the number of Majorana fermion $N\gg 1$, by doing variation with respect to $G$ and $\Sigma$, or equivalently by counting the resummed Feynman diagrams, the solution of SYK model is described by the Schwinger-Dyson (SD) equations in real spacetime as
\beqa
G = (\partial_\tau - \Sigma)^{-1}, \quad \Sigma = J^2  G ^{p-1}. \label{Eq:SD}
\eeqa
where $G=G(\tau,\tau^\prime)$ is the two-point Green's function, $\Sigma=\Sigma(\tau,\tau^\prime)$ is one particle irreducible ($1$PI) self energy.

The first kinetic term in the $G$ represents a conformal breaking term as will be clear in the following. Substituting the full solutions of above classical equations of motion back into the effective action in the partition functions, one obtains the leading large-$N$ saddle point free energy $F$ in low temperature expansion as~\cite{Maldacena:2016hyu,Gu:2016oyy}
\beqa
\frac{F}{N} = -\frac{1}{\beta}\frac{\log{Z}}{N} = \frac{1}{\beta}\bar{S} = e_0 -  s_0 \beta^{-1} - \frac{1}{2} \gamma \beta^{-2}  + \cdots ,  \qquad \quad \label{Eq:F-Seff}
\eeqa
where $e_0$ is the ground state energy density, $s_0$ is te zero temperature entropy density and $ c_v \equiv \gamma \beta^{-1}$ is the specific heat density. $\cdots$ denote terms with higher order in $\beta^{-1}$. 

In the ultraviolet (UV) limit at short distance, $\omega \gg J$, the kinetic term dominates and the four-fermion interactions term is irrelevant so that the theory has $N$ weakly interacting massless Majorana fermions. The fermions have a two point function given by $G_0(\tau) = \text{sgn}(\tau)/2$ regardless of temperature, or $G(\omega)=i\omega^{-1}$ in frequency space, assuming the time translation symmetry is kept. The action is invariant under arbitrary time reparametrizations and consequently the Hamiltonian is zero. While in the low energy IR limit at large distance, the frequency (in the momentum spacetime $\partial_\tau \sim i \omega$) is much smaller than the UV coupling $J$, i.e.,  $\omega \ll J$, means the model becomes strongly interacting at low energies. Consequently, the kinetic term $\partial_\tau$ can be dropped, so that the SD equations in the IR limit is modified to be conformal invariant ones as,
\beqa
\int d\tau'' G(\tau,\tau'') \Sigma(\tau'', \tau') = - \delta(\tau-\tau^\prime),  \quad \Sigma = J^2  G ^{p-1}.  \qquad ~ \label{Eq:SD_CFT1}
\eeqa
In this case, the SD equation in the conformal limit are reparameterization invariant, which means that under an infinitesimal transformation of the time reparameterization $\tau\to \tau + \epsilon(\tau)$, the two point function transforms as $G\to G+ \delta_\epsilon G$ with
\beqa
\delta_\epsilon G =  [ \Delta(  \partial_\tau \epsilon(\tau) + \partial_{\tau'}\epsilon(\tau') ) + \epsilon(\tau) \partial_\tau + \epsilon(\tau') \partial_{\tau'} ] G. \qquad
\eeqa
In this case, $G+\delta_\epsilon G$ still solve the conformal SD equations in Eq.(\ref{Eq:SD_CFT1}).

As a physical consequence, an extra general reparametrization symmetry with an arbitrary function $f(t)$ and conformal invariance is emergent in the IR limit as long as $\omega \ll J$ or equivalently, $J \to \infty$, namely, in strong interactions, as $\tau \to f(\tau)$,
\beqa
&& G \to \abs{f'(\tau)f'(\tau')}^{\Delta} G(f(\tau),f(\tau')), \quad \nn\\
&& \Sigma \to \abs{f'(\tau)f'(\tau')}^{\Delta(p-1)} \Sigma(f(\tau),f(\tau')), \label{Eq:SL(2)_time-reparametrization}
\eeqa
where $\Delta=1/p$ is the conformal dimension of CFT$_1$, which is explicitly broken by the kinetic term $\partial_\tau$ in medium energy range, i.e., $\omega \sim J$, which is the explicit symmetry breaking parameter.
To be brief, both two point function $G$ and self energy $\Sigma$ are conformal invariant in the IR background. Therefore, $\beta J\gg 1$ can also be viewed as the conformal limit of the model.

In the IR limit, the action is reparametrization invariant by dropping the kinetic term $\partial_\tau$ inside the action, while the solution $G$ is only $SL(2,R)$ invariant. Thus one can view reparametrization invariance as an emergent symmetry of the IR theory, which is spontaneously broken by the conformal solution $G$. The emergent full reparameterization symmetry, i.e., the Virasoro group $SL(2)$, is presented by the generators
\beqa
&& D = -\tau \partial_\tau - \Delta, \quad P = \partial_\tau, \quad K = \tau^2 \partial_\tau + 2\tau \Delta, \nn\\
&& [D , P] = P, \quad [D,K] = -K, \quad [P,K] = - 2D.
\eeqa
The zero modes in the effective action can be viewed as Nambu-Goldstone (NG) modes for the spontaneous breaking of the full $SL(2)$ conformal symmetry down to $SL(2,R)$. Since the action is $SL(2,R)$ gauge invariant, in the path integral, one need to dived the the integral by a volume of $SL(2,R)$.

At zero and finite temperature, the two point function has a conformal ansatz form at zero and at finite temperature, respectively, as
\beqa
T=0 : && \quad G(\tau) \overset{}{=} \frac{b}{\abs{\tau}^{2\Delta}} \text{sgn}(\tau),  \qquad  \nn\\
T\ne 0:  && \quad G(\tau) \overset{}{=} b \bigg( \frac{\pi}{\beta\sin\frac{\pi \tau}{\beta}} \bigg)^{2\Delta} \text{sgn}(\tau),  \label{Eq:SYK_two-pt}
\eeqa
where $\Delta$ is the IR conformal dimension, which turns out to be inversely proportional to the d.o.f $p$ of the disordered interaction, i.e., $\Delta = 1/{p}$ and $b^p = {( 2\pi J^2)}^{-1}( 1 - 2 \Delta )\tan{(\pi \Delta)}$ at leading order in $1/N$. The Green's function represent the low frequency behavior of the retarded Green's function for the SYK model in the strong coupling limit. The ansatz form above can be obtained by applying the reparameterization at saddle point $f(\tau)=\tau$ in zero temperature case, while in finite temperature case, the time direction is considered Euclidean and compactified into a thermal circle $f(t)=e^{2\pi i t/\beta}$ or $f(\tau)=\tan{(\pi\tau/\beta)}$ satisfying $f(\tau+\beta)=f(\tau)$. To be brief, the thermal quantum mechanics or quantum statistics can be achieved through the reparameterization of a zero temperature quantum mechanics, by mapping a straight line of imaginary time $\tau$ to a thermal circle, i.e., $\bar{\tau}=\tan{(\pi \tau/\beta)}$ with periodic boundary conditions over a periodic lattice length $\beta$.


\emph{Effective Schwarzian action}--

In the low energy limit, the model can be described by a local effective action proportional to the Schwarzian derivative~\cite{Kitaev:2015} in terms of Schwarzian theory~\cite{Stanford:2017thb}, which can be understood as the dynamics of a Goldstone bosons $f(\tau)$, a near-zero mode for the breaking of reprarametrization invariance~\cite{Maldacena:2016hyu}, with a coefficient of order $(\beta J)^{-1}$ as~\cite{Kitaev:2015}
\beqa
S_{\text{eff}} = - \frac{N\alpha}{J} \int d\tau \text{Sch}(f(\tau),\tau) ,  
\label{Eq:Sch}
\eeqa
where $\text{Sch}(f,\tau)$ is the Schwarzian derivative in Eq.(\ref{Eq:Sch(f,t)}), which is invariant under $SL(2)$ symmetry $f\to (af+b)/(cf+d)$ and it is an exact symmetry at zero temperature since $f(\tau)=\tau$. The prime indicates the derivative with respect to the $\tau$. $f(\tau)$ is the Nambu-Goldstone bosons, or the zero modes involving large diffeomorphisms, which are non-trivial on the boundary. When one move away from the IR fixed point ($\omega \ll J\to \infty$), the NG bosons cease to be zero mode and leads to a non-zero action, i.e., Sch$(f(\tau),\tau)\ne 0$, e.g, $f(\tau)=\tan(\pi\tau/\beta)$, a black hole with finite temperature as a deformed parameter from AdS$_2$. At the finite temperature, the effective action becomes $S_{\text{eff}}=-2\pi^2N\alpha/(J\beta) $

The form of the action itself implies an $SL(2,R)$ invariant solution $f \to (af+b)/(cf+d)$ with $a,b,c,d\in {\mathbb R}$ and $ad-bc=1$, which is the same as $SL(2,C)$ . For instance, at finite temperature, $f(\tau)=\tan(\pi\tau/\beta)$, the Schwarzian is $\text{Sch}(f,\tau)=2\pi^2/\beta^2$. Since the effect coupling of the theory $1/g^{2} \propto N\alpha/(J\beta^2)$, at large $N$ and fixed temperature, the theory is weakly coupled, dominated by fluctuations around the saddle point $\tau$, but is strongly coupled at ultra low temperature, i.e., $g \propto \beta $. It also interesting to consider a reparameterization $f(\tau)\to \tan{(\pi \epsilon(\tau)/\beta)}$, the Schwarzian becomes $\text{Sch}(f(\tau),\tau)= \text{Sch}(f(\tau),\tau)+2(\pi/\beta)^2 f'^2$. Considering a small reparameterization $\tau \to \tau + \epsilon(\tau)$, by using the equivalent form of Schwarzian with terms of the total derivatives, the action can be re-epxressed as a local one as
\beqa
&& S_{\text{eff}} =  \frac{N\alpha}{2J} \int d\tau \bigg( \frac{\epsilon''^2}{\epsilon'^2} - \bigg(\frac{2\pi}{\beta}\bigg)^2 \epsilon'^2 \bigg) , \quad \nn\\
&& \text{Sch}(\epsilon(\tau),\tau)=  \bigg(\frac{\epsilon''}{\epsilon'}\bigg)^\prime - \frac{1}{2}\frac{\epsilon''^2}{\epsilon'^2}, \label{Eq:Sch-2}
\eeqa
where we have dropped the total derivative terms $(\epsilon''/\epsilon)'$, the action has an expression of lowest order in derivatives that vanishes for global $SL(2)$ transformation. Consider a small fluctuation on the fixed parametrization $\epsilon(\tau)=\tau + \hat\epsilon(\tau)$, and expanded up to quadratic order, one obtains the quantum action in terms of Pseudo Nambu-Goldsonte (PNG) boson field $\hat\epsilon$. In this case, not only the $SL(2)$ symmetry is broken by the $SL(2,R)$ invariant IR $G$ solution, but also is explicitly broken, which gives a small Schwarzian action for $\hat\epsilon(\tau)$, which is vanishing in the strong interacting limit $J\to\infty$ at order of large $N$. The effective action is a potential term for the zero mode, thus, the Schwarzian action can be viewed as a mass term for PNG boson. Therefore, the low energy effective Schwarzian action above makes reparameterization modes $\hat\epsilon$ PNG bosons, in terms of soft modes~\cite{Bagrets:2016cdf,Kitaev:2017awl}.


\subsection{Self energy}

With the two point correlation functions $G(t)$, it is also possible to consider the scattering rate of quantum liquid with Schwarzian correction, by doing fourier transformation the imaginary time self-energy. For example, as in SYK model, according to the Schwinger Dyson equations in Eqs.(\ref{Eq:SD}) or (\ref{Eq:SD_CFT1}) with $p=1/\Delta$,  the self energy can be calculated as
\beqa
\Sigma \propto G(\tau)^{p-1} &=&  \frac{1}{i^{2(\Delta-1)}} \frac{1}{[2 \frac{\beta}{2\pi} \sinh\frac{t}{2} \frac{2\pi}{\beta} ]^{2(\Delta-1)}}\nn\\
& \overset{\beta\to \infty}{=} & \frac{1}{i^{2(\Delta-1)}} \frac{1}{ t^{2(\Delta-1)}}
\eeqa
where we have used Eq.(\ref{Eq:GTt-Gt}). Thus, $\Sigma_{\Delta}(\omega)\propto G_{\Delta-1}(\omega)\sim G_{\Delta}''(\omega)$, where the prime denotes the derivative with respect to the frequency $\omega$.

\subsection{Partition functions and free energy}

According to Eq.(\ref{Eq:F-Seff}), by using the thermal parameterization $f(t)=\tan{(\pi t/2)}$, the free energy in low temperature expansion i.e., $\beta\gg 1$, can be obtained from the effective action.

From the effective action of the gravity sector in Eq.(\ref{Eq:S_eff}), one obtains the free energy
\beqa
F_0 &=& -\frac{1}{\beta} \log{Z} = \frac{1}{\beta} S_{\text{eff}} =  \frac{1}{\beta}  C_g \phi_r \int dt \text{Sch}(f(t),t)  \nn\\
&=&     C_g \phi_r  \int dt \frac{1}{2\beta} \to C  \frac{2\pi}{\beta} 2\pi  \frac{1}{2\beta} = C\frac{2\pi^2}{\beta^2},
\eeqa
which leads to the zero temperature entropy $S = N s_0 = \beta F= 2\pi^2 \beta^{-1}$ and the specific heat $C_V=Nc_v= \gamma \beta^{-1} = 4\pi^2C\beta^{-1}$, which are both linear in temperature.

While from the effective action of the scalar matter given in Eq.(\ref{Eq:S-chi_eff}),
\beqa
F_\chi &=& -\frac{1}{\beta}\log{Z_\chi} = \frac{1}{\beta}\bar{S}_{\text{eff}} \propto \int_{\epsilon}^{\beta-\epsilon} dt \bigg( \frac{\pi}{\beta \sin\frac{\pi t}{\beta}} \bigg)^{2\Delta} \nn\\
 &=& \frac{\pi ^{2 \Delta +\frac{1}{2}} \beta ^{1-2 \Delta } \sec (\pi  \Delta )}{\Gamma (1-\Delta ) \Gamma \left(\Delta +\frac{1}{2}\right)}+\frac{2 \epsilon ^{1-2 \Delta }}{2 \Delta -1},
\eeqa
where the first term is a finite one as leading IR correction under the case that $\Delta<3/2$, since this free energy due to matter $F_{\chi}\propto \beta^{1-2\Delta}$ dominates over the free energy due to the gravity $F\propto \beta^{-2}$ at low temperature  limit. While the second term is a UV divergent term, since it is a constant, thus contributes to the ground state energy density $e_0$ as obvious in Eq.(\ref{Eq:F-Seff}).

One can also include the one-loop exact Schwarzian partition function by direct functional path integration of the Schwarzian theoryas~\cite{Stanford:2017thb}
\beqa
Z_{\text{Sch}} = \frac{1}{4\pi g^3} \exp{\bigg( \frac{\pi}{g^2} \bigg) }, \quad g \equiv \frac{\epsilon^2}{C},
\eeqa
from which, the one loop corrections to the free energy is obtained from
\beqa
F_{\text{Sch}} = - \frac{1}{\beta} \log{Z}_{\text{Sch}} 
\sim - \frac{3}{2} \log\frac{\beta}{C} . \qquad \quad
\eeqa
Therefore, the loop corrections of graviton soft mode to the scalar matter field turns to contributes a finite logarithmic temperature term for the free energy.

\section{Wave function in global AdS$_2$}



\subsection{Boson in global AdS$_2$}



Consider the AdS$_2$ metric in global coordinates as
\beqa
ds^2 = \ell^2 ( d\rho^2 - \cosh^2\rho d\tau^2), \quad \ell^2 \equiv - \frac{1}{4c_\mu }, \label{Eq:ds2-rho}
\eeqa
where $\rho\in(-\infty,+\infty)$ and $\tau\in(-\infty,+\infty)$, $c_\mu$ contains UV information of $2$-D gravity.

The global coordinate are within the range $\rho\in(\infty,0)$ for $z\in(0,\infty)$.  Thus, it is expected that the theory is dual to copies of conformal qantum mechanics CQM$_1$ and CQM$_2$ on two boundaries via AdS$_2$/CFT$_1$. The Klein-Gordon equation in this coordinates becomes
\beqa
\phi^{\prime\prime} + \tanh\rho \phi^{\prime} + \bigg( \omega^2 \text{sech}^2\rho - \frac{m^2}{4c_\mu }  \bigg) \phi =0,
\eeqa
which gives the wave functions as
\beqa
\rho(z)  &=&  (z^2-1)^{1/4} [ c_1 P_{i \omega -\frac{1}{2}}^{\nu_1}( z )+c_2 Q_{i \omega -\frac{1}{2}}^{\nu_1}( z ) ] , \qquad
\eeqa
where $z\equiv\tanh\rho$ and
\beqa
\nu_1 = \sqrt{\frac{1}{4}- \frac{m^2}{4c_\mu }} .
\eeqa

%

\subsection{Fermion in global AdS$_2$}
\label{app:Fermion_AdS2}



By using a coordinate transformation
\beqa
\sinh\rho \equiv \tan\sigma, \quad \sigma \in (-\pi/2,\pi/2),
\eeqa
which transforms the the boundary of the spacetime at $\rho=\pm\infty$ into $\sigma=\pm\pi/2$, the global AdS$_2$ metric in Eq.(\ref{Eq:ds2-rho}) becomes equivalently as
\beqa
ds^2 = \ell^2 \frac{-d\tau^2 + d\sigma^2 }{\cos^2\sigma} . \label{Eq:ds2-sigma}
\eeqa

The Dirac equation in $(1+1)$-dimensional spacetime can be expressed as $(\Gamma^\mu D_\mu - m ) \psi =0$,
where $\psi=(\psi_+,\psi_-)^T$ is a two component spinor, $\Gamma^\mu = e^\mu_a \gamma^a$ where $e^\mu_a$ is the inverse of the vielbein $e_\mu^a$, and $\gamma^a$ is the Pauli matrices as defined in the following section. The covariant derivatives are $D_\mu = ( \partial_\mu + \Omega_\mu)$ where $\Omega_\mu\equiv \omega_\mu^{ab}\gamma_{ab}/4$ and $\omega_\mu^{ab}$ is the spin connection $\omega_\mu^{ab}=e_\nu^a(\partial_\mu e^{b \nu }  + e^{b \lambda} \Gamma^\nu_{\lambda \mu} )$, where $\Gamma^\nu_{\lambda\mu}$ is the Christoffel symbols and $\gamma^{ab}=[\gamma^a,\gamma^b]/2$. The vielbein $e_\mu^a$ corresponds to the metric $g_{\mu\nu}= \eta_{ab}e_\mu^a e_\nu^b$ with $\text{diag} (\eta_{ab}) =(-1,1)$. The vielbein and non-vanishing spin connection in the global AdS$_2$ coordinate in Eq.(\ref{Eq:ds2-rho}) are given by
\beqa
e_\tau^{\underline{\tau}} = \ell \cosh\rho, \quad e_\rho^{\underline{\rho}} = \ell^2, \quad \omega_\tau^{\underline{\tau}\underline{\rho}} = - \omega_\tau^{\underline{\rho}\underline{\tau}} = \sinh\rho, \qquad
\eeqa
In the momentum space by assuming the time translation invariance, i.e., $\psi_\pm(t,\rho) \to e^{-i\omega t } \psi_\pm(\omega,\rho)$, one can choose a real representation as in Ref.~\cite{Cai:2016jxd},  in which, $(\gamma^0, \gamma^1) = (\gamma^t, \gamma^z) = (i \sigma^2, \sigma^1) $ so that the Gamma matrix of Dirac fermions in $2$-dimensional Lorentz spacetime satisfy $\{   \gamma^a , \gamma^b \} = 2 \eta^{ab} \textbf{1}_2$, where $\eta^{ab}=\text{diag}(-1,+1)$ is Lorentzian metric for local inertial frame. 

\begin{widetext}
In the representation, the chirality operator is $\gamma^3 =  \gamma^0 \gamma^1 = \sigma^3$, thus the wave function contains two Weyl fermion with opposite chirality as $\phi=(\psi_+,\psi_-)^T$.  The Dirac equation $(\cancel{D} - m ) \psi =0$ in the global AdS$_2$ coordinate in Eq.(\ref{Eq:ds2-rho}), becomes

\beqa
&& \bigg( \partial_\rho  - \frac{i\omega}{\cosh\rho} - \frac{1}{2}\tanh\rho \bigg) \psi_- - m\ell \psi_+ =0, \quad \bigg( \partial_\rho + \frac{i\omega}{\cosh\rho} - \frac{1}{2}\tanh\rho  \bigg) \psi_+ - m\ell \psi_- =0,
\eeqa
which can be combined to be two decoupled equations of motion for $\psi_\pm$, respectively, as
\beqa
  \bigg( \partial_\rho^2 + \tanh\rho \partial_\rho + \frac{1}{4}  + m^2 \ell^2 \mp i \omega \frac{\tanh\rho}{\cosh\rho} + \frac{\omega^2+1/4}{\cosh^2\rho} \bigg) \psi_\pm = 0 .
  \label{Eq:psi_rho}
\eeqa
Alternatively, the Dirac equation in the global AdS$_2$ coordinate in Eq.(\ref{Eq:ds2-sigma}) are
\beqa
\bigg( \partial_\sigma  -i\omega - \frac{\tan\sigma}{2} \bigg) \psi_-  -  \frac{m \ell}{\cos\sigma}\psi_+ =0 , \quad \bigg( \partial_\sigma +i\omega - \frac{\tan\sigma}{2} \bigg) \psi_+ -  \frac{m \ell}{\cos\sigma}\psi_- =0 ,
\eeqa
which can be combined to be two decoupled equations of motion for $\psi_\pm$, respectively, as
\beqa
\bigg( \partial_\sigma^2 \mp i \omega \tan\sigma + \frac{1/4+m^2\ell^2}{\cos^2\sigma} +  \omega^2 + \frac{1}{4} \bigg) \psi_\pm = 0.  \qquad \label{Eq:psi_sigma}
\eeqa
The solutions to Dirac equation in Eqs.(\ref{Eq:psi_rho}) or (\ref{Eq:psi_sigma}) tuns out to be

{ 
\beqa
\psi_+&=& (z+1)^{-i m \ell } \bigg[ c_2 z^{-\omega /2} \, _2F_1\bigg(-i m \ell ,-i m \ell -\omega +\frac{1}{2};\frac{1}{2}-\omega ;-z\bigg) + c_1 \sqrt{z} z^{\omega /2} \, _2F_1\bigg(1-i m \ell ,-i m \ell +\omega +\frac{1}{2};\omega +\frac{3}{2};-z\bigg) \bigg], \nn\\
\psi_- &=& (z+1)^{-i m \ell } \bigg[ c_4 \sqrt{z} z^{-\omega /2} \, _2F_1\bigg(1-i m \ell ,-i m \ell -\omega +\frac{1}{2};\frac{3}{2}-\omega ;-z\bigg) +c_3 z^{\omega /2} \, _2F_1\bigg(-i m \ell ,-i m \ell +\omega +\frac{1}{2};\omega +\frac{1}{2};-z\bigg)\bigg] , \nn
\eeqa
}
where $\psi_\pm=\psi_\pm(z)$ with 
$z=e^{2i\sigma}$.   

\end{widetext}

\section{Higher dimensional generalization}
\label{app:high-D-generalization}


\subsection{Boson in ${d+1}$-dimensional spacetime}

\subsubsection{Klein-Gordon bulk equation of motion}

The equation of motion in coordinate space for $\phi$ with action
\beqa
S  = - \int d^{d+1}x \sqrt{-g} [ g^{MN} (D_M\phi)^\star D_N \phi + m^2 \phi^\star\phi ]. \qquad \quad
\eeqa
where $D_M\equiv \nabla_N-iq A_N$. From the action, one can obtain the equations of motion for the complex scalar,
\beqa
g^{MN}(\nabla_M - iq A_M)(\nabla_N - iq A_N) \phi - m^2\phi = 0 . \quad
\eeqa
Assuming the space-time metric is
\beqa
ds^2 = - g_{tt}(r)dt^2 + g_{rr}(r)dr^2 + g_{xx}(r)dx^2.
\eeqa
Then, the EOM of the charged scalar is
\beqa
&& \frac{1}{g_{rr}}\bigg[
\partial_r^2  +  \frac{1}{2}\bigg(\frac{g_{tt}^\prime}{g_{tt}} - \frac{g_{rr}^\prime}{g_{rr}} + (d-1)\frac{g_{xx}^\prime}{g_{xx}} \bigg) \partial_r
\bigg]\phi \nn\\
& + & \bigg( \frac{1}{g_{xx}} \partial_x^2  + \frac{1}{g_{tt}}\big(\omega + qA_t\big)^2 - m^2 \bigg) \phi = 0,
\eeqa
where $\phi=\phi(t,r,x)$,  and for the briefness, we have dropped $r$ dependence of $g_{tt},g_{rr}, g_{xx}$.
The EOM of the charged scalar can re-expressed as
\beqa
\bigg[ \frac{\partial_r(\sqrt{-g}g^{rr}\partial_r ) }{\sqrt{-g}}+ \bigg(\frac{\partial_x^2 }{g_{xx}} + \frac{(\omega + q A_{t} )^2}{g_{tt}} \bigg)  - m^2 \bigg]\phi = 0. \qquad \quad \label{Eq:EOM-charged-scalar-1}
\eeqa
where we have used the relation
\beqa
\frac{1}{A(r)}\partial_r[ A(r)\partial_r ] = \partial_r^2 + \log{A(r)}^\prime \partial_r ,
\eeqa
so that $ A(r) \equiv  \sqrt{-g}g^{rr} = \sqrt{g_{tt}g^{rr}g_{xx}^{d-1}} $.
By doing Fourier transformation as,
\beqa
\phi(r,x^\mu) = \int \frac{d^dk}{(2\pi)^d}e^{ik_\mu x^\mu}\phi(r,k_\mu) , 
\eeqa
where $k_\mu = (-\omega, \vec{k})$ and $x^\mu=(t,x^i)$. The equation of motion for $\phi = \phi(r,k_\mu)$ is given by
\beqa
-\frac{1}{\sqrt{-g}}\partial_r(\sqrt{-g}g^{rr}\partial_r \phi) + [g^{ii}( \vec{k}^2 - u^2) + m^2]\phi =0, \qquad  \quad \label{Eq:EOM-charged-scalar-2}
\eeqa
where
\beqa
u = \sqrt\frac{g_{xx}}{g_{tt}}[ \omega + q A_t(r) ].
\eeqa

\subsubsection{Charged scalar in RN-AdS$_{d+1}$ spacetime}

In the AdS$_{d+1}$ spacetime in the energy coordinate $z$,
\beqa
ds^2 = \frac{\ell^2}{z^2} \bigg(  - f(z) dt^2 + \frac{dz^2}{f(z)} + dx_{d-1}^2  \bigg),
\eeqa
where the redshift factor and the gauge fields are
\beqa
&& f(z) = 1 + \frac{d}{d-2} \bigg( \frac{z}{z_\star} \bigg)^{2(d-1)} - \frac{2(d-1)}{d-2} \bigg( \frac{z}{z_\star} \bigg)^d, \quad \nn\\
&& A_t = \mu \bigg( 1 - \frac{z^{d-2}}{z_\star^{d-2}}  \bigg).  \label{Eq:RNAdS-f(z)}
\eeqa
The EOM for the charged scalar becomes
\beqa
 && \bigg[ f(z)\partial_z^2  +  \bigg( f^\prime(z)  - (d-1) \frac{f(z)}{z} \bigg) \partial_z  \nn\\
 && -\frac{(\partial_t - iqA_t)^2}{f(z)} + \partial_x^2  - \frac{ m^2\ell^2 }{z^2}  \bigg] \phi = 0,
\eeqa
where $\phi=\phi(t,z,x)$ and
\beqa
u(z) = \frac{1}{\sqrt{f(z)}}\bigg[\omega + q\mu\bigg(1 - \frac{z^{d-2}}{z_\star^{d-2}} \bigg) \bigg] .
\eeqa
Note that, we have chosen the gauge so that the scalar potential is zero \emph{at the horizon} ($r=r_0$,$z=z_0$), as a result
\beqa
A_t(z) \to \mu , \quad u(z) \to \omega + q \mu . \quad (z \to 0) \qquad
\eeqa
This implies that $\omega$ should correspond to the difference of the boundary theory frequency from $q \mu$, thus the low energy limit really means very close to the \emph{effective chemical potential} $q\mu$. In the momentum space ($\partial_t\to i \omega$ and $\partial_x\to -ik$), the EOM can be re-expressed as
\beqa
\bigg[ z^{d-1}\partial_z \bigg( \frac{f(z)}{z^{d-1}}\partial_z \bigg) + \frac{(\omega-qA_t)^2}{f(z)} - k^2  - \frac{m^2\ell^2}{z^2} \bigg] \phi = 0 ,  \qquad \quad \label{Eq:EOM-scalar-UV-AdS(d+1)}
\eeqa
where $\phi=\phi(\omega,z)$.
We will most interested in the case $T=0$, where $f(z)$ is shown in Eq.(\ref{Eq:RNAdS-f(z)}).


\subsubsection{CFT$_1$ correlation functions}


Consider a massive neutral scalar field $\phi$ with mass in the bulk action
\beqa
S_\phi  = - \frac{1}{2}\int d^{2}x\sqrt{-g} [  (\nabla \phi)^2 + m^2 \phi^2 + \cdots  ], \label{S_phi}
\eeqa
where $\cdots$ denotes the interactions terms or high dimensional operator, which is irrelevant at the moment. Consider a pure anti-de Sitter space time with a AdS$_{2}$ metric in conformal coordinate
\beqa
ds^2 
=  \frac{\ell^2}{z^2}  ( - dt^2 + dz^2 ),
\eeqa
where $\mu=(t)$, the signature of the ordinary spacetime is chosen as $(-1,1)$, and $z$ is coordinate of extra-dimension where the gravity is penetrating. In the coordinates, the AdS$_2$ boundary is lying at $z=0$.

The equations of motion is $\nabla^2\phi - m^2 \phi = 0$ in the AdS$_{2}$ metric is
\beqa
\partial_z^2 \phi - \frac{1}{z}\partial_z \phi - \partial_t^2 \phi  - \frac{m^2 \ell^2}{z^2}\phi=0. \label{Eq:EOM-Neutral-Massive-Scalar-AdS(1+1)}
\eeqa
where $\phi= \phi(z,t)$.
In the infinite boundary $z \to 0$, the scalar wave equations is dominated by pole at $z=0$,
\beqa
\partial_z^2 \phi - \frac{1}{z}\partial_z \phi - \frac{m^2 \ell^2}{z^2}\phi=0.
\eeqa
In the infinite boundary, the asymptotic solution to the scalar wave function has the expansion,
\beqa
\phi(t,z) \sim A(\omega) z^{\Delta_-} + B(\omega) z^{\Delta_+}, \quad z \to 0,
\eeqa
where the two exponents are respectively
\beqa
\Delta_\pm = \frac{1}{2} \pm \nu_1, \quad  \nu_1 = \sqrt{\frac{1}{4}+m^2 \ell^2}.  \label{Eq:Delta-AdS(1+1)}
\eeqa
The conformal dimension $\Delta_\pm$ are two roots of the quadratic equations
\beqa
\Delta(\Delta-1) = m^2 \ell,
\eeqa
which are consistent with the results from AdS$2$ gravity~\cite{Strominger:1998yg}.

Let's consider an operator ${\mathcal O}$ the boundary field theory, by considering a bulk scalar field $\phi(t,z)$ with mass $m$ and charge $q$. The boundary field theory in the UV, e.g. CFT$_1$ in the UV is characterized only and completely by the dimension of the operator CFT$_1$, e.g. $\Delta$ which is given in term of bulk quantities as shown in Eq.(\ref{Eq:Delta-AdS(1+1)}) for neutral bulk field. The boundary field theory in the IR, e.g, CFT$_1$, or $(0+1)$-dimensional CQM~\cite{deAlfaro:1976vlx,Chamon:2011xk,Jackiw:2012ur}, is characterized by the scaling dimension of the operator $\Phi$. Then we have the following correspondence to the conformal field theory at the boundary.
\begin{enumerate}
\item $\Delta_+$ is the \emph{conformal dimension} of the dual operator ${\mathcal O}$;
\item $A(\omega)$: the coefficient of the more dominant term($ |z^{\Delta_-}| \gg |z^{\Delta_+}|$, when $z\to 0$) in the infinite boundary condition $z \to 0$, can be identified as the \emph{source} for ${\mathcal O}$, which is equivalent to adding to the Lagrangian of the boundary theory a source term $\delta S_B = \int dt A(t){\mathcal O}(t)$.
\item $B(\omega)$: the coefficient of the sub-dominant term in the infinite boundary condition, can be identified as the \emph{expectation value} of the operator ${\mathcal O}$, e.g., $\langle {\mathcal O}\rangle = 2\nu_1 B(t)$.
\item The \emph{linear response function} in momentum space for ${\mathcal O}$ is $G_R(\omega,\vec{k})=2\nu_1 B(\omega)/A(\omega)$, where the $B$ and $A$ are the quantities after doing Fourier transform along the boundary directions $z$. The ratio is determined by a \emph{regularity condition} at the interior of the space-time, e.g., for the horizon brane, it is the infalling condition at the horizon.
\end{enumerate}
With above identification, we have the following physical consequence:
\begin{enumerate}
\item $A(\omega)=0$ ($\delta S_B=0$) but $B(\omega)\ne 0$: \emph{Spontaneous symmetry breaking} (SSB), the operator ${\mathcal O}$ has developed an expectation value without a source, the linear response function is divergent $G_R(\omega)\to\infty$.
\item $A(\omega)\ne 0$ ($\delta S_B\ne 0$) but $B(\omega) \ne 0$: \emph{Symmetry breaking} (SB), the operator ${\mathcal O}$ has developed an expectation value with a source, the linear response function is finite.
\item $A(\omega)\ne 0$ but $B(\omega)=0$: No expectation value is developed, the linear response function is absent, namely no response at all.
\end{enumerate}

\subsubsection{CFT$_d$ correlation functions}

At the momenta, let's re-visit the asymptotic behavior of the bulk scalar, in the infinite boundary for the AdS$_{d+1}$ spacetime with $d\ge 3$, $z \to 0$ ($f(z)\to 1$, $A_t \to \mu$):
\beqa
\phi(x^\mu,z) & \sim & A(x) z^{\Delta_-} + B(x) z^{\Delta_+}, \quad z\to 0, \nn\\
\phi(k_\mu,r) & \sim & A(k_\mu)r^{- \Delta_-} + B(k_\mu) r^{-\Delta_+} ,
\eeqa
where the two exponents are, respectively,
\beqa
\Delta_\pm = \frac{d}{2} \pm \nu_d, \quad  \nu_d = \sqrt{\frac{d^2}{4}+m^2 \ell^2},  \label{Eq:nu_d-Boson}
\eeqa
and it is worthy of noticing that $\Delta_- = d - \Delta_+$. Thus, the retarded Green function in the infinite boundary condition is
\beqa
G_R(\omega,\vec{k}) = K\frac{B(k_\mu)}{A(k_\mu)},
\eeqa
where we just consider the standard quantization in the discussion and $K$ is a positive constant, which is independent of $k_\mu =  (-\omega,\vec{k})$. The asymptotic behavior is obtained by solving the EOM in the AdS$_{d+1}$ ($d\ge 3$) metric
\beqa
\bigg( \partial_z^2  - \frac{d-1}{z}\partial_z  - \big[ (\partial_t - iq\mu)^2 - \partial_x^2 \big] - \frac{m^2\ell^2}{z^2}  \bigg) \phi = 0 , \qquad \quad
\eeqa
where $\phi=\phi(t,z,x)$.
For neutral massive scalar case $q=0$. The EOMs of the scalar wave equations are dominated by pole at $z=0$,  for $d\ge 3$, in the infinite boundary $z \to 0$, namely,
\beqa
\bigg( \partial_z^2  - \frac{d-1}{z}\partial_z  - \frac{m^2 \ell^2}{z^2} \bigg) \phi(t,z,x)=0.
\eeqa
Assume that $\phi$ is an \emph{in-falling wave} at the horizon ($\partial_x \to i k$, $\partial_t \to -i\omega$), then the EOM in boundary in momentum space becomes
\beqa
\bigg( \partial_z^2   \!-\! \frac{d-1}{z}\partial_z + \big[ (\omega + q\mu)^2 - k^2 \big] \!-\! \frac{m^2\ell^2}{z^2} \bigg) \phi = 0, \qquad\qquad
\eeqa
where $\phi=\phi(\omega,z,k)$.
Therefore the boundary theory energy corresponds to $\omega+ q\mu$, and $\omega$ should be interpreted as measured from the \emph{effective chemical potential} $q\mu>0$. The wave functions becomes
\beqa
\phi(z) \overset{z\to 0}{\sim}  z^{d/2} \left[ c_1 J_{\nu_d}(\bar\omega z)+c_2 Y_{\nu_d}(\bar\omega z)\right] ,
\eeqa
where $\bar\omega=\sqrt{(\omega+q\mu)^2-k^2}$, and the conformal dimension is $\nu_d$ is defined in Eq.(\ref{Eq:nu_d-Boson}).

Take charged scalar in AdS$_2$ vacuum as an example. In this case, $d=1$, $q=0$, $k=0$, and the wave functions becomes
\beqa
\phi(z) = \sqrt{z} \left[c_1 J_{\nu_1}(z \omega )+c_2 Y_{\nu_1}(z \omega )\right],
\eeqa
in the near horizon limit, one obtains the asymptotic behavior of the boson wave function as
\beqa
\phi(z) & \overset{z\to \infty}{\sim} & c_1 \frac{\cos \left[\frac{\pi  \nu_1}{2}- (\omega + q\mu)  z\right]-\sin \left[\frac{\pi  \nu_1}{2}- (\omega + q \mu)  z\right]}{\sqrt{\pi } \sqrt{\omega }} \nn\\
&& - c_2  \frac{\sin \left[\frac{\pi  \nu_1}{2}- (\omega + q\mu)  z\right]+\cos \left[\frac{\pi  \nu_1}{2}-(\omega + q\mu)  z\right]}{\sqrt{\pi } \sqrt{\omega }} \nn\\
&=&  \frac{\left(\frac{1}{2}+\frac{i}{2}\right) (c_1+i c_2)}{\sqrt{\pi } \sqrt{\mu  q+\omega }} e^{-i(\omega+q\mu)z+ i \frac{\pi}{2}\nu_1 } \nn\\
&& + \frac{\left(\frac{1}{2}-\frac{i}{2}\right) (c_1-i c_2)}{\sqrt{\pi } \sqrt{\mu  q+\omega }} e^{i(\omega+q\mu)z - i \frac{\pi}{2}\nu_1 } .
\eeqa
The in-falling wave is $e^{-i\omega t + i (\omega+q\mu)z}$ which entails that $c_1=-ic_2$ so that the outgoing wave are isolated.
On the other hand, in the infinite boundary $z\to 0$, one has
\beqa
\phi(z) & \overset{z\to 0}{\sim} &  \frac{2^{-\nu_1} \left(c_2 \cot (\pi  \nu_1)+c_1\right) (\mu  q+\omega )^{\nu_1}}{\Gamma (\nu_1+1)} z^{\frac{1}{2}+\nu_1} \nn\\
&& -\frac{c_2 2^{\nu_1} \Gamma (\nu_1) (\mu  q+\omega )^{-\nu_1}}{\pi } z^{\frac{1}{2} - \nu_1} \nn\\
 &=& B(\omega) z^{\nu_1} + A(\omega) z^{-\nu_1},
\eeqa
where $A$ and $B$ are identified as source and response, respectively.
The two point Green's function ban be read as
\beqa
G(\omega) &=& \frac{B(\omega)}{A(\omega)} = -\frac{\pi   [\cot (\pi  \nu_1)-i]   }{ 4^{\nu_1} \Gamma (\nu_1) \Gamma (\nu_1+1)} \omega^{2 \nu_1}  \nn\\
&=&  - \frac{\pi  4^{-\nu_1} e^{i \pi  \nu_1}  }{\Gamma (\nu_1) \Gamma (\nu_1+1)}  \frac{\omega^{2 \nu_1}}{   \sin (\pi  \nu_1) } .
\eeqa

\subsection{AdS$_2$ from RN-AdS$_{d+1}$}

The AdS$_2$ spacetime background can be generalized into high dimensional spacetime, e.g., as an embedded boundary in the near horizon boundary of a charged RN black hole in AdS$_{d+1}$.
\beqa
ds^2 =  \bigg(\frac{r}{\ell}\bigg)^2 ( - f(r) dt^2 +  dx^2 ) +  \bigg( \frac{\ell}{r} \bigg)^2  \frac{dr^2}{f(r)} ,  \qquad \label{Eq:ds2-r-AdS(d+1)}
\eeqa
where
\beqa
&& f(r)  =  1 + \frac{d}{d-2}\bigg(\frac{r_\star}{r} \bigg)^{2(d-1)} - \frac{2(d-1)}{d-2} \bigg( \frac{r_\star}{r} \bigg)^{d}, \quad \nn\\
&& A_t = \mu\bigg[ 1 - \bigg(\frac{r_\star}{r} \bigg)^{d-2} \bigg].  \label{Eq:RNAdS-f(r) }
\eeqa
In the IR limit at large distance, we have $f(r) \approx 1$, or energy scale is much larger than the chemical potential $\mu$ (but still much less than the UV scale) is simply a vacuum with conformal symmetry AdS$_{d+1}$.
The near horizon geometry is given by AdS$_2\times {\mathbb R}^{d-1}$, i.e., $ds^2 = ds^2_{AdS_2} + ({r_\star^2}/{\ell^2}) dx^2$, which indicates the boundary system should develop an enhanced symmetry group including \emph{scaling invariance}.
For $T = \frac{1}{2\pi \zeta_0}  $ and $T = 0  $ case, respectively, one has AdS$_2$ in global and local/Poincar\'{e} coorindates, respectively, as
\beqa
 ds_{NAdS_2}^2 &=&  \frac{\ell_2^2}{\zeta^2} \bigg[ - \bigg(  1 - \frac{\zeta^2}{\zeta_0^2} \bigg) dt^2 + \bigg(  1 - \frac{\zeta^2}{\zeta_0^2}  \bigg) ^{-1} d\zeta^2  \bigg] \nn\\
&=& \frac{-d\bar{t}^2 + dz^2}{[\sinh(z/\ell)]^2},  \quad A_t = - \frac{E\ell^2}{\zeta}\bigg(1 - \frac{\zeta}{\zeta_0}\bigg),   \qquad  \label{Eq:ds2-T!=0-AdS2} \\
ds_{AdS_2}^2 &=&  \frac{\ell_2^2}{\zeta^2} ( -  dt^2 + d\zeta^2 ),  \quad A_t =  \frac{e_d}{\zeta},  \label{Eq:ds2-T=0-AdS2}
\eeqa
where $\ell_2$ is the curvature radius of AdS$_2$ given by
\beqa
\ell_2 = \frac{1}{d(d-1)}\ell .
\eeqa
For the second equality, we have used the coordinate transformation as
\beqa
  \zeta  = \zeta_0 \tanh\frac{z}{\ell_2}, \quad  \bar{t} = \frac{\ell_2}{\zeta_0}  t , \quad A_{\bar{t}} = E\ell \bigg(\coth\frac{z}{\ell_2} - 1\bigg). \qquad \quad
\eeqa
In the finite temperature limit $\zeta\to\zeta_0$, the metric and the vector potential of Maxwell field are dominated by
\beqa
ds^2 &\to & \frac{\ell_2^2}{\zeta^2} \bigg(1-\frac{\zeta^2}{\zeta_0^2} \bigg)^{-1} d\zeta^2  + \frac{r_\star^2}{\ell^2} dx^2, \quad A_t(\zeta) \to 0. \qquad
\eeqa
The time direction also shrinks to zero, and the the spatial direction approaches a constant, the Maxwell field approaches zero.
The RG scales is flowing from AdS$_{d+1}$ with scale $z$ to near horizon boundary with scale $z_0$, a single scale of the boundary theory. In the boundary field theory aspect, the theory flows from CFT$_d$ in the UV to the IR-CFT$_1$, which is a conformal symmetries of a $(0+1)$-dimensional conformal quantum mechanics (CQM)~\cite{deAlfaro:1976vlx,Chamon:2011xk,Jackiw:2012ur}, including the scaling symmetry in the time direction. Therefore, the corresponding IR fixed point is just a IR-CFT$_1$, which is a conformal symmetry only in the time direction. The new conformal symmetry is emergent and has relation with the collective motion of the large number of charged excitation. Of course one has to take a notice that the spatial direction can also have important physical consequence.

In the zero temperature limit $\zeta \to 0$, the metric and the gauge field in Eq.(\ref{Eq:ds2-T!=0-AdS2}) reduce to the $T=0$ case in Eq.(\ref{Eq:ds2-T=0-AdS2}). It is worthy of emphasizing here that the central charge is \emph{infinite}, since it is proportional to the \emph{volume} of the $d$-dimensional transverse space ${\mathbb R}^{d-1}$. To have a \emph{finite} central charge one could replace ${R}^{d-1}$ by other manifold, i.e., a torus.

The time direction shrinks to zero, and the the spatial direction approaches a constant, the Maxwell field approaches zero. The AdS$_2$ symmetry is emergent in the near-horizon region. The AdS$_2$ is isomorphic to a full $SL(2,R)$ symmetry, which own the scaling isometry:
\beqa
t\to \lambda t, \quad \zeta \to \lambda \zeta, \quad x \to x,
\eeqa
where only the time sector scales. The \emph{finite} $t$ corresponds to the \emph{long time limit} of the original time coordinate, meanwhile the \emph{short distance limit} of the original spatial coordinates. Thus, the metric obtained above should apply to the \emph{low frequency limit}, since $\omega$ is the frequency conjugate to $t$
\beqa
 \omega \sim T \ll \mu.
\eeqa
In the \emph{low frequency limit}, the d-simensional boundary theory at finite charge density should be described by a CFT$_1$, in terms of \emph{IR CFT of the boundary theory, which is an emergent conformal symmetry due to collective behavior of a large number of degrees of freedom}. It is not related to the \emph{microscopic conformal symmetry in the UV}, which is broken by \emph{finite charge density}.

The metric is the AdS$_2$ slice of the high dimensional RN-AdS$_{d+1}$, the letters with bar above is associated with $\bar{t}$, $T$ and $\bar{T}$ are Hawking temperature with respect to the coordinates $t$ and $\bar{t}$, respectively. In this case, the frequency conjugated to the rescaled time $\bar{t}$ becomes
\beqa
\bar\omega = \frac{\zeta_0}{\ell}\omega, \quad  \frac{\bar\omega}{2\pi \bar{T}} = \frac{\omega}{2\pi T},  \quad \bar{T} = \frac{1}{2\pi \ell_2} .
\eeqa

\subsubsection{Near horizon field equations of motion}

Consider a massive scalar field $\phi$ with mass in the bulk action as
\beqa
S  = - \int d^{d+1}x \sqrt{-g} [ g^{MN}(D_M \phi)^\star (D_N \phi) + m^2 \phi^\star\phi ],  \qquad \quad \label{Eq:S_phi}
\eeqa
where the covariant derivative is defined as $D_N \equiv \nabla_N- iq A_N$.
From the action, one can obtain the equations of motion for the complex scalar,
\beqa
[g^{MN}(\nabla_M - iq A_M)(\nabla_N - iq A_N) - m^2]\phi = 0 .  \qquad \label{Eq:EOMs_phi}
\eeqa
Let's consider a charged scalar field in the background metric and gauge field at zero temperature $T = 0$, which  is given by Eq.(\ref{Eq:ds2-T=0-AdS2}),
\beqa
ds^2 &=&  \frac{\ell_2^2}{\zeta^2} ( -  dt^2 + d\zeta^2 ) + \frac{r_\star^2}{\ell^2} dx^2 ,  ~ A_t =  \frac{e_d}{\zeta}, \quad
\eeqa
which recovers the asymptotic AdS$_2$ spacetime in Eq.(\ref{Eq:AdS2-bd}), by setting $\ell_2=1$ and $e_d = \mu$. To be explicit,  by considering that $g^{xx}= 1/r_\star^2$, $A_x = 0 $, the field equations become,
\beqa
\bigg( \partial_\zeta^2 -\partial_t^2 + \frac{\ell^2\ell_2^2}{r_\star^2 \zeta^2}\partial_x^2  + \frac{2iq e_d}{\zeta}\partial_t   + \frac{q^2 e_d^2}{\zeta^2}  - \frac{m^2\ell_2^2}{\zeta^2} \bigg) \phi = 0, \qquad \quad
\eeqa
where $\phi= \phi(t,\zeta)$.
Alternatively, one can expand $\phi$ in momentum space by doing Fourier transformation,
\beqa
\phi(t, \vec{x}, \zeta) = \int \frac{d\omega d\vec{k}}{(2\pi)^d}e^{-i \omega t + i\vec{k}\cdot\vec{x}} \phi(\omega,k,\zeta),
\eeqa
one obtains
\beqa
S  = - \int d^{2}x \sqrt{-g} [ g^{ab}(D_a\phi)^\star_{\vec{k}}D_b\phi_{\vec{k}} + m_k^2 \phi^\star_{\vec{k}}\phi_{\vec{k}} ] , \qquad
\eeqa
where $D_a= \nabla_a - iq A_a$ and
\beqa
m_k^2 \equiv m^2 + \frac{k^2\ell^2}{r_\star^2} = m^2 + \frac{k^2 z_\star^2}{\ell^2} \equiv m^2 + \tilde{m}^2 .  \label{Eq:mk2}
\eeqa
The indices $a$, $b$ only run over $t$ and $\zeta$. From the action, one can obtain the equations of motion for the complex scalar,
\beqa
[g^{ab}(\nabla_a - iq A_a)(\nabla_b - iq A_b) - m_{\vec{k}}^2]\phi_{\vec{k}}=0,
\eeqa
which can be written as
\beqa
\bigg( \partial_\zeta^2  - \partial_t^2   + \frac{2iqe_d}{\zeta}\partial_t   + \frac{q^2e_d^2}{\zeta^2}   - \frac{m_{\vec{k}}^2\ell_2^2}{\zeta^2} \bigg) \phi  = 0 ,
\eeqa
where $\phi=\phi(t,\zeta)$.
For pure AdS$_2$ case as in Eq.(\ref{Eq:AdS2-bd}), one has $\tilde{m}$.
By doing Fourier transformation from the time coordinate to the frequency space, the equation becomes,
\beqa
&&\bigg[ \partial_\zeta^2 -  \frac{m_{\vec{k}}^2\ell_2^2}{\zeta^2} + \bigg( \omega \pm \frac{q e_d}{\zeta} \bigg)^2   \bigg] \phi(\omega,\zeta)   = 0,   \label{Eq:EOM-AdS2-R(d-1)}
\eeqa
where the $\pm$ sign corresponds to out/in going waves ($\partial_t \to \mp i\omega$ and $\partial_x \to \pm i k$), respectively,

\subsubsection{IR correlation functions of scalars}
Firstly, let's consider the case in the infinite boundary conditions,
\beqa
d\ge 2: \zeta \sim \frac{z_\star}{d(d-1)} \to 0, \quad d =1 : z  \to 0.
\eeqa
The EOM is dominated by the singularity at $z_\star \to 0$,
\beqa
\bigg( \partial_\zeta^2 - \frac{m_{\vec{k}}^2\ell_2^2 - q^2 e_d^2 }{\zeta^2} \bigg)  \phi(\omega,\zeta)  = 0. \label{Eq:EOM-scalar-IR}
\eeqa
The asymptotic behavior of the solution are
\beqa
\phi(\omega, \zeta) &=&  A(\omega) \zeta^{\frac{1}{2}-\nu_k}[1+O(\zeta)] + B(\omega) \zeta^{\frac{1}{2}+\nu_k}[1+O(\zeta)]  \nn\\
&\sim & A(\omega) \zeta^{\Delta_-^{\text{IR}}} + B(\omega) \zeta^{\Delta_+^{\text{IR}}}, \quad \zeta\to 0,  
\label{Eq:phi-IR-A-B}
\eeqa
where the two exponents are respectively
\beqa
\Delta_\pm^{\text{IR}} = \frac{1}{2} \pm \nu_k, \quad  \nu_k = \sqrt{\frac{1}{4}+m_k^2\ell_2^2- q^2 e_d^2} 
. \quad \label{Eq:nu_k}
\eeqa
For pure AdS$_2$ case in $(1+1)$-dimensional spacetime, one has $\ell_2=1$, $k=0$, and $e_d=\mu$. Thus,
\beqa
\Delta_\pm^{\text{IR}} = \frac{1}{2} \pm \sqrt{\frac{1}{4} + m^2 - q^2 \mu^2}.  \label{Eq:Delta_IR-AdS2}
\eeqa

It is worthy of emphasizing that it is possible that the $\nu_k$ become pure imaginary, once the electric charge $q$ becomes sufficiently large.
Then, $\nu_k$ becomes imaginary for sufficiently small $k^2<k_0^2$ (for a given $m$, this always occurs for a sufficiently large $q$) with
\beqa
\nu_k= - i \lambda_k, \quad && d\ge 2: \lambda_k = \frac{\ell_2\ell}{r_\star}\sqrt{k_0^2-k^2} , \quad \nn\\
&& d =1 : \lambda= \sqrt{\mu_L^2 - m^2 - \frac{1}{4}},
  \label{Eq:lambda-k}
\eeqa
where
\beqa
 k_0^2 & \equiv  & \frac{r_\star^2}{\ell^2} \bigg( \frac{q^2 e_d^2 -\frac{1}{4}}{\ell_2^2} - m^2  \bigg) \nn\\ 
&=& \frac{d(d-1)}{z_\star^2}\bigg(  ( \mu_L^2  - m^2 ) \ell_2^2 - \frac{1}{4} \bigg)>0,   \label{Eq:k0}
\eeqa
where $z_\star\equiv \ell^2/r_\star$ and the local chemical potential $\mu_L$ is defined as in Eq.(\ref{Eq:mu_L}). By observing the conformal dimension in neutral background as in Eq.(\ref{Eq:Delta-AdS(1+1)}). It can be viewed that the background electric field acts through the charge as an \emph{effective negative mass square}, which make it possible that the total mass square $m^2-q^2\mu^2$ becomes negative and resulting in an imaginary conformal dimension.

For a neutral scalar operator with $q=0$ in AdS$_{d+1}$ spacetime , one can obtain the oscillatory region mass window,
\beqa
- \frac{d^2}{4} <  m^2 \ell^2 < - \frac{d(d-1)}{4},
\eeqa
Where the \emph{lower limit} comes form the \emph{stability of vacuum} theory, i.e., Breitenlohner-Freedman (BF) bound
of AdS$_{d+1}$ and the \emph{upper limit} comes from the condition $k_0^2>0$.
For AdS$_2$ case, one has
\beqa
\text{AdS}_2:  - \frac{1}{4} <  m^2 \ell^2 <  0 .
\eeqa
By using Eq.(\ref{Eq:ds2-T=0-AdS2}), we obtain the chemical potential $\mu_L$ for a \emph{local observer} with charge $q$ in the bulk ({the spatial part is flat}),
\beqa
&& d \ge 2:  \mu_L = q\sqrt{g^{tt}}A_t = q \frac{\zeta}{\ell_2}\frac{e_d}{\zeta} = \frac{q e_d}{\ell_2} , \quad \nn\\
&& d = 1 :  \mu_L = q  \mu,  \label{Eq:mu_L}
\eeqa
when this \emph{local chemical potential} excedds the mass of a charged particles
\beqa
&& d \ge 2: \mu_L^2 = \frac{q^2 e_d^2}{\ell_2^2} > m^2, \quad \nn\\
&& d =1:  \mu_L^2 = q^2 \mu^2 > m^2,
\eeqa
the system will be the Bose-Einstein condensation. This gives the bulk origins of the scalar instabilities in the parameter region in Eq.(\ref{Eq:k0}). Physically, in the near horizon region of the extremal black brane geometry ($T=0$), at each point in the bulk geometry, e.g., AdS$_2 \times{\mathbb R}^{d-1}$, there is a \emph{local $d$-dimensional Fermi surface} with Fermi momentum $k_0$, which upon projection to the boundary theory, would result in a $(d-1)$-dimensional \emph{Fermi disc}, in which there are \emph{gapless excitations} at each point in the \emph{interior of the disc} in the $(d-1)$-dimensional momentum space. For a pure AdS$_2$ bulk vacuum spacetime in $(1+1)$-dimensional spacetime, if $\mu_L=q\mu>m$ there will be Bose-Einstein condensation for scalar bosons, while for fermions there is no Fermi surface, since by definition $k_0=0$, while there is still a critical chemical potential
\beqa
\mu_L^c = \sqrt{m^2 + \frac{1}{4}},
\eeqa
at which, the NFL phase (since $\nu_k<1/2$) of CFT$_1$ is separated into steady and oscillatory due to the charge instability.

It is worthy of noticing that at $\nu_k=0$, or equivalently
\beqa
m_k = \frac{1}{\ell_2}\sqrt{q^2 e_d^2-\frac{1}{4}} = \sqrt{\mu_L^2 - \frac{d(d-1)}{4\ell^2} }, ~ \Delta_\pm^{\text{IR}} = \frac{1}{2}, \qquad \quad
\eeqa
which is impossible for neutral black brane with $q=0$, since $m_k>0$ is always positive unless $m=0$ and $k=0$(no scalar at all). As a result, the original two independent solutions are degenerate. In this case, the equation of motion of charged scalar in the infinite boundary conditions at IR fixed point in Eq.(\ref{Eq:EOM-scalar-IR}) become
\beqa
\bigg( \partial_\zeta^2 - \frac{\nu_k^2 - 1/4}{\zeta^2} \bigg) \phi = \bigg( \partial_\zeta^2 + \frac{1}{4\zeta^2} \bigg) \phi = 0,  \label{Eq:EOM-scalar-IR-nu-k=0}
\eeqa
where $\phi=\phi(\omega,\zeta)$.
The asymptotic behavior of the solution are
\beqa
\phi(\omega, \zeta) \sim  A(\omega) \zeta^{\frac{1}{2}}\log\sqrt{\zeta} + B(\omega) \zeta^{\frac{1}{2}} , ~ \zeta\to 0, \quad \label{Eq:phi-IR-A-B-nu-k=0}
\eeqa
where $\zeta>0$.
Secondly, let's consider a more generic case of the solution. In \emph{frequency space} $\phi(t,\zeta)=e^{-i\omega t }\phi(\omega,\zeta)$, the equation of the motion for a charged scalar $\phi$ can be written as
\beqa
[ \partial_\zeta^2  - V(\zeta)  ]  \phi  = 0,  ~ V(\zeta) \equiv \bigg[ \frac{m_{\vec{k}}^2\ell_2^2}{\zeta^2} - \bigg( \omega \pm \frac{q e_d}{\zeta} \bigg)^2  \bigg], \qquad \quad
\eeqa
where the plus and minus sign corresponds to the {out/in going} waves ($\partial_t \to \mp i\omega$).
Note that $\omega$ can be scaled away by redefining $\zeta$, reflecting the \emph{scaling symmetry} of the background solution. the potential can be expressed as a function depending on the dimension of operators in the infinite boundary conditions,
\beqa
V(\zeta) 
= \frac{\nu_k^2 - \frac{1}{4}}{\zeta^2}  - \omega^2 \mp \frac{2q e_d}{\zeta}\omega,    
\eeqa
where in the last identity, we have used the Eq.(\ref{Eq:nu_k}). According to the asymptotic behavior of the solution in Eq.(\ref{Eq:phi-IR-A-B}) and the EOMs above, it is obviously that the frequency $\omega$ dependence in the potential $V(\zeta)$, can be scaled away from by redefining the $\zeta\to qe_d/\omega$ so that the EOMs recovers Eq.(\ref{Eq:EOM-scalar-IR-nu-k=0}). Thus, the solution to the generic case will be of form
\beqa
\phi = B \bigg(\frac{\zeta}{\omega}\bigg)^{\frac{1}{2}+\nu_k}[1+ O (\zeta)] + A \bigg(\frac{\zeta}{\omega}\bigg)^{\frac{1}{2}-\nu_k}[1+ O (\zeta)], \qquad \quad
\eeqa
where as $\zeta \to 0$, one has
\beqa
 B \sim \omega^{\frac{1}{2}+\nu_k}, \quad A\sim \omega^{\frac{1}{2}-\nu_k} .
\eeqa
Thus after imposing the infalling boundary condition on $\phi$, the correlated functions are expected to be of form
\beqa
{\mathcal G}_R \propto \frac{B}{A}\sim \omega^{2\nu_k}.
\eeqa
This implies a \emph{coordinate space correlation function} by doing an inverse Fourier transformation\footnote{Note: for ${\mathcal G}_R \sim |\omega|^{2\nu_k}$, the inverse Fourier transformation gives the result $-\sqrt{2/\pi}|t|^{-1-2\nu_k}\Gamma[1+2\nu_k]\sin(\pi\nu_k) \sim t^{-2\Delta}$.},
\beqa
{\mathcal G}_R(t) = - \frac{1+\text{sgn}(t)}{|t|^{1+2\nu_k}}\frac{e^{i\pi\nu_k}}{\sqrt{2\pi}}\Gamma[1+2\nu_k]\sin(2\pi\nu_k)\sim \frac{1}{t^{2\Delta}}, \qquad \quad
\eeqa
with the \emph{conformal dimension} $\Delta$ of the boson operator ${\mathcal O}_B$ given by $\Delta = \frac{1}{2} + \nu_k$, which is nothing but $\Delta_+^{\text{IR}}$ defined in Eq.(\ref{Eq:nu_k}). It is worthy of noticing that the dimension $\Delta$ depends on the charge $q$ through $\nu_k$. In particular, it is possible for $\nu_k$ to become imaginary when the charge $q$ becomes \emph{sufficiently large}. This implies that in the constant electric field $A_t=e_d/\zeta$ with sufficiently large charge $q$ can be \emph{pair produced}, which cause an \emph{instability for scalars}. When $\nu_k$ is imaginary, there is an ambiguity in specifying ${\mathcal G}_R$ since one can choose either $A$ or $B$ as the source term.

\subsubsection{Zero temperature case} 

\begin{widetext}

The solution turns out to be \footnote{We choose the solution $\phi(\zeta) = C_1 M_{- iq e_d,\nu_k}(2i\omega\zeta) + C_2 W_{- iq e_d,\nu_k}(2i\omega\zeta) , \quad \text{in/out going}\, (\partial_t \to \pm i\omega)$, which corresponds to $\omega+q e_d/\zeta$ in the potential.}
\beqa
\phi(\zeta) &=& C_1 M_{ iq e_d,-\nu_k}(2i\omega\zeta) + C_2 W_{ iq e_d,-\nu_k}(2i\omega\zeta) , \quad \text{in going}\,  e^{ +i \omega t - i k x}\, (\partial_t \to + i\omega) \nn\\
\phi(\zeta) &=& C_1 M_{- iq e_d,\nu_k}(2i\omega\zeta) + C_2 W_{- iq e_d,\nu_k}(2i\omega\zeta) , \quad \text{out going}\, e^{ -i \omega t + i k x}\, (\partial_t \to - i\omega)
\eeqa
where $M_{k,\nu}(z)$ and $W_{k,\nu}$ are both the Whittaker functions, which can be converted to be Kummer confluent hypergeometric function $F_1(a,b;z)$ and confluent hypergeometric function $U(a,b;z)$ respectively as below,
\beqa
M_{k,\nu}(z) = e^{-z/2}z^{1/2 + \nu}F_1(\frac{1}{2}+ \nu - k,1+2\nu,z), \quad W_{k,\nu}(z) = e^{-z/2}z^{\nu + 1/2}U(\frac{1}{2} + \nu - k ,1+2\nu,z).
\eeqa
$M_{k,m}(z)$ is zero at $z=0$ for $m>0$, while $W_{k,m}(z)$ is infinite at $z=0$ for integer $m>0$, and has a branch cut discontinuity in the complex $z$ plane running from $-\infty$ to $0$. The Kummer confluent hypergeometric function $F_1(a,b,z)\equiv {}_1F_1(a,b;z)$, has the series expansion
\beqa
F_1(a,b,z) = \sum_{k=0}^{\infty}\frac{(a)_k}{(b)_k}\frac{z^k}{k!} = 1 + \frac{a}{b}z + \frac{a(a+1)}{2b(b+1)}z^2 + O(z^3),
\eeqa
and the confluent hypergeometric function $U(a,b,z)$ has the integral representation
\beqa
U(a,b,z) &=& \frac{1}{\Gamma[a]}\int_0^\infty e^{-zt}t^{a-1}(1+t)^{b-a-1}dt = \frac{1}{z^a}\bigg(1 - \frac{a(1+a-b)}{z} + \frac{a(1+a)(1+a-b)(2+a-b)}{2z^2} + O(\frac{1}{z^{3}})  \bigg). \qquad \qquad
\eeqa
Therefore in the $z\to \infty$ expansion, one has $M_{k,\nu}(z) \sim e^{\frac{z}{2}}z^{-\nu}$, $W_{k,\nu}(z) \sim e^{-\frac{z}{2}} z^\nu$, which basically implies that Whittaker $M_{k,\nu}(z)$ is divergent in the large $z$.
In the \emph{near horizon region} $\zeta\to \infty$ for the \emph{out going wave} $ e^{-i\omega t+ i k x} \, (\partial_t  \to - i \omega)$, the asymptotic
\beqa
M_{-iq e_d,\nu_k}(2i\omega\zeta)  \sim  e^{i\omega \zeta} (2i\omega\zeta)^{iq e_d}, \quad
W_{-iq e_d,\nu_k}(2i\omega\zeta) \sim  e^{-i\omega\zeta}(2i\omega\zeta)^{-iq e_d} + \cdots .
\eeqa
Thus, in the near horizon region,
\beqa
\phi &\to & C_1 e^{-i\omega t} e^{i\omega \zeta} \zeta^{i q e_d} + C_2 e^{-i\omega t} e^{-i\omega\zeta}\zeta^{-iq e_d} \sim  C_1 e^{-i\omega ( t-\zeta - \frac{qe_d}{\omega}\log\zeta)} + C_2 e^{-i(\omega+\zeta +\frac{q e_d}{\omega}\log\zeta )},  \quad (\zeta \to \infty)
\eeqa
and it is obviously that the $M$-Whittaker function associated with $C_1$ is an out-going solution, while the $W$-Whittaker function associated with $C_2$ is always an in-going solution. Therefore, in the following, we will only keep the in-going part from $W$ by setting $C_1=0$. In the \emph{infinite boundary condition} $\zeta\to 0$, the Whittaker function can be expanded as
\beqa
M_{-i q e_d, \nu_k}( 2i\omega\zeta) &\sim  & (2i\omega\zeta)^{\frac{1}{2}+\nu_k}, \nn\\
W_{-i q e_d, \nu_k}( 2i\omega\zeta) &\sim &  (2i\omega\zeta)^{\frac{1}{2}-\nu_k} \frac{\Gamma[2\nu_k]}{\Gamma[\frac{1}{2}+\nu_k + i q e_d]} + (2i\omega\zeta)^{\frac{1}{2}+\nu_k} \frac{\Gamma[-2\nu_k]}{\Gamma[\frac{1}{2}-\nu_k + i q e_d]}
\eeqa
From which we have
\beqa
{\mathcal G}_k (\omega) \equiv \frac{B(\omega)}{A(\omega)} = (2i\omega)^{2\nu_k} \frac{\Gamma[-2\nu_k]}{\Gamma[2\nu_k]}   \frac{\Gamma[\frac{1}{2} + \nu_k +i q e_d ]}{\Gamma[\frac{1}{2} - \nu_k + i q e_d]},  \quad (\partial_t \to -i \omega),
\eeqa
which corresponds to the \emph{advanced Green's function} associated with the \emph{out-going modes}.  If one choose \emph{in-going wave} conversion, $(\partial_t \to i \omega)$ (  $ \phi \sim e^{i\omega t - i kx} $ ), then in the \emph{infinite boundary condition} $\zeta\to 0$, the Whittaker function can be expanded as
\beqa
M_{iq e_d,\nu_k}(2i\omega\zeta)  &=&  (2i\omega\zeta)^{\frac{1}{2}+\nu_k} \bigg( 1 - \frac{2i\omega\zeta}{1+2\nu_k}(2i\omega\zeta) + \cdots   \bigg), \\
W_{iq e_d,\nu_k}(2i\omega\zeta) &=& (2i\omega\zeta)^{\frac{1}{2}+\nu_k} \frac{\Gamma[-2\nu_k]}{\Gamma[\frac{1}{2}-iq e_d - \nu_k]} + (2i\omega\zeta)^{\frac{1}{2}-\nu_k} \frac{\Gamma[2\nu_k]}{\Gamma[\frac{1}{2}-iq e_d + \nu_k]} + \cdots ,
\eeqa
$M$-Whittaker function is always vanishing at infinite boundary condition, while $W$-Whittaker has both positive and negative branches labeled by dimensions of operator $\Delta_{\pm}^{\text{IR}}$. Thus, the asymptotic behavior of the complex scalar $\phi$ at $\zeta\to 0$ is given by
\beqa
\phi &\sim & (2i\omega\zeta)^{\Delta_+^{\text{IR}}} \frac{\Gamma[-2\nu_k]}{\Gamma[\frac{1}{2}-iq e_d - \nu_k]} + (2i\omega\zeta)^{\Delta_-^{\text{IR}}} \frac{\Gamma[2\nu_k]}{\Gamma[\frac{1}{2}-iq e_d + \nu_k]}  \sim  B \zeta^{\nu_k+\frac{1}{2}} + A \zeta^{-\nu_k+\frac{1}{2}}.
\eeqa
\end{widetext}
Then by comparing with Eq.(\ref{Eq:phi-IR-A-B}), one can read \emph{scalar correlation function at the IR fixed point}.
\beqa
{\mathcal G}_k (\omega) \equiv \frac{B(\omega)}{A(\omega)} = (2i\omega)^{2\nu_k} \frac{\Gamma[-2\nu_k]}{\Gamma[2\nu_k]}   \frac{\Gamma[\frac{1}{2} + \nu_k - i q e_d ]}{\Gamma[\frac{1}{2} - \nu_k - i q e_d ]},  \qquad \quad \label{Eq:Corr-fun-IR-AdS2-R(d-1)}
\eeqa
where we have used the notation $ (\partial_t \to i \omega)$, it corresponds to the \emph{retarded Greens' function} associated with the \emph{in-falling modes}. In summary, the \emph{retarded scalar function} of the IR CFT is given by
\beqa
&& {\mathcal G}_k^R (\omega) = (-i)^{2\nu_k} \frac{\Gamma[-2\nu_k]}{\Gamma[2\nu_k]}   \frac{\Gamma[\frac{1}{2}-iq e_d + \nu_k]}{\Gamma[\frac{1}{2}-iq e_d - \nu_k]}(2\omega)^{2\nu_k}, \quad \nn\\
&& (-i)^{2\nu_k}  = e^{-i\pi\nu}, \qquad \label{Eq:GR-AdS2-T=0}
\eeqa
which has the form of the retarded two-point function of a \emph{scalar operator} in a \emph{$(1+0)$-dimensional conformal quantum mechanics CFT$_1$} with transformation law
\beqa
\langle {\mathcal O}^\dag(x) {\mathcal O}(x^\prime) \rangle = \bigg( \frac{dy}{dx}\bigg)^\Delta \bigg( \frac{dy^\prime}{dx^\prime}\bigg)^\Delta \langle {\mathcal O}^\dag(y){\mathcal O}(y^\prime) \rangle, \qquad
\eeqa
with \emph{left/right-moving dimensions} and momenta of the IR CFT operator,
\beqa
\Delta_L = \Delta_R = \frac{1}{2} + \nu_k, \quad p_L = k, \quad p_R = \omega,
\eeqa
and with \emph{left/right temperatures} given by
\beqa
T_L = \frac{1}{4\pi e_d}, \quad T_R = 0.
\eeqa
The \emph{advanced function} is given by
\beqa
&& {\mathcal G}_k^A (\omega) = (i)^{2\nu_k}  \frac{\Gamma[-2\nu_k]}{\Gamma[2\nu_k]}   \frac{\Gamma[\frac{1}{2}+iq e_d + \nu_k]}{\Gamma[\frac{1}{2}+iq e_d - \nu_k]}(2\omega)^{2\nu_k}, \qquad \nn\\
&& (i)^{2\nu_k} = e^{i\pi\nu_k}.
\eeqa
For a scalar, from the correlator functions above, one find that
\beqa
\frac{{\mathcal G}^R(\omega)}{{\mathcal G}^A(\omega)} &=& e^{-2\pi i \nu_k} \frac{\cos[\pi(\nu_k + i q e_d) ]}{\cos[\pi(\nu_k - i q e_d)]} \nn\\
&=& \frac{e^{-2\pi \nu_k i} + e^{-2\pi q e_d}}{e^{2\pi \nu_k i} + e^{-2\pi q e_d}}. \label{Eq:GRS-GAS}
\eeqa
By introducing that
\beqa
{\mathcal G}^R(\omega) \equiv c(k)\omega^{2\nu_k}, \quad c(k) \equiv |c(k)|e^{i\gamma_k},
\eeqa
where $c(k)$ denotes the \emph{prefactor},
\beqa
c(k) = c(\nu_k)(-i)^{2\nu_k}, \quad c^\star(k) = c^\star(\nu_k)i^{2\nu_k}.
\eeqa
Thus, one has
\beqa
\frac{{\mathcal G}^R(\omega)}{{\mathcal G}^A(\omega)} = \frac{c(k)}{c^\star(k)} = e^{2i\gamma_k} = \frac{(e^{-2\pi \nu_k i} + e^{-2\pi q e_d})^2}{|e^{2\pi \nu_k i} + e^{-2\pi q e_d}|^2}. \qquad \quad
\eeqa
By writing ${\mathcal G}^R(\omega) = c(\nu_k) (-i\omega)^{2\nu_k}$ and
${\mathcal G}_A(\omega)= c^\star(\nu_k)(i\omega)^{2\nu_k}$, then
\beqa
\frac{{\mathcal G}^R(\omega)}{{\mathcal G}^A(\omega)} = e^{-2\pi i\nu_k }\frac{c(\nu_k)}{c^\star(\nu_k)}
\eeqa
for \emph{real} $\nu_k$, the equation in Eq.(\ref{Eq:GRS-GAS}) give the \emph{phase} of $c(\nu_k)$, and for \emph{imaginary} $\nu_k$, the equation give the \emph{modulus} of $c(\nu_k)$.
\begin{enumerate}
\item For real $\nu_k$: The ratios $G^R(\omega)/G^A(\omega)$ become a pure phase and one find that
\beqa
\gamma_k = \text{arg}[ \Gamma[-2\nu_k](e^{-2\pi \nu_k i} + e^{-2\pi q e_d})  ].
\eeqa
The factor $e^{i\gamma_k}$ and thus $c(k)$ always lies in the \emph{upper-half complex plane}, while for scalars $e^{i\gamma_k + 2\pi \nu_k i}$ always lies in the \emph{lower-half complex plane}. Namely, for $\nu_k\in(0,1/2)$,
\beqa
\gamma_k + 2\pi \nu_k > \pi, \quad \Rightarrow \quad \pi - \gamma_k < 2\pi \nu_k.
\eeqa
\item For pure imaginary $\nu_k = - i\lambda_k(\lambda_k>0)$, the ratio becomes real and give the modulus of $c(k)$.
\beqa
 e^{-4\pi \lambda_k} < \frac{e^{-2\pi \lambda_k } + e^{-2\pi q e_d}}{e^{2\pi \lambda_k} + e^{-2\pi q e_d}} = e^{2i\gamma_k} <1.
\eeqa
\item For generic $\nu_k$, ${\mathcal G}_k(\omega)$ and $G^R(\omega,k)$ have a \emph{logarithmic branch point} at $\omega=0$. One can choose the branch cut along the negative imaginary axis, i.e., the physical sheet to be $\theta\in(-\pi/2,3\pi/2)$, which resolves into a line of poles along the branch cut when going to finite temperature.
\end{enumerate}

\subsubsection{Finite Temperature case}  

For a charged scalar field at finite temperature, the near horizon region is a charged black brane in NAdS$_2\times {\mathbb R}^{d-1}$ space-time at finite charge density, the background metric and gauge field are given by Eq.(\ref{Eq:ds2-T!=0-AdS2})
\beqa
&& ds^2 = \frac{\ell_2^2}{\zeta^2}\bigg[  -\bigg(1-\frac{\zeta^2}{\zeta_0^2} \bigg)dt^2 + \bigg(1-\frac{\zeta^2}{\zeta_0^2} \bigg)^{-1} d\zeta^2  \bigg] + \frac{r_\star^2}{\ell^2} dx^2, \quad \nn\\
&& A_t(\zeta) = e_d \bigg( \frac{1}{\zeta} - \frac{1}{\zeta_0} \bigg) = \frac{e_d}{\zeta}\bigg(1 - \frac{\zeta}{\zeta_0} \bigg),
\eeqa
where the physical coordinates are listed as below
\beqa
\text{IR horizon} \quad && r_\star \le r_0 \le r < \infty, \quad \text{UV boundary} \nn\\
&& z_\star \ge z_0 \ge z > 0, \nn\\
&& \zeta_\star \ge \zeta_0 \ge \zeta >0, \nn
\eeqa
where the coordinates with $\star$, $0$ and none subscript corresponds to those radius for zero black brane(extreme one), finite temperature black brane and ordinary black brane, respectively. By definition, we have
\beqa
\zeta \equiv \frac{z_\star^2}{d(d-1)(z_\star - z)} >0, \quad \zeta_0 \equiv \frac{z_\star^2}{d(d-1)(z_\star - z_0)}, \nn
\eeqa
where $ z \equiv {\ell^2}/{r}$.
In this case, the finite temperature with respect to $\tau$ is defined by
\beqa
T= \frac{1}{2\pi\zeta_0}.  \label{Eq:T-zeta0}
\eeqa
The scalar action in the background is given by Eq.(\ref{Eq:S_phi}), from which, one can obtain the equations of motion for the complex scalar as in Eq.(\ref{Eq:EOMs_phi}),
which explicitly become
\begin{widetext}
\beqa
&&\bigg[\bigg( 1- \frac{\zeta^2}{\zeta_0^2} \bigg)^2\partial_\zeta^2 -\partial_t^2  \!+\! \frac{\ell^2\ell_2^2}{r_\star^2 \zeta^2}\bigg( 1 - \frac{\zeta^2}{\zeta_0^2}\bigg)\partial_x^2  \!+\! \frac{2iq e_d}{\zeta}\bigg(1-\frac{\zeta}{\zeta_0}\bigg)\partial_t  \!-\! \frac{2\zeta}{\zeta_0^2}\bigg( 1 - \frac{\zeta^2}{\zeta_0^2} \bigg)\partial_\zeta  \!+\! \frac{q^2 e_d^2}{\zeta^2} \bigg( 1 - \frac{\zeta}{\zeta_0}\bigg)^2  \!-\! \frac{m^2\ell_2^2}{\zeta^2}\bigg( 1 - \frac{\zeta^2}{\zeta_0^2}\bigg)  \bigg] \phi = 0. \qquad \quad \nn
\eeqa
where $\phi=\phi(\zeta,t,x)$.
In the momentum space(out/in going:$\partial_t\to \mp i \omega$, $\partial_x\to \pm ik$), the EOM can be re-expressed as
\beqa
\bigg( 1- \frac{\zeta^2}{\zeta_0^2} \bigg)\partial_\zeta\bigg[ \bigg( 1- \frac{\zeta^2}{\zeta_0^2} \bigg) \partial_\zeta\phi  \bigg]  + \bigg[ \omega \pm \frac{q e_d}{\zeta}\bigg(1-\frac{\zeta}{\zeta_0}\bigg)\bigg]^2 \phi - \bigg(  m^2 + \frac{k^2 \ell^2}{r_\star^2}\bigg)\frac{\ell_2^2}{\zeta^2}\bigg( 1 - \frac{\zeta^2}{\zeta_0^2}\bigg)\phi = 0, \qquad
\eeqa
where $\phi=\phi(\zeta,t,x)$.
In the infinite boundary conditions $z\to 0$, $\zeta \to {z_\star}/{{d(d-1)}} \ll 1$, the EOM is dominated by the singularity at $z_\star \to 0$,
\beqa
\bigg( \partial_\zeta^2  - \frac{m_{\vec{k}}^2\ell_2^2 - q^2 e_d^2 }{\zeta^2}  \bigg) \phi(\zeta) = 0. \label{Eq:EOM-scalar-IR-T!=0}
\eeqa
where for simplicity we have denoted $\phi(\zeta)=\phi(\zeta,\omega,k)$ and the equation of motion is just the same as that in Eq.(\ref{Eq:EOM-scalar-IR}). The asymptotic behavior of the solution are the same as that in Eq.(\ref{Eq:phi-IR-A-B}) with exponents indexes as those in Eq.(\ref{Eq:nu_k}). The full EOM above can be simplified as
\beqa
  \partial_\zeta^2 \phi(\zeta)  + \frac{2\zeta}{\zeta^2 - \zeta_0^2}\partial_\zeta \phi(\zeta)  + \Bigg( - \frac{m_k^2\ell_2^2}{\zeta^2\big( 1- \frac{\zeta^2}{\zeta_0^2} \big)} +  \frac{\big[\omega \mp q e_d \big(\frac{1}{\zeta}-\frac{1}{\zeta_0}\big)\big]^2}{\big(1- \frac{\zeta^2}{\zeta_0^2}\big)^2} \Bigg) \phi(\zeta)  = 0,
\eeqa
where $m_k^2$ is define in Eq.(\ref{Eq:mk2}) and $\nu_k$ is defined in Eq.(\ref{Eq:nu_k}).
In the following, we will chose the out-going wave conversion $\partial_t \to -i\omega$.
The two linearly independent solutions turns out to be 
\beqa
\phi(\zeta) &=& \bigg( \frac{\zeta+\zeta_0}{\zeta-\zeta_0}\bigg)^{\frac{i\omega\zeta_0}{2}} \bigg[ C_1 \bigg( 1+ \frac{\zeta_0}{\zeta} \bigg)^{-\frac{1}{2}-\nu_k}
{}_2F_1\bigg(\frac{1}{2}+\nu_k-iq e_d , \frac{1}{2}+\nu_k + iq e_d - i\omega\zeta_0 ; 1+2\nu_k; \frac{2\zeta}{\zeta+\zeta_0}\bigg), \nn\\
&+& C_2 (-2)^{-2\nu_k}\bigg( 1 + \frac{\zeta_0}{\zeta}\bigg)^{-\frac{1}{2}+\nu_k}{}_2F_1\bigg(\frac{1}{2}-\nu_k-iq e_d , \frac{1}{2}-\nu_k + iq e_d - i\omega\zeta_0 ; 1-2\nu_k; \frac{2\zeta}{\zeta+\zeta_0}\bigg) \bigg], \label{Eq:phi-scalar-T!=0}
\eeqa
Or, the two linearly independent solutions turns out to be 
\beqa
\phi(\zeta) &=& C_1 \bigg(1-\frac{\zeta_0}{\zeta}\bigg)^{-\frac{1}{2}+\nu_k}\bigg( \frac{\zeta+\zeta_0}{\zeta-\zeta_0} \bigg)^{i q e_d-\frac{i\omega\zeta_0}{2}}
{}_2F_1\bigg(\frac{1}{2}-\nu_k+iq e_d , \frac{1}{2}-\nu_k + iq e_d - i\zeta_0\omega ; 1-2\nu_k; \frac{2\zeta}{\zeta-\zeta_0}\bigg) \nn\\
&+& C_2 \bigg(1-\frac{\zeta_0}{\zeta}\bigg)^{-\frac{1}{2}-\nu_k} \bigg( \frac{\zeta+\zeta_0}{\zeta-\zeta_0} \bigg)^{iq e_d-\frac{i\omega\zeta_0}{2}}  {}_2F_1\bigg(\frac{1}{2}+\nu_k+iq e_d , \frac{1}{2}+\nu_k + iq e_d - i\zeta_0\omega ; 1+2\nu_k; \frac{2\zeta}{\zeta-\zeta_0}\bigg). \nn
\eeqa
where the hypergeometric function is defined by
\beqa
{}_2F_1(a,b,;c,;z)=\sum_{k=0}^{\infty}\frac{(a)_k(b)_k}{(c)_k}\frac{z^k}{k!} = 1 + \frac{ab}{c}z + \frac{a(1+a)b(1+b)}{2c(1+c)}z^2 + O(z^3).
\eeqa
In the near horizon limit, $\zeta\to \zeta_0$, the special function associated with the coefficients $C_1$ and $C_2$ in Eq.(\ref{Eq:phi-scalar-T!=0}), respectively have the following expansion
{ 
\beqa
&& {}_2F_1\bigg(\frac{1}{2}+\nu_k-iq e_d , \frac{1}{2}+\nu_k + iq e_d - i\omega\zeta_0 ; 1+2\nu_k; \frac{2\zeta}{\zeta+\zeta_0}\bigg) \approx \frac{i\pi}{\sinh(\pi\omega\zeta_0)}\Gamma[1+2\nu_k]  \nn\\
& \times & \bigg( \frac{\big(\frac{2\zeta_0}{\zeta_0-\zeta}\big)^{-i \omega\zeta_0} }{\Gamma[\frac{1}{2}+\nu_k - i q e_d]\Gamma[\frac{1}{2}+\nu_k + i q e_d - \omega\zeta_0 ] \Gamma[ 1+ i\omega\zeta_0   ]} - \frac{1}{\Gamma[ \frac{1}{2}+\nu_k + i q e_d  ]  \Gamma[ \frac{1}{2} + \nu_k - i q e_d + i \omega\zeta_0  ] \Gamma[ 1 - i \omega\zeta_0 ] }     \bigg),  \nn \\
&& {}_2F_1\bigg(\frac{1}{2}-\nu_k-iq e_d , \frac{1}{2}-\nu_k + iq e_d - i\omega\zeta_0 ; 1-2\nu_k; \frac{2\zeta}{\zeta+\zeta_0}\bigg) \approx \frac{i\pi}{\sinh(\pi\omega\zeta_0)}\Gamma[1-2\nu_k]  \nn\\
& \times & \bigg(  \frac{\big(\frac{2\zeta_0}{\zeta_0-\zeta}\big)^{-i \omega\zeta_0}}{\Gamma[\frac{1}{2}-\nu_k - i q e_d]\Gamma[\frac{1}{2}-\nu_k + i q e_d - \omega\zeta_0 ] \Gamma[ 1+ i\omega\zeta_0   ]}  - \frac{1}{\Gamma[ \frac{1}{2}-\nu_k + i q e_d  ]  \Gamma[ \frac{1}{2} - \nu_k - i q e_d + i \omega\zeta_0  ] \Gamma[ 1 - i \omega\zeta_0 ] }     \bigg).  \nn
\eeqa
}
Considering an out wave conversion $\partial_t \to -i\omega$, then it is worthy of noticing that without the expansion wave factor $\big(\frac{2\zeta_0}{\zeta_0-\zeta}\big)^{-i \omega\zeta_0} $, the solution are pure \emph{in-falling wave}, since for fixed phase,
\beqa
e^{-i\omega t} \bigg( \frac{\zeta_0 + \zeta}{\zeta_0-\zeta}\bigg)^{\frac{i\omega\zeta_0}{2}} \sim e^{-i\omega\big[ t - \frac{1}{2}\zeta_0 \log\big(\frac{\zeta_0 + \zeta}{\zeta_0 -\zeta} \big)  \big]} \quad \Rightarrow \quad \zeta = \zeta_0\tanh\bigg( \frac{t}{\zeta_0}\bigg).
\eeqa
While the solution associated with the expansion wave factor will become \emph{out-going wave}, since
\beqa
e^{-i\omega t} \bigg( \frac{\zeta_0+\zeta}{\zeta_0-\zeta}\bigg)^{\frac{i\omega\zeta_0}{2}}\big(\frac{2\zeta_0}{\zeta_0-\zeta}\big)^{-i \omega\zeta_0}  \sim e^{-i\omega [ t - \frac{1}{2}\zeta_0 \log{(\zeta_0 + \zeta)}{(\zeta_0 -\zeta)} ]} \quad \Rightarrow \quad \zeta = \sqrt{\zeta_0^2 - e^{\frac{2t}{\zeta_0}}}.
\eeqa
Thus, one needs to choose the proper coefficients so that the expansion wave factors are completely canceled and only the in-falling wave solution are present in the near horizon region. It turns out that one has to choose
\beqa
\frac{C_1}{C_2} &=& - (-1)^{2\nu_k} \frac{\Gamma[1-2\nu_k]}{\Gamma[1+2\nu_k]}\frac{\Gamma[\frac{1}{2}+\nu_k-i qe_d]}{\Gamma[\frac{1}{2}-\nu_k - i q e_d  ]} \frac{\Gamma[ \frac{1}{2} + \nu_k + i qe_d - i\omega\zeta_0  ]}{\Gamma[ \frac{1}{2} - \nu_k + i qe_d - i\omega\zeta_0  ]} \nn\\
&=& \frac{\Gamma[-2\nu_k]}{\Gamma[2\nu_k]}\frac{\Gamma[\frac{1}{2}+\nu_k-i q e_d]}{\Gamma[\frac{1}{2}-\nu_k - i q e_d]}\frac{\Gamma[\frac{1}{2}+\nu_k+i q e_d-i\omega \zeta_0]}{\Gamma[\frac{1}{2}-\nu_k + i q e_d - i \omega \zeta_0]}, \label{Eq:C1-C2-scalar}
\eeqa
where we have used that $\Gamma [1+\alpha]=\alpha \Gamma[\alpha]$ and $(-1)^{2\nu_k}=1$. In this case, the final solution of bulk equation in Eq.(\ref{Eq:phi-scalar-T!=0}) with pure in-falling wave near the horizon becomes,
{ 
\beqa
\phi(\zeta) &=& \bigg( \frac{\zeta+\zeta_0}{\zeta-\zeta_0}\bigg)^{\frac{i\omega\zeta_0}{2}} C_2 (-2)^{-2\nu_k} \bigg[
 \bigg( 1 + \frac{\zeta_0}{\zeta}\bigg)^{-\frac{1}{2}+\nu_k}{}_2F_1\bigg(\frac{1}{2}-\nu_k-iq e_d , \frac{1}{2}-\nu_k + iq e_d - i\omega\zeta_0 ; 1-2\nu_k; \frac{2\zeta}{\zeta+\zeta_0}\bigg) \nn\\
&+& \frac{C_1}{C_2} (-2)^{2\nu_k} \bigg( 1+ \frac{\zeta_0}{\zeta} \bigg)^{-\frac{1}{2}-\nu_k}
{}_2F_1\bigg(\frac{1}{2}+\nu_k-iq e_d , \frac{1}{2}+\nu_k + iq e_d - i\omega\zeta_0 ; 1+2\nu_k; \frac{2\zeta}{\zeta+\zeta_0}\bigg) \bigg].
\eeqa
}
In the infinite boundary condition($z\to 0$, $\zeta \to z_\star/(d(d-1))\ll 1$), the EOM is dominated by
\beqa
\partial_\zeta^2 \phi(\zeta)  + \frac{2\zeta}{\zeta^2 - \zeta_0^2}\partial_\zeta \phi(\zeta)  + \bigg( - \frac{m_k^2\ell_2^2}{\zeta^2 \big( 1- \frac{\zeta^2}{\zeta_0^2} \big)} +  \frac{\big[q e_d \big(\frac{1}{\zeta}-\frac{1}{\zeta_0}\big)\big]^2}{\big(1- \frac{\zeta^2}{\zeta_0^2}\big)^2} \bigg) \phi(\zeta)  = 0,
\eeqa
which give the asymptotic behavior of the solution
\beqa
\phi(\zeta)
&\approx& C_3 \bigg(\frac{\zeta}{\zeta_0}\bigg)^{\frac{1}{2}-\nu_k} \!+\! C_4 (-2)^{2\nu_k} \bigg(\frac{\zeta}{\zeta_0}\bigg)^{\frac{1}{2}+\nu_k} \ii \sim (-2)^{-2\nu_k}C_2 \bigg(\frac{\zeta}{\zeta_0}\bigg)^{\frac{1}{2}-\nu_k} +C_1 \bigg(\frac{\zeta}{\zeta_0}\bigg)^{\frac{1}{2}+\nu_k}
\ii =  
A(\omega) \zeta^{\Delta_-^{\text{IR}}} + B(\omega) \zeta^{\Delta_+^{\text{IR}}}, \qquad \quad
\eeqa
where $\Delta_\pm^{\text{IR}} = {1}/{2} \pm \nu_k$.
On the other hand, in the infinite boundary condition, Eq.(\ref{Eq:phi-scalar-T!=0}) becomes
\beqa
\phi(\zeta)  =   (-1)^{\frac{i\omega\zeta_0}{2}} \bigg[ C_1  \bigg(\frac{\zeta}{\zeta_0} \bigg)^{\frac{1}{2}+\nu_k} \!+\! C_2  (-2)^{-2\nu_k}  \bigg( \frac{\zeta}{\zeta_0} \bigg)^{\frac{1}{2}-\nu_k} \bigg]
=  (-1)^{\frac{i\omega\zeta_0}{2}} \big[ C_1 \zeta_0^{-\Delta_+^{\text{IR}}}  \zeta^{\Delta_+^{\text{IR}}} + C_2  (-2)^{-2\nu_k}  \zeta_0^{-\Delta_-^{\text{IR}}}  \zeta^{\Delta_-^{\text{IR}}} \big] , \qquad \quad
\eeqa
from which, we have $C_3 \sim (-2)^{-2\nu_k} C_2$, $(-2)^{2\nu_k} C_4 \sim C_1$. Thus, one can read the correlator functions
\beqa
&&{\mathcal G}^{A}(\omega,k) = \frac{B(\omega)}{A(\omega)} = \frac{C_4}{C_3}(-2)^{2\nu_k} \zeta_0^{(-\Delta_+^{\text{IR}}+\Delta_-^{\text{IR}})} = \frac{C_1}{C_2} (-2)^{2\nu_k} \zeta_0^{(\Delta_-^{\text{IR}}-\Delta_+^{\text{IR}})} = \frac{C_1}{C_2}\bigg(\frac{\zeta_0}{2}\bigg)^{-2\nu_k}.
\eeqa
By using Eq.(\ref{Eq:C1-C2-scalar}), we obtain the retarded Green's function as
\beqa
{\mathcal G}^{T}(\omega,k) = \bigg( \frac{\zeta_0}{2} \bigg)^{-2\nu_k} \frac{\Gamma[-2\nu_k]}{\Gamma[2\nu_k]}\frac{\Gamma[\frac{1}{2}+\nu_k-i q e_d]}{\Gamma[\frac{1}{2}-\nu_k - i q e_d]}\frac{\Gamma[\frac{1}{2}+\nu_k+i q e_d-i\omega \zeta_0]}{\Gamma[\frac{1}{2}-\nu_k + i q e_d - i \omega \zeta_0]}.
\eeqa
By using that finite temperature definition $T$ as in Eq.(\ref{Eq:T-zeta0}),
we have
\beqa
{\mathcal G}^{T}(\omega,k) = (4 \pi T)^{2\nu_k} \frac{\Gamma[-2\nu_k]}{\Gamma[2\nu_k]}\frac{\Gamma[\frac{1}{2}+\nu_k-i q e_d]}{\Gamma[\frac{1}{2}-\nu_k - i q e_d]}\frac{\Gamma[\frac{1}{2}+\nu_k+i q e_d-i \frac{\omega}{2\pi T}]}{\Gamma[\frac{1}{2}-\nu_k + i q e_d - i \frac{\omega}{2\pi T} ]} \equiv T^{2\nu_k}g_b(\nu_k, \frac{\omega}{2\pi T}),  \label{Eq:GR-AdS2-T!=0}
\eeqa
where $g_b$ is a scaling function given by
\beqa
g_b(\nu_k, x) \equiv (4 \pi )^{2\nu_k} \frac{\Gamma[-2\nu_k]}{\Gamma[2\nu_k]}\frac{\Gamma[\frac{1}{2}+\nu_k-i q e_d]}{\Gamma[\frac{1}{2}-\nu_k - i q e_d]}\frac{\Gamma[\frac{1}{2}+\nu_k+i q e_d-i x]}{\Gamma[\frac{1}{2}-\nu_k + i q e_d - i x ]}.
\eeqa
It is worthy of noticing that  at zero temperature the original branch point at $\omega = 0$ disappears and the branch cut is replaced at \emph{finite temperature} by a \emph{line of poles} parallel to the next imaginary axis. In the zero temperature limit ($T\to 0$), these pole line emerges as a branch cut. At finite low temperature, the near horizon geometry is a black brane in AdS$_2$. This \emph{IR geometry} results in the Green's functions at the finite temperature. The fermion \emph{self energy} at finite temperature becomes
\beqa
\Sigma(\omega,T) = T^{2\nu_k} g_b(\frac{\omega}{T})\sim (4\pi T)^{2\nu_k}\frac{\Gamma[\frac{1}{2}+\nu_k+ i q e_d - i \frac{\omega}{2\pi T}]}{\Gamma[\frac{1}{2}-\nu_k+ i q e_d - i \frac{\omega}{2\pi T}]}  \xrightarrow[]{T\to 0}   (4\pi T)^{2\nu_k} (-i \frac{\omega}{2\pi T})^{2\nu_k} \sim c_k \omega^{2\nu_k}.
\eeqa
In the zero temperature limit, i.e., $T\to 0$, the line of \emph{discrete poles} of the Gamma function at finite temperature emerges as a branch cut for $\omega^{2\nu_k}$ at $T=0$.
\end{widetext}

\subsection{Quantum liquid and disordered state}

At the moment, let's discuss the physical properties of the boundary field theory in the IR, namely CFT$_1$, or $(0+1)$-dimensional conformal quantum mechanics. The physics is totally characterized by the dimension of the operator $\Phi$ at the IR fixed point, which should be matched by the operator of the boundary field theory ${\mathcal O}$ from the bulk scalar field $\phi(t,x,z)$. Thus, the retarded function of $\phi_k$ at the IR fixed point can be written as
\beqa
&& {\mathcal G}_k(\omega) = c(\nu_k) (i\omega)^{2\nu_k}, \quad \nn\\
&& c(\nu_k) = 2^{2\nu_k}\frac{\Gamma[-2\nu_k]}{\Gamma[2\nu_k]}   \frac{\Gamma[\frac{1}{2}-iq e_d + \nu_k]}{\Gamma[\frac{1}{2}-iq e_d - \nu_k]},  \label{Eq:Gk-IR-scalar}
\eeqa
where $\nu_k$ is given in Eq.(\ref{Eq:nu_k}), which can be re-expressed as
\beqa
&& d\ge 2: \nu_k 
=  \sqrt{\frac{1}{4} + \bigg(m^2 + \frac{k^2 \ell^2}{r_\star^2}- \frac{q^2 e_d^2}{\ell_2^2} \bigg) \ell_2^2}, \quad \nn\\
&&  d=1: \nu_1 = \sqrt{\frac{1}{4} + m^2 - \mu^2} .
\eeqa
There are some physical consequence as listed below:
\begin{enumerate}
\item The dimension index $\nu_k$ depends on the momentum $k$, as a result the operators with larger momentum $k$ become less important in the IR.
\item The dimension index $\nu_k$ is decreasing with the charge $q$, an operator with larger $q$ will have more significant IR fluctuations.
\item As one approach the infinite AdS$_2$ boundary, the electric field \emph{linearly} blows up and becomes strong field
\beqa
A_t(\zeta)= \frac{e_d}{\zeta} \to \infty, \quad \zeta \to 0, \quad z \to \infty.
\eeqa
\item The spectrum weight scales with $\omega$ as a power for any momentum $|k|$
\beqa
\text{Im}{\mathcal 	G}_k(\omega) = (-1)^{\nu_k}\text{Im}[c(\nu_k)]  \omega^{2\nu_k} \sim \omega^{\frac{\ell \abs{k}}{\sqrt{d(d-1)}r_\star}}, \qquad
\eeqa
which indicates that \emph{the presence of low energy excitation for all momenta} (including those at larger momenta, although it will more suppressed due to larger scaling dimension).
\item The dimension index $\nu_k$ can be rewritten as
\beqa
&& \nu_k 
\equiv \frac{1}{\sqrt{d(d-1)}}\frac{\ell^2}{r_\star}\sqrt{ \frac{1}{\xi^2} + k^2},  \qquad \nn\\
&& \nu_{k=0} = \frac{1}{\sqrt{d(d-1)}}\frac{\ell^2}{r_\star}\frac{1}{\xi}. \qquad
\eeqa
where $\nu_k$ has a branch point at $k = i\xi^{-1}$, and $\xi$
is the \emph{correlation length}. By using the definition that $r_\star\equiv \ell^2/z_\star$ in energy coordinate, 
the dimension index and correlation length can be expresses as
\beqa
&& \nu_k = \frac{z_\star}{\sqrt{d(d-1)}}\sqrt{ \frac{1}{\xi^2} + k^2}, \quad \nn\\
&& \xi = \frac{z_\star}{\sqrt{d(d-1)}} \bigg[\sqrt{\frac{1}{4} + (m^2 \ell_2^2 - q^2 e_d^2)} \bigg]^{-1} . \label{Eq:nuk-xi}
\eeqa
\begin{enumerate}
\item In the limit that $|\vec{x}| \ll \xi$, $|\vec{k}|\ll \xi^{-1}$, $\nu_k \approx \nu_{k=0}$, one only need to focus on the time component of the Fourier transform from $\omega$ to $t$, the retarded function of $\phi_k$ at the IR fixed point is
\beqa
{\mathcal G}^R(t,k) & \sim & \frac{1}{t^{1+2\nu_k}} \approx 
\frac{1}{t^{2\Delta_{k=0}}} \nn \\
& \sim & t^{-1 - \frac{1}{\sqrt{ d(d-1) } } \frac{\ell^2}{r_\star}\frac{1}{\xi}  } .  \label{Eq:IR-tau-scalar}
\eeqa
\item In the limit that $|\vec{x}| \gg \xi$, $|\vec{k}|\gg \xi^{-1}$, one only needs to focus on the spatial component of the Fourier transform from $\vec{k}$ to $\vec{x}$, the correlation function decays \emph{at least} exponentially as
\beqa
{\mathcal G}^R_E(t, x) & \sim & e^{-k_0 |\vec{x}|} = e^{- \frac{|\vec{x}|}{\xi}} \nn\\
&\sim & e^{-\frac{(d-2)}{\sqrt{2}}\sqrt{\frac{1}{4} + m^2 \ell_2^2 - q^2 e_d^2}\mu|\vec{x}|}, \label{Eq:IR-xi-scalar}
\eeqa
where $E$ indicates it is in Euclidean spacetime.
\end{enumerate}
Intuitively speaking, the system are separated into independent domains of size of order $\xi$, 
according to Eq.(\ref{Eq:IR-xi-scalar}), domains separated by distances larger than $\xi$ are uncorrelated with one another. Within each of the domain, the dynamics of the domain are controlled by CFT$_1$, namely a conformal quantum mechanics in the time direction with a power law correlation, according to the Eq.(\ref{Eq:IR-tau-scalar}). Given the system has a nonzero entropy density, each cluster has a non-zero entropy that counts the number of degrees of freedom inside the domain.
\end{enumerate}

In summary, the correlation functions and the scaling dimension index in Eq.(\ref{Eq:Gk-IR-scalar}) describes a \emph{disordered state}, or a \emph{quantum liquid phase}, where the space factorizes into independent domains of correlation length $\xi$. \emph{Within each domain, one has scale invariance along the time direction}. However, it is worthy of emphasizing that the scaling behavior within each cluster here describes not the behavior of a single site, but the collective behavior of a large number of sites over a size of order $\xi$. The correlation function ${\mathcal G}_k$ and scaling dimension index $\nu_k$ depend nontrivially on $k$. Generally speaking, a generic point in parameter space the dependence of $\nu_k$ and ${\mathcal G}_k$ on $k$ is analytic and only through $k/z_\star \sim kr_\star$. Near certain \emph{quantum critical points} (QCP), the dependence of $\nu_k$ and ${\mathcal G}_k$ on $k$ at $k=0$ becomes \emph{non-analytic}.

\section{Susceptibility}
\label{app:rho-chi}

To appreciate quantum phase transition with order parameter, an important observable is \emph{susceptibilities}, which characterize the dynamical nature of the quantum phase transition. Suppose that the \emph{order parameter} is given by the \emph{expectation value of some bosonic operator ${\mathcal O}$ }, then the corresponding \emph{susceptibility $\chi(\omega,\vec{k})$} is given by the \emph{retarded function of ${\mathcal O}$}. To be concrete, the uniform/zero momentum/dynamical susceptibility~\cite{Parcollet:1999,Grempel:2001} $\chi(\omega)$ and  {momequandependent static} susceptibility $\chi(\vec{k})$ can be defined, respectively, as
\beqa
 \chi(\omega) \equiv G^R(\omega,\vec{k}=0), \quad \chi(\vec{k}) \equiv G^R(\omega=0,\vec{k}),
\eeqa
and the {full dynamical susceptibility}  $\chi(\omega,\vec{k})$ is defined as
\beqa
\quad \chi(\omega,\vec{k}) \equiv G^R(\omega,\vec{k}).
\eeqa
Then, the existence of growing modes with instability are reflected in the singularities of susceptibility in the upper complex momentum space. Thus, it can be indicated by the divergence of uniform static susceptibility $\chi(0)$, The critical behavior of quantum phase transition, is different from the Landau's phase transition where one expects that the uniform susceptibility always diverges wen approaching a critical point, e.g, near the critical point, $\chi(0)$ is characterized by a critical exponent $\gamma$, $\chi(\omega) \sim |g-g_c|^{-\gamma}$ where $g$ is the tuning parameter and $g_c$ is the value at critical point. Instead, at the critical point, the uniform static susceptibility remains  finite, i.e., $\chi(0)\sim \text{const}.$, while the singularity behavior of $\chi(0)$ around $\omega\to 0$ can be indicated by taking a derivative with respect to $\omega$, thus one finds that $\chi'(0)$ is divergent.

In momentum space, the advanced and retarded Green's function as well as the positive/negative Wightman functions are defined as~\cite{Weinberg:1995mt,Parcollet:1997ysb}
\beqa
G^A(\omega) &=& i  \int_{-\infty}^{+\infty} dt \theta(-t) e^{i\omega t} G(t), \nn\\
G^R(\omega) &=& -i  \int_{-\infty}^{+\infty} dt \theta(t) e^{i\omega t} G(t), \quad \nn\\
G^+(\omega) &=& \int_{-\infty}^{+\infty} dt  e^{i\omega t} \langle {\mathcal O}(t)  {\mathcal O}(0)  \rangle, \quad \nn\\
G^-(\omega) &=& \int_{-\infty}^{+\infty} dt e^{i\omega t} \langle {\mathcal O}(0)  {\mathcal O}(t)  \rangle.  \label{Eq:GRA-Gpm}
\eeqa
Except for the Pauli-Jordan commutator function $G(t) \equiv \langle [ {\mathcal O}(t) , {\mathcal O}(0)  ] \rangle$, there are also other real-time two-point functions such as Hadamard two point function $ G^H(t) \equiv \langle  \{ {\mathcal O}(t) , {\mathcal O}(0)  \}  \rangle$ and the Feynman two point function $ G^F(t) \equiv \theta(t) \langle  {\mathcal O}(t)  {\mathcal O}(0)  \rangle + \theta(-t) \langle  {\mathcal O}(0)  {\mathcal O}(t)  \rangle$ can be expressed by linear combinations of above Green's functions. In momentum space, the finite temperature generalization  is Fenman's Green's function as $G^F_s(\omega,\beta) \equiv (1-n_s(\omega))G^R(\omega) + n_s(\omega) G^{R\star}(\omega)$ where $n_s(\nu)$ statistic distribution functions, e.g., $s=F$ to denote Fermi-Dirac statistics for fermions, or $s=B$ to denote Bose-Einstein statistics for bosons. In imaginary time, the retarded Green's functions and the Wightman functions are defined as $G^R(\tau) = [G(\tau+\epsilon) - G(\tau-\epsilon)] \theta(t)$ and $G^\pm(\tau) = G(\tau \pm \epsilon)$, $G^\pm(t) \overset{T\ne 0}{=} G(\tau \pm \beta/2) $ with $\tau=it$. The spectral functions of quantum liquid is defined as imaginary part of retarded Green's function as
\beqa
 \rho(\omega) \propto -\frac{1}{\pi} \text{Im}G^R(\omega) . \quad \label{Eq:rho-ImGR}
\eeqa


%
The local spin-spin correlation function is directly related to the Green's function and defined as
\beqa
\chi_{\text{loc}}(\omega) \propto \int_{-\infty}^{+\infty} dt e^{i\omega t}  G^R(t) G^R(-t), \qquad  \label{Eq:chi_loc}
\eeqa
and the {local dynamical susceptibility} is defined by~\cite{Parcollet:1999}
\beqa
\chi_{\text{loc}}''(\omega) = \pi \int_{-\infty}^{+\infty} d\nu \rho_s(\nu)\rho_s(\nu-\omega)[n_s(\nu-\omega)-n_s(\nu)], \qquad
\eeqa
where $\rho_s(\nu)$ is spectral function with spin index $s$ and frequency $\nu$. The static local susceptibility $\chi_{\text{loc}}'(\omega=0)$ is defined as
\beqa
\chi_{\text{loc}}'(\omega=0) = \int d\omega \frac{\chi_{\text{loc}}''(\omega)}{\omega}.
\eeqa
In dealing with Fourier transformation for retarded Green's function with loop corrections, it is useful to introduce a set of integral formulas as follows
~\cite{Gradshteyn:2007,Parcollet:1997ysb}
\beqa
 I^{(0,1,2)}_{\Delta-\frac{i \beta  \omega }{2 \pi }}    & \equiv & \int_0^{+\infty}    \frac{ {(2\pi/\beta)} dt e^{i\omega t}}{[ \frac{\beta}{2\pi}  \sinh{\frac{t}{2}   \frac{2\pi}{\beta}}  ]^{2\Delta}} \bigg\{ 1 , \frac{2\pi}{\beta} t , \bigg( \frac{2\pi}{\beta} \bigg)^2 t^2 \bigg\} \nn\\
&=&   I^{(0)}_{\Delta-\frac{i \beta  \omega }{2 \pi }} \bigg\{ 1, - \psi^{(0)}_{\Delta-i\frac{\beta\omega}{2\pi}} , \psi^{(1)}_{\Delta-i\frac{\beta\omega}{2\pi}}+\left(\psi^{(0)}_{\Delta-i\frac{\beta\omega}{2\pi}}\right){}^2\bigg\}, \nn\\
&& \qquad 0<\Delta < \frac{1}{2}, \quad \beta>0, \quad \text{Im}\omega>0. \label{Eq:integ1-chi} \label{Eq:I-even_chi}
\eeqa
where
\beqa
I^{(0)}_{\Delta-\frac{i \beta  \omega }{2 \pi }} = \bigg( \frac{4\pi}{\beta} \bigg)^{2\Delta} \frac{ \Gamma (1-2 \Delta ) \Gamma \left(\Delta -\frac{i \beta  \omega }{2 \pi }\right)}{\Gamma \left(1-\Delta -\frac{i \beta  \omega }{2 \pi }\right)}, \label{Eq:I0}
\eeqa
and we have defined a series of new functions though digamma function $\psi(z)$,
\beqa
\psi^{(n)}_{\Delta-i\frac{\beta\omega}{2\pi}} \equiv \psi ^{(n)}\left(\Delta -\frac{i \beta  \omega }{2 \pi }\right) -\psi ^{(n)}\left(1-\Delta -\frac{i \beta  \omega }{2 \pi }\right), \qquad\quad \label{Eq:psi_n-new}
\eeqa
where $\psi ^{(n)}(z)=d^n\psi(z)/dz^n$ is the $n$-th derivative of the digamma function $\psi(z) = \psi^{(0)}(z)=\Gamma'(z)/\Gamma(z)$ which satisfy the reflection formula, $\psi(1-z) - \psi(z) = \pi  \cot (\pi  z)$.

At low frequency limit ($\omega\ll 1$), or large temperature limit ($\beta\ll 1$), the leading order behavior of the function
\beqa
\psi^{(0,1)}_{\Delta-i\frac{\beta\omega}{2\pi}} =\{ -\pi  \cot (\pi  \Delta ),\psi ^{(1)}(\Delta )-\psi ^{(1)}(1-\Delta ) \} + O(\omega\beta), \nn
\eeqa
which are constants and independent of frequency.
If one shifted $\Delta\to \Delta+1$, then
\beqa
&& I^{(0,1,2)}_{\Delta+1-\frac{i \beta  \omega }{2 \pi }}     \equiv  \int_0^{+\infty}    \frac{ {(2\pi/\beta)} dt e^{i\omega t}}{[ \frac{\beta}{2\pi}  \sinh{\frac{t}{2}   \frac{2\pi}{\beta}}  ]^{2\Delta+2}} \bigg\{ 1 , \frac{2\pi}{\beta} t , \bigg( \frac{2\pi}{\beta} \bigg)^2 t^2 \bigg\} \nn\\
&&=   I^{(0)}_{\Delta + 1 -\frac{i \beta  \omega }{2 \pi }} \bigg\{ 1, - \psi^{(0)}_{\Delta+1-i\frac{\beta\omega}{2\pi}}, \psi^{(1)}_{\Delta+1-i\frac{\beta\omega}{2\pi}}+\left(\psi^{(0)}_{\Delta+1-i\frac{\beta\omega}{2\pi}}\right){}^2\bigg\}, \nn\\
&& \qquad \Delta < \{ -1/2, 0, 1/2 \}, \quad \beta>0, \quad \text{Im}\omega>0.
\eeqa
For $\Delta\in (0,1/2)$ case, one has $I^{(0,1)}_{\Delta+1-\frac{i \beta  \omega }{2 \pi }}=0$, but $I^{(2)}_{\Delta+1-\frac{i \beta  \omega }{2 \pi }}\ne 0$. It is intuitive to observe that the high point integral can be obtained through
\beqa
I^{(n)}_{\Delta+1-\frac{i \beta  \omega }{2 \pi }}     & \equiv & \int_0^{+\infty}    \frac{ {(2\pi/\beta)} dt e^{i\omega t}}{[ \frac{\beta}{2\pi}  \sinh{\frac{t}{2}   \frac{2\pi}{\beta}}  ]^{2\Delta+2}}  \bigg( \frac{2\pi}{\beta} \bigg)^n t^n \ \nn\\
&=&  \bigg( \frac{2\pi}{\beta} \bigg)^n \frac{1}{i^n}\frac{d^n}{d\omega^n} I^{(0)}_{\Delta + 1 -\frac{i \beta  \omega }{2 \pi }} ,
\eeqa
where $n\ge 3$. As would be expected, the higher order derivatives to the digamma function, i.e., $\psi^{(n)}_{\Delta-i\beta\omega/(2\pi)}$ will be present. Similarly, one also obtains another series of non-vanishing integrals as
\beqa
&& \int_0^{+\infty} \frac{2\pi}{\beta} dt   \frac{ e^{i\omega t}  \cosh{\big(\frac{t}{2} \frac{2\pi}{\beta}\big)}}{[ \frac{\beta}{2\pi}  \sinh{\big(\frac{t}{2}   \frac{2\pi}{\beta}\big)}  ]^{2\Delta+1}} \bigg\{ 1 , \frac{2\pi}{\beta} t , \bigg( \frac{2\pi}{\beta} \bigg)^2 t^2 \bigg\} \nn\\
&=&   \frac{i\omega}{\Delta}   I^{(0)}_{\Delta-\frac{i \beta  \omega }{2 \pi }} \bigg\{ 1, -\frac{i}{\omega} \frac{2\pi}{\beta} - \psi^{(0)}_{\Delta-i\frac{\beta\omega}{2\pi}} , \nn\\
&& + \frac{2i}{\omega} \frac{2\pi}{\beta} \psi^{(0)}_{\Delta-i\frac{\beta\omega}{2\pi}}  +\big(\psi^{(0)}_{\Delta-i\frac{\beta\omega}{2\pi}}\big)^2 + \psi^{(1)}_{\Delta-i\frac{\beta\omega}{2\pi}} \bigg\}. \nn\\
&& \qquad  \Delta < \{0,1/2,1/2\}, \quad \beta>0, \quad \text{Im}\omega> 0 . 
\label{Eq:integ1-chi} \label{Eq:I-odd_chi}
\eeqa
To simplify Eq.(\ref{Eq:I-odd_chi}), we have also used the relation that
\beqa
\psi ^{(n)}(-x)-\psi ^{(n)}(1-x) = \frac{n!}{x^{n+1}},
\eeqa
as well as the relation as below
\beqa
\psi^{(0)}_{\Delta-i\frac{\beta\omega}{2\pi}}  = H_{\Delta -1 -\frac{i \beta  \omega }{2 \pi }} - H_{-\Delta -\frac{i \beta  \omega }{2 \pi }},  \nn 
\eeqa
where $H_n=\sum_{i=1}^n 1/i$ is the $n$-th harmonic number. This is due to the identity $H_z = \psi(z+1) + \gamma_E $, where $\gamma_E\approx 0.577216$ is Euler-Mascheroni constant.
In low temperature limit ($\beta \gg 1$), or large frequency limit ($\omega \gg 1$), one has
\beqa
H_{\Delta -\frac{i \beta  \omega }{2 \pi }} \! \overset{}{=} \!\gamma_E \!+ \! \log \left(-\frac{i \beta  \omega }{2 \pi }\right)+\frac{i (2\Delta + 1)\pi }{\beta  \omega }+O(\beta^{-2}) . \qquad \quad
\eeqa
In the low frequency limit ($\omega \ll 1$) (but with finite temperature), one has
\beqa
H_{\Delta -\frac{i \beta  \omega }{2 \pi }} \overset{\omega \to \infty }{=} H_{\Delta }-\frac{i \beta  \omega  \psi ^{(1)}(\Delta +1)}{2 \pi }+O\left(\omega
   ^2\right),  \quad
\eeqa
take $\Delta=\pm 1/2$ and $\Delta=\pm 1/4$ as an examples, one has
\beqa
&& H_{\frac{1}{2} -\frac{i \beta  \omega }{2 \pi }} =  2 - \log{4} -\frac{1}{4} i \pi  \beta  \omega +\frac{2 i \beta  \omega }{\pi } + O(\omega^2) , \nn\\
&& H_{-\frac{1}{2} -\frac{i \beta  \omega }{2 \pi }} = -\log (4)-\frac{1}{4} i \pi  \beta  \omega + O(\omega^2), \quad \nn\\
&& H_{\frac{1}{4} -\frac{i \beta  \omega }{2 \pi }} = -\frac{\pi }{2}+4-\log (8)  -\frac{i \beta  \omega }{2 \pi }  \left(8 C-16+\pi ^2\right) + O(\omega^2) , \nn\\
&& H_{-\frac{1}{4} -\frac{i \beta  \omega }{2 \pi }} = \frac{\pi }{2}-\log (8) -\frac{i \beta   \omega }{2 \pi } \left(\pi ^2-8 C\right) + O(\omega^2) ,
\eeqa
where $C \equiv \sum_{n=0}^\infty (-1)^n/(2n+1)^2  \approx 0.915966$ is Catalan's constant.

\end{CJK}



\end{document}